\documentclass[twoside,12pt]{article}
\usepackage{epsfig}  
\usepackage{pst-plot}
\usepackage{amssymb}
\usepackage{bm}
\newcommand{\Fd}{F^{\dagger}}

\newcommand{\GG}{\bm G \!\!\!\!G}
\newcommand{\SSigma}{\bm \Sigma \!\!\!\!\,\Sigma}
\newcommand{\ome}{\omega_{e}}
\newcommand{\anu}{\bar\nu}
\newcommand{\omnu}{\omega_{\nu}}
\newcommand{\sla}{\! \not \!}
\newcommand{\be}{\begin{equation}}
\newcommand{\ee}{\end{equation}}
\newcommand{\bea}{\begin{eqnarray}}
\newcommand{\eea}{\end{eqnarray}}

\newcommand{\ep}{\varepsilon}
\def\Re{\mathop{\rm Re}}
\def\Im{\mathop{\rm Im}}

\newcommand {\veck} {\bm k}
\newcommand {\vecq} {\bm q}
\newcommand {\vecr} {\bm r}
\newcommand {\vecP} {\bm P}
\newcommand {\vecp} {\bm p}
\newcommand {\vecj} {\bm j}
\newcommand {\vecnabla} {\bm \nabla}
\newcommand {\vecdelta} {\bm \delta}
\newcommand {\vecsigma} {\bm \sigma}
\newcommand {\vectau} {\bm \tau}
\newcommand {\vecL} {\bm L}
\newcommand{\Tr}{{\rm Tr}}
\newcommand{\Cos}{{\rm cos~}}
\newcommand{\Sin}{{\rm sin~}}

\begin{document}
\title{\bf The physics of dense hadronic matter and compact stars}
\author{Armen~Sedrakian \\ \\
Institut f\"ur Theoretische Physik,\\
Universit\"at T\"ubingen,\\
D-72076 T\"ubingen, Germany}

\maketitle

\vskip 2truecm

\begin{abstract}
This review describes the properties of hadronic phases of dense
matter in compact stars. The theory is developed within the  method 
of real-time Green's functions and is applied to study of 
baryonic matter at and above the saturation density. The non-relativistic 
and covariant theories based on continuum Green's functions and  the  
$T$-matrix and related approximations to the self-energies are reviewed.
The effects of symmetry energy, onset of hyperons and meson condensation
on the properties of stellar configurations are demonstrated on 
specific examples. Neutrino  interactions with baryonic matter 
are introduced within a  kinetic theory. We concentrate
on the classification, analysis and first principle derivation of 
neutrino radiation processes from unpaired and superfluid hadronic phases.
We then demonstrate how neutrino radiation rates from various microscopic 
processes affect the macroscopic cooling  of neutron stars and how 
the observed  $X$-ray fluxes from pulsars constrain the properties 
of dense hadronic matter.
\end{abstract}

%\vfill To appear in {Prog.\ in Particle and Nuclear Physics} (2006).
%\eject \tableofcontents
%\clearpage

\tableofcontents

\section{Introduction}\label{sec:INTRO}

Neutron stars are born in a gravitational collapse of luminous stars
whose core mass exceeds the Chandrasekhar limit for a self-gravitating 
body supported by degeneracy pressure of electron gas~\cite{CHANDRA}. 
Being the densest observable bodies in our universe they open a window 
on the physics of matter under extreme conditions of high densities, 
pressures and strong electromagnetic and gravitational fields.
Most of the known pulsars are
isolated objects which emit radio-waves at frequencies
$10^8-10^{10}$ Hz, which are pulsed at the rotation frequency of the 
star. Young objects, like the pulsar in the Crab nebula, 
are also observed through $X$-rays that are emitted from their
surface as the star radiates away its thermal energy.
Relativistic magnetospheres of young pulsars emit 
detectable non-thermal optical and gamma radiation; they could be 
sources of high-energy elementary particles. Neutron stars
in the binaries are powered by the energy of matter accreted from 
companion star.

The radio observations of pulsars stretch back to 1967 when the first pulsar 
was discovered. Since then, the observational 
pulsar astronomy has been extremely 
important to the fundamental physics and astrophysics, which is
evidenced by two Nobel awards, one for the discovery of
pulsars (Hewish 1974)~\cite{HEWISH}, 
the other for the discovery of the first neutron star
-- neutron star binary pulsar (Hulse and Taylor, 1993)~\cite{HULSE_TAYLOR} 
whose orbital decay confirmed the gravitational radiation in full 
agreement with Einstein's General Theory of Relativity. 
The measurements of neutron star (NS) masses in binaries provide 
one of the most stringent constraint on properties of the 
superdense matter.  The timing observations 
of the millisecond pulsars set an upper limit on the angular momentum that 
can be accommodated by a stable NS  - a limit that can 
potentially constrains the  properties of dense matter. 
Noise and rotational anomalies  that are superimposed on 
the otherwise highly stable rotation of these objects provide
a clue to the superfluid interiors of these stars.  

The orbiting $X$-ray satellites allow astronomers to explore the 
properties of the NS's surface, in particular its composition through   
the spectrum of thermal radiation and the transients
like thermonuclear burning of accreted material on the surface of 
an accreting NS.  
The currently operating Chandra $X$-ray satellite,  Newton $X$-ray Multi 
Mirror Mission (XMM),  Rossi $X$-ray timing explorer (RXTE) continue 
to provide new insight into compact objects, their environment and 
thermal histories. Confronting the theoretical models
of NS cooling with the $X$-ray observations constrains the properties of dense 
hadronic matter, its elementary particle content and its condensed 
matter aspects such as superfluidity and superconductivity. 

The currently operating gravitational wave observatories VIRGO, LIGO, 
GEO and TAMA are expected to detect gravity waves from various 
compact objects, the NS-NS binaries being one of the most important
targets of their search. Higher sensitivity will be achieved by the
future space-based observatory LISA, currently under construction. 
The global oscillations of isolated NS and NS in binaries with compact objects 
(in particular with NSs and black holes) are believed to be  
important sources of gravity waves, which will have the potential 
to shed light on the internal structure and composition of a NS.

The theory of neutron stars has its roots in the 1930s when it was realized 
that self-gravitating matter can support itself against gravitational collapse
by the degeneracy pressure of fermions (electrons in the case of white 
dwarfs and neutrons and heavier baryons in the case of neutron stars). The
underlying mechanism is the Pauli exclusion principle. Thus,
unlike the ordinary stars, which are 
stabilized by their thermal pressure, neutron stars owe their very existence 
to the quantum nature of matter. The theory of neutron stars has been 
rapidly developing during the past four decades since the 
discovery of pulsars. The progress in this field was 
driven by different factors:  
the studies of elementary particles and their strong and weak interactions 
at terrestrial accelerators and the parallel developments in the fundamental 
theory of matter deeply affected our current understanding of neutron stars. 
The concepts of condensed matter, such as superfluidity and superconductivity, 
play a fundamental role in the dynamical manifestation of pulsars, 
their cooling and transport properties. Another factor is the steady 
increase of computational capabilities.

This review gives a survey of the many-body theory of dense matter in 
NS. It concentrates on the uniform phases and hadronic degrees of freedom
which cover the density range $\rho_0 \le \rho \le 3\rho_0$, where $\rho_0$ is 
the nuclear saturation density. The theory is developed within the framework 
of  continuum Green's function technique at finite temperature and 
density~\cite{SCHWINGER61,KADANOFF62,KELDYSH64,ABRIKOSOV75}. 
Such an approach allows for a certain degree of coherence of
the presentation. Subsection~\ref{other_CHAP2} at the end of Chapter~2 
gives a brief summary of  methods and results omitted in the 
discussion.  It is virtually impossible to cover all the aspects 
of the NS theory even when restricting to a certain domain of densities 
and degrees of freedom.  The topics included in this review are 
not surprisingly aligned with the research interests of the author. 
It should be noted that there are very good 
textbooks~\cite{SHAPIRO_TEUKOLSKY,ZELDOVICH_NOVIKOV},
 monographs~\cite{GLEN_BOOK,WEBER_BOOK,SAAKYAN}
and recent lecture notes~\cite{BGS} 
that cover the basics of compact star theory much 
more completely than it is done in this review. Furthermore there 
are several recent comprehensive reviews that cover more specialized
topics of NS theory such as the role of strangeness in compact 
stars~\cite{WEBER_REVIEW}, cooling of neutron 
stars~\cite{PAGE_REVIEW,PETHICK}, neutrino propagation~\cite{VOLKAS},
phases of quark matter at high densities~\cite{RISCHKE,ALFORD,RAJAGOPAL}
and the equations of state of hadronic 
matter~\cite{BALDO_BOOK,LATTIMER_PRA,HH}.

The remainder of this introduction gives a brief overview of phases 
of dense matter in neutron stars. Chapter 2 starts with an introduction 
to the finite temperature non-equilibrium Green's functions theory.
Further we give a detailed account of the $T$-matrix theory in the 
background medium and discuss the precursor effects in the vicinity 
of critical 
temperature of superfluid phase transition. Extensions of the $T$-matrix 
theory to describe the superfluid phases and three-body correlations are
presented. A virial equation of state is derived, which includes the 
second and the third virial coefficients for degenerate Fermi-systems.
The formal part of this Chapter includes further a discussion of the 
Brueckner-Bethe-Goldstone theory of nuclear matter, the covariant version 
of the $T$-matrix theory (known as the Dirac-Brueckner theory). 
We further discuss isospin asymmetric nuclear matter, onset of hyperons
and meson condensates, 
and the conditions of charge neutrality and $\beta$-equilibrium that 
are maintained in compact stars. We close the chapter by a discussion of 
stellar models constructed from representative equations of state.

Chapter~3 discusses the weak interactions in dense matter 
within the covariant extension of the real-time Green's functions 
formalism described in Chapter~2. We  discuss the derivation 
of neutrino emissivities from quantum transport equations for 
neutrinos and the dominant processes that contribute to the 
neutrino cooling rates of hadronic matter. Special attention is 
paid to processes that occur in the superfluid phases of neutron stars.

Chapter~4 is a short overview of cooling simulations of compact stars. 
It contains several example simulations with an emphasis on the processes
that were discussed in Chapter~3.

\subsection{\it A brief overview of neutron star structure}

\begin{figure}[tb] 
\begin{center}
\epsfig{figure=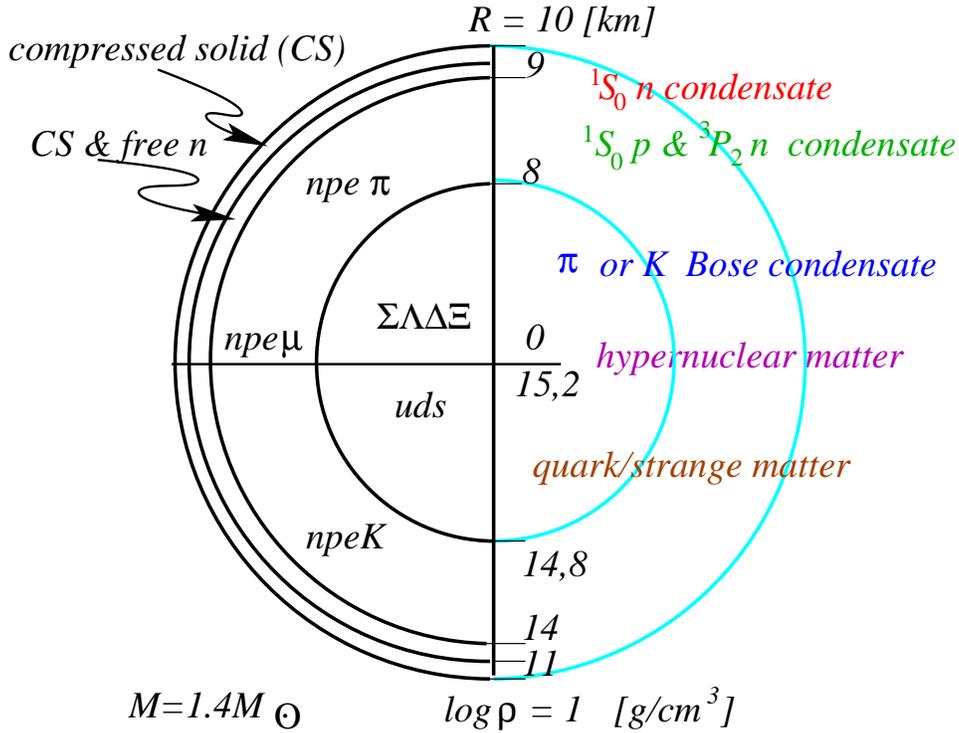,width=13.0cm,angle=0}
\begin{minipage}[t]{16.5 cm}
\caption{ Schematic cross section of a 
$M = 1.4 M_{\odot}$ mass neutron star.
The different stable phases and their constituents 
(using the standard notations)  are shown as a function of 
the radius in the upper half of the diagram and as a function 
of the logarithm of the density $\rho$ in the lower half. Note that 
the both scales are strongly expanded in the 
low-density/large-radius domain. For details see the text.
}
\label{fig:NS_CROSS}
\end{minipage}
\end{center}
\end{figure}

We now adopt a top to bottom
approach and review briefly the sequence(s) of the phases of matter 
in neutron stars as the density is increased. A schematic picture of 
the interior of a $M=1.4 M_{\odot}$ mass neutron star is shown in
Fig.~\ref{fig:NS_CROSS}. The low-density region of 
the star is a highly compressed fully ionized matter at densities about 
$\rho = 10^6$ g cm$^3$ composed of electrons and ions of $^{56}$Fe. 
This phase could be covered by several cm thick `blanket' material composed 
of H, He, and other light elements, their ions and/or molecules. 
The composition of the surface material is 
an important ingredient of the photon spectrum of the
radiation, which is used to infer the surface photon luminosities of NS.  
Charge neutrality and equilibrium with respect to  weak processes 
imply that the matter becomes neutron rich as the density is increased.
In the density range $10^7\le \rho \le 10^{11}$ g cm$^{-3}$
a typical sequence of nuclei that are stable in the ground state   
is $^{62}$Ni, $^{86}$Kr, $^{84}$Se, $^{82}$Ge, $^{80}$Zn, $^{124}$Mo,
$^{122}$Zr, $^{120}$Sr and their neutron rich 
isotopes. The matter is 
solid below the melting temperature $T_m\sim 10^7$ K, the electron 
wave functions are periodic Bloch states, and the elementary 
excitations are the electron quasiparticles and phonons. The lattice 
may also contain impurities, i.~e. nuclei with mass numbers 
different from the predicted stable nucleus, as the time-scales 
for relaxation to the absolute ground state via weak interactions 
could be very large. The transport properties of the highly 
compressed solid (CS) are fundamental to the understanding of the way the 
thermal energy is transported from the core to the surface  
and the way the magnetic fields evolve in time. 

Above the density $\rho \simeq 4\times 10^{11}$ g cm$^{-3}$ 
not all the neutrons can be bound into clusters, and those 
which are free to form a continuum of states
fill a Fermi-surface characterized by a positive chemical potential.
Thus, the ``inner crust" is a CS featuring a neutron fluid.
The sequence of the nuclei in the inner crust are neutron rich 
isotopes of Zr and Sn, with the number of protons $Z=40$ and 
$50$ respectively and mass number in the range $100 \le A \le 1500 $. 
Below the critical temperature $T_c \sim 10^{10}$~K, which corresponds
to 1~MeV (1~MeV $= 1.602\times 10^{10}$ K) neutrons in the continuum 
undergo a phase transition to the superfluid state. At low temperatures,
the electron quasiparticles and the lattice phonons 
are the relevant degrees of freedom which control the thermal and 
magnetic properties of the matter. At non-zero temperatures
neutron excitations out of condensate can play an important 
role in mass transport and weak neutral current processes.

At about half of the nuclear saturation density,
$\rho_0 = 2.8\times 10^{14}$ g  cm$^{-3}$ the clusters merge into 
continuum leaving behind a uniform fluid of neutrons, protons and 
electrons. The uniform neutron ($n$), proton ($p$) and 
electron ($e$) and possibly muon ($\mu$)   
phase extends up to densities of a few $\rho_0$; the $p$, $e$ and $\mu$ 
abundances are  in the range 5-10$\%$. The many-body theory 
which determines the energy density of matter in this density range
(as well as at higher densities) is crucial for the structure of the 
neutron stars, since most of the mass of the star resides above the nuclear
saturation density. The neutrons and protons condense in
superfluid and superconducting states below critical temperatures of the order 
$10^9$~K. Because of their low density the protons pair in the 
relative $^1S_0$ state; neutron Fermi-energies lie in the energy range 
where attractive interaction between neutrons is in the $^3P_2-^3F_2$
tensor spin-triplet channel. 
The relevant quasiparticle excitations of the $npe\mu$-phase
are the electrons and muons at low temperatures; at moderate 
temperatures the neutron and proton excitations out of the 
condensate can be important.

The actual state of matter above the densities 
$2-3\times \rho_0$ is unknown. The various possible phases are shown 
in Fig~\ref{fig:NS_CROSS}. At a given density the largest energy scale 
for charge neutral and charged particles are the Fermi-energies of 
neutrons and electrons, respectively. Once these scales become of the 
order of the rest mass of strangeness carrying heavy baryons, 
the $\Sigma^{\pm,0}$,  $\Lambda$, $\Xi^{\pm,0}$ hyperons nucleate 
in matter. Their abundances are again controlled by the equilibrium 
with respect to weak interactions and charge neutrality. Since the 
densities reached in the center of a massive NS are about 
10$\times\rho_0$, it is likely that the critical deconfinment
density at which the baryons lose their identity
and disintegrate into  up ($u$), down ($d$), and possibly strange ($s$) 
quarks is reached inside massive compact stars.
The critical density for the deconfinement transition 
cannot be calculated reliably since it lies in a range where 
quantum chromodynamics (QCD) is non-perturbative. 
NS phenomenology is potentially useful for testing 
the conjecture of high-density cold quark matter in compact stars.

At densities $\rho \ge 2\rho_0$ Bose-Einstein 
condensates (BEC) of pions ($\pi$), kaons ($K$), and heavier mesons
can arise under favorable assumptions about the effective meson-nucleon and 
nucleon-nucleon interactions in matter. For example, the
pion condensation arises because of the instability of the particle-hole 
nucleonic excitations in the medium with quantum numbers of pions. 
This instability depends on the details of 
the nuclear interaction in the particle-hole channel 
and is uncertain. There are very distinct signatures associated with
the pion and kaon BEC in the physics of NS featuring such a condensate, which
includes softening of the equation of state and fast neutrino cooling. 
 
\section{The nuclear many-body problem }\label{sec:sec2}

\subsection{\it Real-time Green's functions}\label{sec:RTG}

\begin{figure}[tb] 
\begin{center}
\epsfig{figure=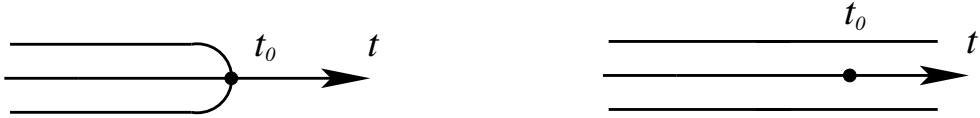,width=13.cm,angle=0}
\begin{minipage}[t]{16.5 cm}
\caption{The Schwinger (left) and Keldysh (right) contours; the axis
corresponds to the real time, $t_0$ is the observation time. 
The time flows from left to the right on the upper branch 
and from right to the left on the lower branch of the contour.
}
\label{fig:contour}
\end{minipage}
\end{center}
\end{figure}
Consider a non-relativistic Fermi system interacting via two-body forces. 
The Hamiltonian in the second quantized form is
\be\label{eq:HAMILTON}
{\cal H}
=\frac{1}{2m}\sum_{\sigma}\int d^4 x \vecnabla\psi_{\sigma}^{\dagger}(x)
\vecnabla\psi_{\sigma}(x)+\frac{1}{2}\sum_{\sigma\sigma'}
\int d^4x d^4x' \psi_{\sigma}^{\dagger}(x)
\psi_{\sigma'}^{\dagger}(x') 
V_{\sigma,\sigma'}(x,x')
\psi_{\sigma'}(x') \psi_{\sigma}(x),
\ee
where $\psi_{\sigma}(x)$ are the Heisenberg field operators, 
$x=(\vecr , t)$ is the space-time four vector,  $\sigma$ stands for 
the internal degrees of freedom (spin, isospin, etc.), 
$m$ is the fermion mass and $V(x,x')$ is the interaction 
potential, which we assume to be local in time $V(x,x') 
= V(\vecr,\vecr')\delta(t-t')$. The creation and annihilation 
operators obey the fermionic anti-commutation rules, 
$\{\psi_{\sigma'}(x'),\psi_{\sigma}^{\dagger}(x)\} = \delta(x-x')$ and 
$\{\psi_{\sigma'}(x'),\psi_{\sigma}(x)\} = 0$, and their 
equation of motion is given by 
$ 
i\partial_t \psi(x) = \left[\psi(x), {\cal H}\right],
$
where $[,]$ and $\{,\}$  stand for a commutator and anti-commutator 
(here and below $\hbar = c = 1$). The fundamental object of the theory 
is the path-ordered correlation function
\be\label{G1}
{\bm G}(1,1') = -i \langle {\cal P}\psi(1) \psi^{\dagger}(1') \rangle ,
\ee
where the path-ordering operator arranges the fields along the contour
such that that time arguments of the operators increase from left to right
as one moves along the contour shown in Fig.~\ref{fig:contour}
(here and below the boldface characters
stand for functions that are ordered on the contour). The 
unitary time evolution  operator propagates the fermionic wave 
function according to $\psi(t_2) = \hat S(t_1,t_2)\psi (t_1)$ along 
the Schwinger contour from $-\infty \to t_0 \to +\infty$, 
where $t_0$ is the observation time (Fig.~\ref{fig:contour}). The Keldysh
contour is obtained from the above one by inserting a piece
that propagates from $ t_0 \to +\infty $ and back. The Keldysh formalism 
is based on a minimal extension of the usual diagrammatic rules, where 
depending on whether the time argument lies on the upper or lower branch of 
the contour a correlation function is assigned a + or -- sign 
(per time argument). 
Below we  follow a different path, which is based on mapping 
correlation function defined on the Schwinger contour 
on a set of alternative functions which obey tractable 
transport equations.

Starting from the equation of motion for the field $\psi(x)$ and its
conjugate one can establish a hierarchy of coupled equations of motions for 
correlations functions involving increasing number of fields 
(in equilibrium this hierarchy is know as the Martin-Schwinger
hierarchy~\cite{MARTIN59}). 
For the single particle propagator the equation of motion is
\be\label{MS_HIERACHY}
{\bm G}_0(1)^{-1}{\bm G}(1,1') 
= \vecdelta (1-1')-i\int_C d2 d3 d4~V(12;34)~{\bm G}_2(34,1'2^+),
\ee
where ${\bm G}_0(1) = (i\partial_t+{\vecnabla^2}/{2m}-\mu)^{-1}$
and  ${\bm G}_2$ are the free single particle and
the two-particle propagators, $\mu$ is the chemical potential,
the notation $1^+ \equiv (x_1, t_1+0$),
the time integration goes over the contour and the counter-ordered 
delta function is defined as $\vecdelta (t-t') = \delta(t-t')$ 
if $t,t' \in C_+$,  
$\vecdelta (t-t') = -\delta(t-t')$ if $t,t' \in C_-$  and 
$\vecdelta (t-t') = 0$ otherwise; here $C_{+/-}$ refer to the 
upper and lower branches of the contour in Fig.~\ref{fig:contour}. 
To solve Eq. (\ref{MS_HIERACHY})  we need an equation of 
motion for the two-particle propagator which 
in turn depends on the three-particle propagator and so on. The
hierarchy is (formally) decoupled by defining the contour self-energy 
as
\be\label{SIGMA}
 {\bm \Sigma} (1,3) {\bm G}(3,1') 
= -i \int_C d2d4~V(12;34)~ {\bm G}_2 (34,1'2).
\ee
This leads to a closed equation for the single particle propagator, which
upon subtracting its complex conjugate takes the form 
\be\label{KIN1}
\left[{\bm G}^*_0(1') - {\bm G}_0(1)\right]
{\bm G}(1,1')=\int_C d2\left[{\bm G}(1,2)~{\bm\Sigma}(2,1')-
{\bm\Sigma}(1,2)~{\bm G}(2,1') \right].
\ee
If the time arguments of the contour ordered propagators are
constrained to the upper/lower branches of the contour we 
obtain the causal/acausal propagators of the ordinary
propagator time perturbation theory
\be\label{CAUSAL}
G^c(1,1') = -i \langle {\cal T} \psi(1)\psi^{\dagger}(1')\rangle ,
\quad 
G^a(1,1') = -i \langle {\cal A} \psi(1)\psi^{\dagger}(1')\rangle ,
\ee
where ${\cal T}$ and ${\cal A}$ are the time ordering and anti-ordering 
operators. The fundamental difference to the ordinary theory is the 
appearance of the propagators with fixed time arguments (which can be
located on either  branch of the contour)
\be\label{KB1}
G^<(1,1') = -i \langle \psi(1)\psi^{\dagger}(1')\rangle ,
\quad 
G^>(1,1') = i \langle \psi^{\dagger}(1')\psi(1)\rangle .
\ee
The propagators (\ref{CAUSAL}) and (\ref{KB1}) are not independent
\bea\label{Gc}  
G^c(1,1') &=& \theta(t_1-t_1') G^>(1,1') + \theta(t_1'-t_1)G^<(1,1'),\\
\label{Ga} 
G^a(1,1') &=& \theta(t_1'-t_1) G^>(1,1') + \theta(t_1-t_1') G^<(1,1'),
\eea
where $\theta(t)$ is the Heaviside step function.
The equilibrium properties of the system are most easily described 
by the retarded and advanced propagators, which obey integral equations 
in equilibrium. These are defined as
\bea\label{GR} 
G^R(1,1') &=& \theta(t_1-t_1')[ G^>(1,1') - G^<(1,1')],\\
\label{GA} 
G^A(1,1') &=& \theta(t_1'-t_1)[ G^<(1,1')  - G^>(1,1')].
\eea
There are six different self-energies associated with 
each propagator in Eqs. (\ref{KB1})-(\ref{GA}). 
The components of any two-point function defined 
on a time contour (in particular the single-time Green's functions 
and self-energies) obey the following relations
\bea\label{F1}
{\cal F}^R(1,1')  - {\cal F}^A(1,1') &=& {\cal F}^>(1,1')
 - {\cal F}^<(1,1'),\\
\label{F2}
{\cal F}^{R/A}(1,1') &=& \Re {\cal F}(1,1') \pm i\Im {\cal F}(1,1'),
\eea
from which, in particular, we obtain a useful relation
$2i \Im {\cal F}(1,1') = {\cal F}^>(1,1')  - {\cal F}^<(1,1').$
In practice, systems out of equilibrium are described by  
the time evolution of the distribution function which, as we shall see, 
is related to the propagator $G^{<}(1,1')$. The propagator
$G^{>}(1,1')$ is related to the distribution function of holes. 
To obtain an equation of motion for 
these functions from the equation of motion of the path ordered 
Green's function (\ref{KIN1}) we shall use algebraic relations 
for a convolution of path-ordered functions, 
\be 
C(1,1') = \int_C d2 A(1,2) B(2,1'),
\ee
known as the Langreth-Wilkins rules~\cite{LANGRETH}. These rules are stated as
\bea \label{LW1}
C^{>,<}(1,1') &=& \int d2
\left[A^R(1,2) B^{>,<}(2,1')+A^{>,<}(1,2)B^A(2,1')\right] ,\\
\label{LW2}
C^{R/A}(1,1') &=& \int d2 
\left[A^{R/A}(1,2) B^{R/A}(2,1')\right].
\eea
Upon applying the rule (\ref{LW1}) to Eq. (\ref{KIN1}) and using 
the relations (\ref{F1}) and (\ref{F2}) one obtains the Kadanoff-Baym 
transport equation~\cite{KADANOFF62}
\bea \label{KIN2}
&&\left[G_0^{-1}(1') - \Re \Sigma(1,1'), G^<(1',1) \right] -
\left[\Re G(1,1'), \Sigma (1',1) \right]  \nonumber\\
&&\hspace{4cm} = \frac{1}{2}\left\{ G^>(1,1'), \Sigma^<(1',1) \right\}
-\frac{1}{2}\left\{ \Sigma^>(1,1'), G^<(1',1) \right\}.
\eea
The first term on the l.~h. side of Eq.~(\ref{KIN2}) is the counterpart 
of the drift term of the Boltzmann equation; the second term does
not have an analog in the Boltzmann equation and vanishes in the limit 
where the particles are treated on the mass-shell. The r.~h. side of 
Eq.~(\ref{KIN2}) is the counterpart of the collision integral in the 
Boltzmann equation, whereby $\Sigma^{>,<}(1,1')$ are the collision
rates. An important property of the collision term is its symmetry 
with respect to the exchange $>~\leftrightarrow ~<$, which 
means that the collision term is invariant under the exchange 
of particle and holes. Before turning to the evaluation of the 
self-energies we briefly outline the reduction of Eq. (\ref{KIN2}) 
to the Boltzmann's quantum kinetic equation.

If the characteristic inter-collision distances are much greater 
than the inverse momenta of particles and the relaxation times are 
much larger than the inverse particle frequencies, quasiclassical 
approximations is valid. This means that the dynamics of slowly varying 
center-of-mass four-coordinate $x = (x_1+x_2)/2$ separates from the 
the dynamics of rapidly varying relative coordinate
$\xi = x_1-x_2$. One performs a Fourier transform 
with respect to the relative coordinates and expands the two-point 
functions with respect to (small) gradients of the slowly varying
center-of-mass coordinates~\cite{MALFLIET90}. 
Upon keeping the first order gradients
one obtains
\bea \label{KIN3}
i\{ \Re G^{-1}(p,x), G^<(p,x) \}_{P.B.} 
+i\{\Sigma^<(p,x), \Re G(p,x)\}_{P.B.}= 
\Sigma^<(p,x) G^>(p,x) - \Sigma^>(p,x) G^<(p,x),
\eea
where P. B. stands for the Poisson bracket defined as 
\be 
\{f,g\}_{P.B.} =
\partial_{\omega}f~\partial_t g -\partial_{t}f~\partial_{\omega} g
-\partial_{\vecp}f~ \partial_{\vecr} g +  \partial_{\vecr}
f~ \partial_{\vecp} g.
\ee
Instead of working with the functions $G^{>,<}(p,x)$ we introduce 
two new functions $a(p,x)$ and $f(p,x)$ defined by the relations, 
known as the Kadanoff-Baym (KB) ansatz, 
\bea\label{ANSATZ}
-iG^{<}(p,x) = a(p,x)f(p,x), \quad
iG^{>}(p,x) = a(p,x)\left[1-f(p,x)\right]
\eea
The KB ansatz is motivated by the Kubo-Martin-Schwinger (KMS) boundary 
condition on the Green's functions  
$G^{<}(p) = - {\rm exp}[\beta (\omega -\mu)]G^{>}(p)$, 
where $\beta$ is the inverse temperature, which is valid in {\it equilibrium}.
The KMS boundary condition is consistent with Eqs. (\ref{ANSATZ}) 
if we define $a(p) = i\left[ G^{>}(p) - G^{<}(p)\right]$
and identify the function $f(p)$ with the Fermi-Dirac distribution function 
$f_F = \left\{1+ {\rm exp}[\beta (\omega -\mu)]\right\}^{-1}$.
Thus, the KB ansatz extrapolates the exact equilibrium relations
(\ref{ANSATZ})  to non-equilibrium case, whereby 
the Wigner function $f(p,x)$ plays the role of non-equilibrium distribution
function which should be determined from the solution of appropriate 
kinetic equation. Eq. (\ref{F2}) implies that in equilibrium 
\be \label{SPEC}
a(p) =i\left[ G^{R}(p) - G^{A}(p)\right] = 
-\frac{2\Im \Sigma (p)}{[\omega - \epsilon(p) - \Re\Sigma(p)]^2 
+ [\Im \Sigma(p)]^2},
\ee
which is just the ordinary spectral function,
where $\epsilon(p)$ is the free single particle spectrum. 
In non-equilibrium the spectral function need not have the 
form (\ref{SPEC}). Furthermore, the self-energies $\Sigma(p)$ are 
functionals of the Green's functions $G^{>,<}(p,x)$ and a complete
solution of the problem requires simultaneous treatment of
functions $f(p,x)$ and $a(p,x)$. The spectral function is 
determined by the following (integral) Dyson equation
\be\label{RET}
G^{R/A}(p)  = \left[\omega -\epsilon(p) - \Sigma^{R/A}(p) \right]^{-1}.
\ee
The level of sophistication of the kinetic equation depends on the
the spectral function of the system, i.e. the form of the excitation 
spectrum. The spectral function of nuclear systems could be rather complex
especially in the presence of bound states; for many practical purposes the 
quasiparticle approximation supplemented by small damping corrections
is accurate. In this approximation~\cite{ZIMMERMAN85,SCHMIDT90}
\be\label{ASMALL}
a(p) = 2\pi z(\vecp,x) \delta(\omega-\epsilon(p)) - \gamma (p,x) 
\partial_{\omega} \left(\frac{\cal P}{\omega-\epsilon(p)}\right),
\ee
where ${\cal P}$ stands for principal value.
The first term corresponds to the quasiparticle approximation, 
while the second terms is the next-to-leading order expansion with 
respect to small $\Im\Sigma^R(p,x)$ or equivalently small damping 
$\gamma (p,x) = i\left[ \Sigma^R(p,x)-\Sigma^A(p,x)\right]$.
The wave function renormalization, within the same approximation,  
is defined as
\be\label{ZFACTOR} 
z(\vecp,x) = 1+\int \frac{d\omega'}{2\pi} 
\gamma(\omega',\vecp,x) \partial_{\omega}
\left(\frac{\cal P}{\omega'-\omega}\right)\vert_{\omega = \epsilon(p)}
= 1 -\delta\Sigma(\vecp).
\ee
Note that the approximation (\ref{ASMALL}) fulfils the spectral sum rule
\be \label{SUM_RULE}
\int \frac{d\omega}{(2\pi)} A(p,x) = 1.
\ee
We shall use below the small damping approximation (\ref{ASMALL}) 
to establish the second and third virial corrections to the 
equation of state of Fermi-systems. The kinetic equation is 
obtained upon decomposing  the Green's functions in the 
leading and next-to-leading terms in $\gamma(p,x)$~\cite{SEDRAKIAN_ROEPKE}
\be
\label{GSMALL}
G^{>,<}(p,x) = G_{[0]}^{>,<}(p,x) + G_{[1]}^{>,<}(p,x).
\ee
Substituting this decomposition in Eq. (\ref{KIN3}) one obtains 
two kinetic equations~\cite{SEDRAKIAN_ROEPKE,SLM,SL}
\bea\label{KIN4}
i\left\{{\Re G^{-1}}, G_{[0]}^{<}(p,x) -
\delta\Sigma(p,x)G_{[0]}^{<}(p,x)\right\}_{P.B.} 
&=& \Sigma^>(p,x) G^<(p,x)  -  G^>(p,x) \Sigma^< (p,x), \\
\label{KIN5}
i\left\{{\Re G^{-1}}, G_{[1]}^{<}(p,x) + 
\delta\Sigma(p,x)G_{[0]}^{<}(p,x)\right\}_{P.B.}  &=&
- i\left\{ \Sigma^<(p,x),\Re G(p,x) \right\}_{P.B.}.
\eea
We see that the second term in Eq. (\ref{KIN3}) drops out of 
the quasiparticle kinetic equation 
(\ref{KIN4}). The frequency dependence of the drift term 
of the kinetic equation (\ref{KIN4}) is now constrained 
to have a single value corresponding to the quasiparticle energy;
an integration over the frequency gives 
\be\label{KIN6}
\left[\partial_t + \partial_{\vecp}\epsilon(p,x) \partial_{\vecr}
 + \partial_{\vecr}\epsilon(p,x) \partial_{\vecp}\right] f(\vecp, x) = 
\int \frac{d\omega}{(2\pi)} 
\left[ \Sigma^>(p,x) G^<(p,x)  -  G^>(p,x) \Sigma^< (p,x)\right].
\ee
The drift term on the l. h. side has the familiar form of the 
quasiparticle Boltzmann equation; the r. h. side is an expression for the 
gain and loss terms of the collision integrals in terms of the self-energies. 
The conservation laws for particle number, momentum and energy now can 
be recovered from the kinetic equation (\ref{KIN6}); e.g. integrating 
over the momentum we obtain the particle number conservation as
\be 
\frac{dn}{dt} = 
\partial_{t} \int\frac{d^3p}{(2\pi)^3}f(\vecp,x)  
+\vecnabla\int\frac{d^3p}{(2\pi)^3}
\partial_{\vecp} \epsilon(p)f(\vecp,x)= 0.
\ee
The collision integrals must vanish in equilibrium. This constrains the form
of the self-energies $\Sigma^{>,<}(p,x)$ to be symmetric under the exchange 
$>~\leftrightarrow~<$. A fundamental requirement that follows from the 
conservation laws is that the self-energies must be 
symmetric with respect to the interchange of particles to holes. 
In other words, the kinetic theory implies that any many-body approximation 
to the self-energies needs to be particle-hole symmetric.

\subsection{\it The ladder T-matrix theory}\label{sec:TMATRIx}

\begin{figure}[tb] 
\begin{center}
\epsfig{figure=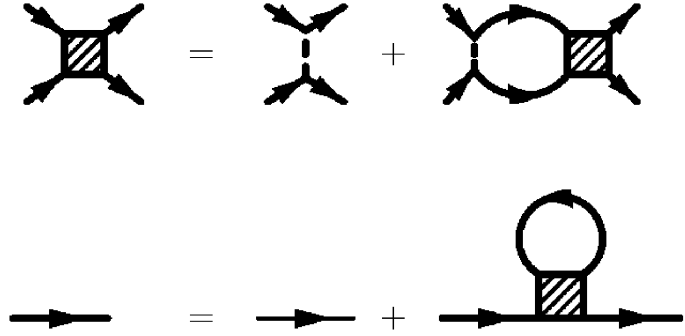,width=9.0cm,angle=0}
\begin{minipage}[t]{16.5 cm}
\caption{Coupled equations for the $T$-matrix (upper line) and the 
self-energy (lower line). The $T$-matrix is represented by the square, 
the bare interaction $V$ by a vertical dashed line, the solid lines 
correspond to single-particle Green's functions.}
\label{fig:T2}
\end{minipage}
\end{center}
\end{figure}

The nuclear interactions, which are fitted to the experimental
phase shifts and the binding energy of the deuteron, 
are characterized by a repulsive core which precludes perturbation 
theory with respect to the bare interaction. The existence of 
low-energy bound state in the isospin singlet and spin 
triplet $^3S_1-^3D_1$ state - the deuteron -
implies further that the low-energy nuclear 
interactions are non-perturbative. The $T$-matrix (or ladder) 
approximation, which sums successively the ladder diagrams
of perturbation theory to all orders, provides a good starting point 
for treating the repulsive component of nuclear interaction. The obvious
reason is that the free-space interactions are fitted to reproduce the 
experimental phase-shifts below the laboratory energies 350 MeV and the 
deuteron binding energy by adjusting the on-shell free-space $T$-matrix. 
The contour-order counterpart of the free space $T$-matrix reads 
(Fig.~\ref{fig:T2}, first line)
\be \label{TMAT1}
{\bm T}(12;34)  = {\bm V}(12;34) 
  + i\int_C d5d6 ~{\bm V}(12;34)~{\bm G}(35)~{\bm G}(46)~{\bm T}(56;34).
\ee
The time dependence of the $T$-matrix is constrained by the fact that the 
interaction is time-local ${\bm V}(12;34) = V(x_1,x_2;x_3,x_4)\delta(t_1-t_2)
\delta(t_3-t_4)$; therefore we can write 
$
{\bm T}(12;34)  = {\bm T}(x_1,x_2,t_1;x_3,x_4,t_3).
$
For the same reason, the time-structure of the propagator product 
in the kernel of Eq. (\ref{TMAT1}) is that of a single two-particle 
propagator ${\bm G}(12)~{\bm G}(23) = {\bm G_2}(12;34)$. The retarded/advanced 
components of the $T$-matrix are obtained by applying the 
Langreth-Wilkins rule~(\ref{LW2}) to Eq. (\ref{TMAT1}). 
The Fourier transform of the resulting equation is 
\be \label{T1}
{T}^{R/A}(\vecp,\vecp';P)  = {V}(\vecp,\vecp') 
  + \int\!\!\frac{d\vecp''}{(2\pi)^3} ~V(\vecp,\vecp'') G_2^{R/A}(\vecp'';P) 
  T^{R/A}(\vecp'',\vecp';P),
\ee
with the two-particle propagator
\bea\label{G2}
G_2^{R/A}(\vecp;P) &=& \int\!\!\frac{d^4P'}{(2\pi)^4}\int\!\!\frac{d\omega}{(2\pi)}
                       \left[G^>(p_+)G^>(p_-)- G^<(p_+)G^<(p_-)
                       \right] \frac{\delta(\vecP-\vecP')}{\Omega-\Omega'\pm i\eta},
\eea
where $p_{\pm} = {P}/{2}\pm p$ and the four-vector $P = (\vecP,\Omega)$ is the 
center of mass four-momentum. The remaining components 
of the $T$-matrix are given by the relations 
\be \label{OPTICAL}
T^{>,<}(\vecp,\vecp';P) = i\int\frac{d^4p_2}{(2\pi)^4}\frac{d^4p_3}{(2\pi)^4}
    T^{R}(\vecp,\frac{\vecp_2-\vecp_3}{2};P) G^{>,<}(p_2)G^{>,<}(p_3)
    T^{A}(\frac{\vecp_2-\vecp_3}{2},\vecp';P),
\ee
which can be interpreted as a variant of the optical theorem. Indeed
due to the property $T^A = [T^R]^*$ the product $T^A T^R = \vert T\vert ^2$ 
on the r. h. side of Eq. (\ref{OPTICAL})
(we use operator notations for simplicity).  
At the same time $T^> -T^< = 2i \Im T^R$ [see Eqs. (\ref{F1}) and (\ref{F2})] 
which implies that $\Im T^R \propto G_2\vert T\vert ^2$, where $G_2$ is 
defined by Eq. (\ref{G2}). Thus computing $T^> -T^<$ and comparing the result
to Eq. (\ref{OPTICAL}) we arrive at another form of the optical theorem
\bea\label{OPTICAL1} 
(2i)^{-1}T^{<}(\vecp,\vecp';P) &=& 
f(\omega)f(\omega')[1-f(\omega)-f(\omega')]^{-1}
\Im T(\vecp,\vecp';P) \nonumber\\
&=& g_B(\omega+\omega') \Im T(\vecp,\vecp';P),
\eea
where the second relation follows in equilibrium limit with $g_B(\omega) = 
[1-{\rm exp}(\beta\omega)]^{-1}$ being the Bose distribution function.
The contour ordered self-energy in the $T$-matrix approximation is defined 
as  (Fig.~\ref{fig:T2}, second line)
\be
{\bm \Sigma}(1,2) = i\int_C d3d4~{\bm T}(12;34) {\bm G}(43^+). 
\ee
Since the time dependence of the $T$-matrix is constrained by the time-locality 
of the interaction, we can immediately write-down the two components
\bea\label{SIGMA1}
i\Sigma^{>,<}(p) &=& \int\frac{d^4p'}{(2\pi)^4} T^{>,<}
\left(\frac{\vecp-\vecp'}{2}, \frac{\vecp-\vecp'}{2}; 
p+p'\right)_A G^{<,>}(p').
\eea
where the index $A$ stands for the anti-symmetrization of final 
states. Explicit expressions for the retarded and advanced components of 
the self-energy can be obtained, e.g., from the relation
$\Sigma^> -\Sigma^< = 2i \Im \Sigma^R$ and the Kramers-Kronig relation 
between the real and imaginary parts of the 
self-energies~\cite{ABRIKOSOV75}. Alternatively we can use 
the Langreth-Wilkins rules to obtain (in operator form)
\be\label{SIGMA2}
i\Sigma^{R,A} = T^{R,A} G^{<} + T^{<} G^{A,R}.
\ee
When the self-energies (\ref{SIGMA1}) are substituted in the kinetic equation 
(\ref{KIN6}) one finds the Boltzmann transport equation where the collision 
integrals are evaluated in the $T$-matrix 
approximation~\cite{KADANOFF62,MALFLIET90,SEDRAKIAN_ROEPKE}. The 
on shell scattering $T$-matrix can be directly expressed through the 
differential scattering cross-section 
\be\label{CROSS}
\frac{d\sigma}{d\Omega}(p,P) = \frac{m^{*2}}{(4\pi\hbar^2)^2}
\vert T(p,P)_A \vert^2,
\ee
where $m^*$ is the effective mass of the particle. Thus, in the dilute limit
and at not too high energies the collision integrals can be evaluated 
in a model independent way in terms of experimental elastic scattering
cross-sections. In dense, correlated systems one needs to take into account the
modifications of the scattering by the environment, in this case 
the drift and collision terms are coupled through the self-energies.

Consider now the equilibrium limit. In this limit the fermionic distribution 
function reduces to the Fermi-Dirac form. 
The number of unknown correlations functions is reduced from two to 
one because  one of the equations (\ref{ANSATZ}) 
is redundant.  For a complete description of the 
system the coupled equations for 
the $T$-matrix and self-energy need to be solved. These are given by
Eq. (\ref{T1}) where the retarded two-particle Green's function is now 
defined as  
\bea\label{G22}
G_2^{R}(\vecp;P) &=& \int\!\!\frac{d^4P'}{(2\pi)^4}
\int\!\!\frac{d\omega}{(2\pi)} a(p_+)a(p_-) Q_2(p_+,p_-)
\frac{\delta(\vecP-\vecP')}{\Omega-\Omega'+ i\eta},
\eea
where
$Q_2(p_+,p_-)  = \left[1 - f_F(p_+) - f_F(p_-)\right]$ is the Pauli-blocking
function. The retarded self-energy is given by the equilibrium limit of 
Eq. (\ref{SIGMA2}) which, upon using the optical theorem
(\ref{OPTICAL1}), becomes
\bea \label{SIGMAR}
\Sigma^R(p) &=& \int\frac{d^4p'}{(2\pi)^4} \Biggl[
T^{R}(\frac{\vecp-\vecp'}{2},\frac{\vecp-\vecp'}{2};p+p') a(p') f(\omega')
\nonumber\\
&&\hspace{3cm}+ 2g(\omega+\omega')\Im T^{R}(\frac{\vecp-\vecp'}{2},\frac{\vecp-\vecp'}{2};p+p')
\int \frac{d\bar\omega}{2\pi}\frac{a(\vecp',\bar\omega)}{\omega'-\bar\omega}
\Biggr].
\eea
Eqs. (\ref{T1}), (\ref{G22}) and (\ref{SIGMAR}) form a closed 
set of coupled  integral equations. If the interaction 
between the fermions is known these equations can be solved 
numerically by iteration. In the context  of nuclear physics 
this scheme is known also as 
the {\it Self-Consistent Green's Functions} (SCGF) 
method~\cite{KADANOFF62,RAMOS89,RAMOS91,BALDO92,DICKHOFF92,POLLS94,MUETHER95,MUETHER96,SCHNELL96,ALM,BOZEK1,BOZEK2,DEWULF01,DEWULF04,MUETHER04,RIOS}. 
Once the single particle Green's function 
(or equivalently the self-energy) is determined, the free energy of the 
system can be computed from the thermodynamic relation 
\be \label{FREE_ENERGY}
F = E -\beta^{-1}S ,
\ee
where the internal energy  is 
\be\label{ENERGY}
E =   g\int\frac{d^4p}{(2\pi)^4} \frac{1}{2}
\left[\omega + \epsilon(p)\right] a(p)f_F(\omega),
\ee  
and the entropy $S$ is given by the combinatorial 
formula
\be \label{ENTROPY}
S =  g \int \frac{d^4p}{(2\pi)^4}a(p) \left\{
f_F(\omega){\rm ln} f_F(\omega) + [1-f_F(\omega)]{\rm ln} [1-f_F(\omega)] 
\right\}.
\ee
Here $g$ is the spin-isospin degeneracy factor; $g=2$ for (unpolarized) 
neutron matter and $g=4$ for isospin symmetric nuclear matter.
An important feature of the $T$-matrix theory is that it preserves the 
particle hole-symmetry which, as we have seen, is fundamental for the
conservation laws to hold. These can be verified by integrating 
Eq. (\ref{KIN6}) with appropriate weights to recover the flow equations 
for the energy and momentum. Another attractive feature of this theory is that 
its low-density (high-temperature) limit is the free-space scattering 
theory. The latter can be constrained by experiments.
The structure of the theory and the numerical effort needed for its solution 
is simplified in this limit, since instead of 
working with the full spectral function  (\ref{SPEC}) one can approximate
it with the $\gamma(p) =  0$ limit, i.e. a $\delta$-function. 
Another interesting limit is that of low temperatures. 
If the damping is dropped, but the renormalization 
of the on-shell self-energies is retained (i.e. the real part of the
self-energy is expanded with respect to small deviations from the 
Fermi-momentum $p_F$) the spectral function reduces to
\be\label{SPEC_QP}
a(p) = 2\pi Z(\vecp)\delta(\omega-\xi(\vecp)),
\quad\quad
\xi(p) = p_F(p-p_F)/m^* - \mu^*,
\ee
where $\mu^*\equiv -\epsilon(p_F)+\mu - \Re \Sigma(p_F)$ is the 
effective chemical potential; the effective
mass and the wave function renormalization are defined as 
\bea
\frac{m^*}{m}  = 
\left(1+\frac{m}{p_F}\partial_{p}\Re\Sigma(p)\vert_{p=p_F}\right)^{-1},
\quad \quad 
Z(\vecp) = 
\left(1 -\partial_{\omega}\Re\Sigma(\omega,p_F)\vert_{\omega=\xi}\right)^{-1}.
\eea
With these approximations one recovers the elementary excitations of 
the Landau Fermi-liquid theory - the dressed quasiparticles.
Retaining the quasiparticle damping, i.e. using the small $\gamma(p)$
approximation, Eq. (\ref{ASMALL}), leads to virial corrections to 
the quasiparticle pictures. We shall discuss  these corrections in more detail
in Subsec.~{\ref{sec:VIRIAL}}. 

The $T$-matrix approximation to self-energy leads to a model which 
satisfies the conservation laws (it is said that the model is 
conserving).
In addition to being conserving any model that is based on an 
certain approximation to the self-energy needs to be thermodynamically 
consistent. The thermodynamic consistency refers to the fact that 
thermodynamic quantities like free energy or pressure computed from 
different expressions agree. An example is the Hugenholtz-van Hove
theorem~\cite{HUGENHOLZ},
which relates the single particle energy at the Fermi-surface 
to the binding energy $E_B$ at the zero temperature 
\be 
\varepsilon(p_F) = \frac{P}{\rho}+E_B,
\ee
where the pressure is defined as $P = \rho^2\partial_{\rho} E_B$.
Another example is the equivalence of the thermodynamic pressure 
defined above and the virial pressure, the latter
being the pressure calculated from the energy momentum tensor.
\begin{figure}[tb]
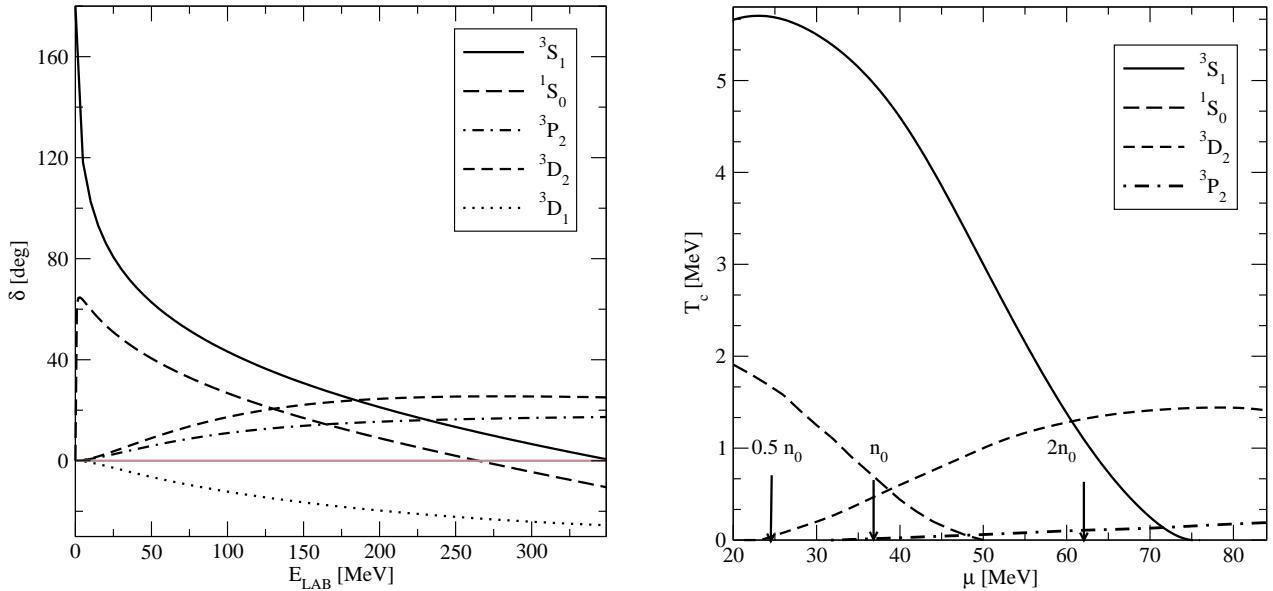
 
\begin{center}
\epsfig{figure=ppnp_fig4a.eps,width=8.cm,angle=0}
\hspace{0.7cm}
\epsfig{figure=ppnp_fig4b.eps,width=7.8cm,angle=0}
\begin{minipage}[t]{16.5 cm}
\caption{{\it Left panel.} Dependence of the
experimental scattering phase shifts in the 
$^3S_1$, $^3P_2$, $^3D_2$, and $^3D_1$ partial waves on the laboratory 
energy. {\it Right panel.} The dependence of the 
critical temperatures of superfluid phase 
transitions in the attractive channels 
on the chemical potential. The corresponding densities are indicated by arrows.
The critical temperatures $T_c$ are computed from the $T$-matrix 
instability~\cite{SEDRAKIAN95}.  
}
\label{fig:TC}
\end{minipage}
\end{center}
\end{figure}

\subsubsection{\it Pairing instability and precursor phenomena}

We have seen that in the low-density and high-temperature domain the
$T$-matrix  theory is well defined in terms of free-space parameters
and it can be used at arbitrary temperatures and densities.
In the opposite limit of high densities and low temperatures
its validity domain is restricted to the temperatures above the
critical temperature $T_c$ of superfluid phase transition.
The physical reason is that at $T_c$
there appears a bound state in the particle-particle channel - the Cooper
pair. This has far reaching consequences, since the onset of macroscopic 
coherence implies that the average value of correlation function 
$\langle \psi(x)\psi(x)\rangle \neq 0 $, which requires a doubling of the 
number of Green's function needed to describe the superfluid 
state. At temperatures  $T\ge T_c$ the $T$-matrix is strongly enhanced for 
particle scattering with equal and opposite momenta and it diverges 
at $T=T_c$.

Partial wave analysis of the nucleon-nucleon scattering allows us to 
identify the attractive channels (which feature positive phase shifts). 
The critical temperature in each channel is determined from the condition 
that the $T$-matrix, Eq.~(\ref{T1}), develops a pole for parameter values
$\tilde P = (\Omega, \vecP) = (2\mu, 0)$~\cite{ABRIKOSOV75,SCHMIDT90,ALM93,SEDRAKIAN95,ALM96}. 
To illustrate this feature assume a 
rank-one separable interaction $V(\vecp , \vecp') 
= \chi(\vecp)\chi(\vecp')$ and the quasiparticle approximation.
The solution of Eq.~(\ref{T1}), which parametrically
depends on the chemical potential and the temperature is
\be
T(\vecp, \vecp',\tilde P) = V(\vecp, \vecp')
\left[1-\int\frac{d^3p}{(2\pi)^3}  \chi^2(\vecp) G_2(\vecp,\tilde P)\right]^{-1}.
\ee
At the critical temperature both the real and imaginary parts of the 
expression in braces vanish; the zero of the real part 
determines the critical temperature $T_c$.
Fig.~\ref{fig:TC}  shows the neutron-proton 
scattering phase shifts which are relevant for the pairing 
pattern in the isospin symmetric nuclear matter 
(left panel) and the associated critical temperatures (right panel) 
determined from the $T$-matrix instability~\cite{SEDRAKIAN95}. 
For isospin symmetric systems the most attractive 
channels are the tensor channel $^3S_1-^3D_1$ and the 
$^3D_2$ channel where only the neutrons and protons interact. 
For small isospin asymmetries, which correspond to 
$\alpha = (\rho_n-\rho_p)/(\rho_n+\rho_p)\le \alpha_c \simeq 0.1$, 
where $\rho_n$ and $\rho_p$ are the neutron and proton densities, 
the mismatch in the Fermi-surfaces of neutrons and protons 
suppresses the pairing~\cite{ALM96,SEDRAKIAN97,SEDRAKIAN00,SEDRAKIAN01,LOMBARDO01,AKHISEZER,SEDRAKIAN02,ISAYEV02,SEDRAKIAN03,ALFORD05}. 
\begin{figure}[tb] 
\begin{center}
\epsfig{figure=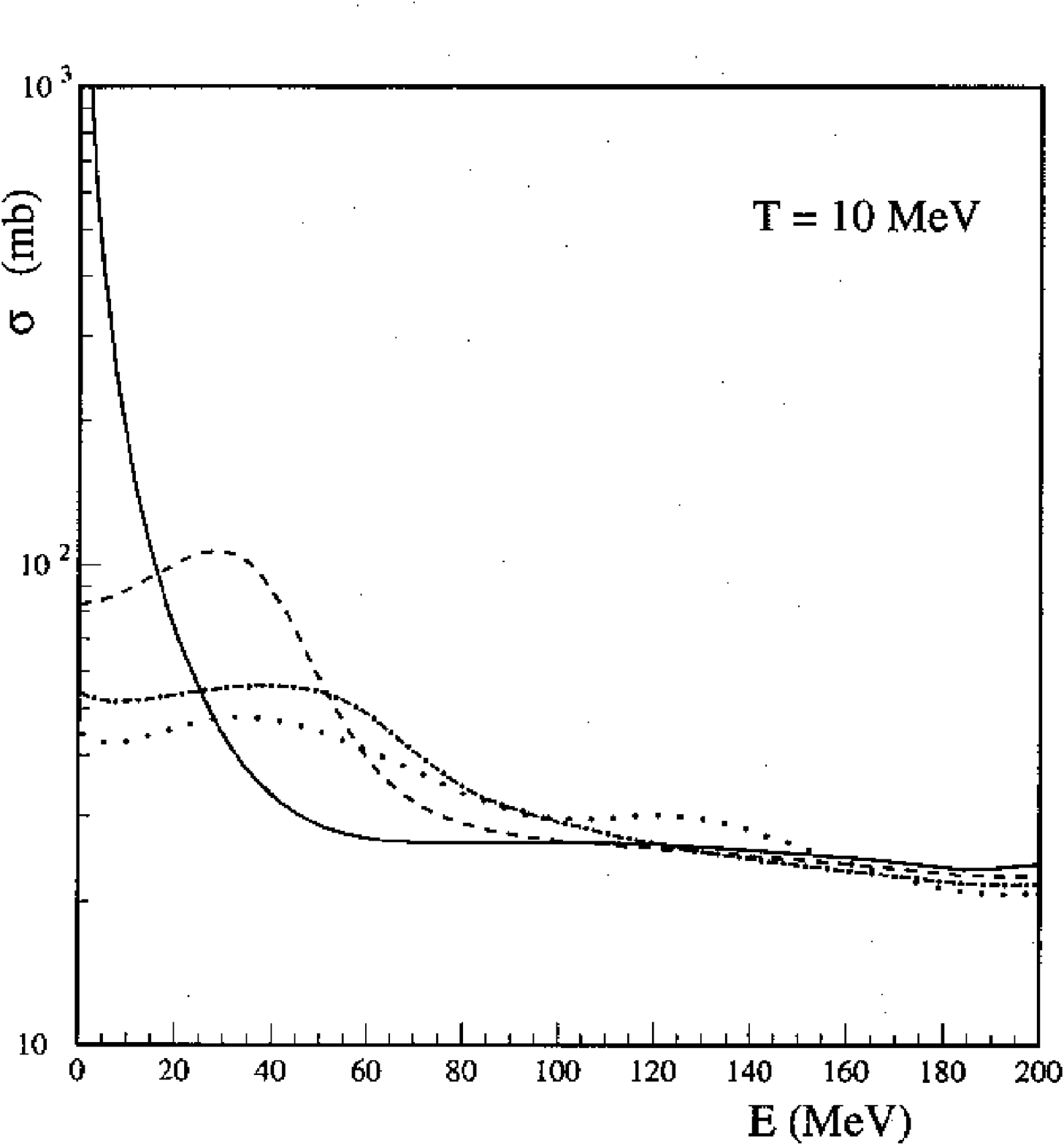,width=7.8cm,angle=0}
\hspace{0.7cm}
\epsfig{figure=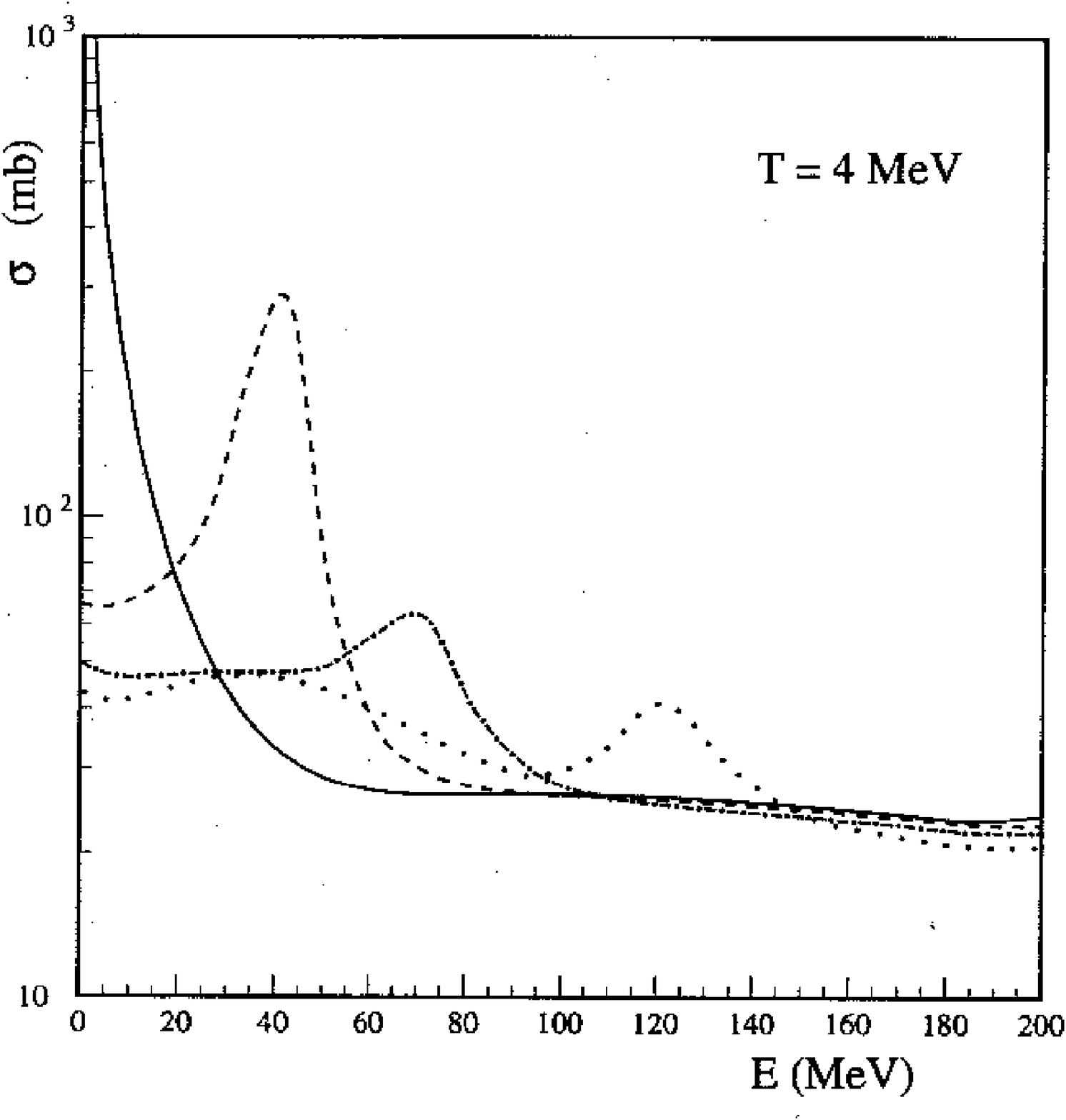,width=8.cm,angle=0}
\begin{minipage}[t]{16.5 cm}
\caption{ The cross section in neutron matter
as a function of relative energy of particles ($E = E_{\rm LAB}/2$)
for temperatures $T=10$ MeV ({\it left panel}) 
and  $T=4$ MeV ({\it right panel})
and densities $\rho = 0$ (solid line), $\rho = \rho_0/5$ (dashed line), 
$\rho = \rho_0/2$ (dashed-dotted line) 
and $\rho = \rho_0$ dotted line.
}
\label{fig:CROSS}
\end{minipage}
\end{center}
\end{figure}
For large asymmetries typical
for compact stars the pairing is among same isospin particles in the $^1S_0$ 
and $^3P_2$ channels. Because of the smallness of the charge symmetry breaking
effects, the critical temperatures in the  $^1S_0$ 
and $^3P_2$ shown in Fig.~\ref{fig:TC} channels are representative for 
neutron star matter as well (note however that the relation between the
density and the chemical potential changes)~\cite{LOMBARDO_SCHULZE,DEAN}. 
Some models of neutron star
matter which predict kaon condensation at high densities feature isospin symmetric
nucleonic matter, in which case the high-density $D$-wave neutron-proton
paring will dominate the $P$-wave neutron-neutron pairing~\cite{ALM96}.  
The scattering characteristics of the system such as the phase-shifts and
the scattering cross-sections are affected by the pairing instability, since
these are directly related to the on-shell $T$-matrix. The phase-shifts in 
a pairing channel change by $\pi/2$ at the critical temperature when the energy 
equals $2\mu$. According to the Levinson theorem, this corresponds to 
the appearance of bound state (Cooper pair). For many practical applications
the cross-section is the relevant quantity. 
According to Eq. (\ref{CROSS}) the 
cross-section being proportional to the $T$-matrix 
will diverge at 
$T_c$~\cite{SCHMIDT90,SEDRAKIAN95,SEDRAKIAN94,SCHULZE97,SCHNELL98,DICKHOFF99,BAO}.
The precursor effect of the superfluid phase transition on the neutron-neutron 
scattering cross-section in the low temperature neutron matter 
is shown in Fig~\ref{fig:CROSS}~\cite{SEDRAKIAN94}. The cross-section 
develops a spike for lower temperature 
as a precursor of the onset of superfluid in neutron matter in the $^1S_0$ 
interaction channel. 
The largest enhancement is seen for the density $\rho= 0.5 \rho_0$
which is closest to the maximum of the critical temperature as a function of 
density. The above precritical behavior of the cross-section has a significant
effect on the transport and radiation processes in matter, for example,
it could lead to a critical opalescence in the transport phenomena. 

\subsubsection{\it  $T$-matrix theory in the superfluid phase}
\label{sec:SUPT}
We have seen in the previous section that at the critical temperature of 
superfluid phase transition the two-body scattering $T$-matrix develops
a singularity, which is related to the instability of the normal 
state with respect to formation of Cooper pairs; this is manifested in 
pole of the two-body $T$-matrix when the relative energy of interacting 
fermions is twice their chemical potential. Thus, the $T$-matrix theory 
described above breaks down at the temperature $T = T_c$. A $T$-matrix 
theory appropriate for temperatures below $T_c$ can be formulated in 
terms of the normal and anomalous Green's functions~\cite{ABRIKOSOV75}.
To account for pair correlation we represent each Green's function 
in the Keldysh-Schwinger formalism as a $2\times 2$ matrix in 
the Gor'kov space:
\be
{\GG} (x,x') = \left( \begin{array}{cc}
 {\bm G}_{\alpha\beta}(x,x') & {\bm F}_{\alpha\beta}(x,x')\\
-{\bm F}^{\dagger}_{\alpha\beta}(x,x') &
\tilde {\bm G}_{\alpha\beta}(x,x')\\
\end{array}
\right) =
\left( \begin{array}{cc}
-i\langle T\psi_{\alpha}(x)\psi_{\beta}^{\dagger}(x')\rangle
&\langle \psi_{\alpha}(x)\psi_{\beta}(x')\rangle \\
\langle \psi_{\alpha}^{\dagger}(x)\psi_{\beta}^{\dagger}(x')\rangle
&-i\langle \tilde T\psi_{\alpha}(x)\psi_{\beta}^{\dagger}(x')\rangle\\
\end{array}
\right),
\ee
where ${\bm G}_{\alpha\beta}(x,x')$ and ${\bm
  F}^{\dagger}_{\alpha\beta}(x,x')$  are referred to as
the normal and anomalous propagators.  The $4\times 4$ matrix
Green's function satisfies the familiar Dyson equation
\be\label{DYSON}
{\GG}_{\alpha\beta}(x,x') =
{\GG}^0_{\alpha\beta}(x,x')
+ \sum_{\gamma , \delta}\int\!\!d^4x'' d^4x'''
{\GG}^0_{\alpha\gamma}(x,x''')
{\SSigma}_{\gamma\delta}(x''',x'')
{ \GG}_{\delta\beta} (x'',x'),
\ee
where the free propagators ${\GG}^0_{\alpha\beta}(x,x')$
are diagonal in the Gor'kov space.  We consider below
uniform fermionic systems; the propagators now depend only 
on the difference of their arguments due to translational symmetry.  
A Fourier transformation of Eq.~(\ref{DYSON}) 
with respect to the difference of the space arguments of the 
two-point correlation functions leads to on- and off-diagonal 
Dyson equations
\begin{figure}[tb] 
\begin{center}
\epsfig{figure=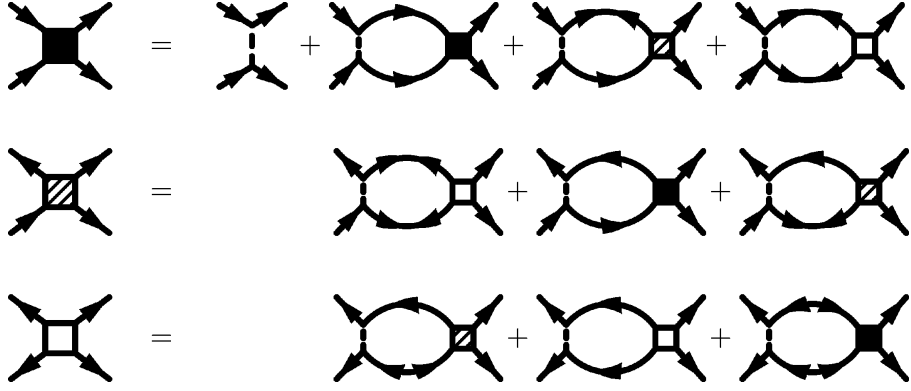,width=12.cm,angle=0}
\begin{minipage}[t]{16.5 cm}
\caption{The ladder series in a superfluid fermionic system. 
The dashed lines correspond to the driving interaction, single 
arrow lines to normal propagators and double arrow lines to 
anomalous propagators. The filled square is the counterpart 
of the unpaired state $T$-matrix, the shaded and empty squares
are specific to superfluid state.
}
\label{fig:SUP_LADDERS}
\end{minipage}
\end{center}
\end{figure}
\bea\label{eq:G1}
 {\bm G}_{\alpha\beta}(p) &=&    {\bm G}_{0\alpha\beta}(p) +
 {\bm G}_{0\alpha\gamma}(p) \left[{\bm\Sigma}_{\gamma\delta}(p)
{\bm G}_{\delta\beta}(p)+{\bm\Delta}_{\gamma\delta}(p)
{\bm F}d_{\delta\beta}(p) \right],\\
\label{eq:G2}
{\bm\Fd}_{\alpha\beta} (p) &=&   {\bm G}_{0\alpha\gamma}(-p)\left[
{\bm\Delta}^{\dagger}_{\gamma\delta}(p){\bm G}_{\delta\beta}(p)
+ {\bm\Sigma}_{\gamma\delta}(-p)
{\bm F}{\bm \Delta}_{\delta\beta} (p) \right],
\eea
where $p$ is the four-momentum, ${\bm G}_{0\alpha\beta}(p)$ is the
free normal propagator, and ${\bm \Sigma}_{\alpha\beta}(p)$ and
${\bm \Delta}_{\alpha\beta}(p)$ are the normal and anomalous
self-energies.  Summation over repeated indices is understood.
Specifying the self-energies in terms of the propagators
closes the set of equations consisting of (\ref{eq:G1}) and (\ref{eq:G2}) and
their time-reversed counterparts [$\tilde {\bm G}_{\alpha\beta}(p)$ and 
${\bm F}_{\alpha\beta}(p)$]. 
The particle-particle scattering in the superfluid state is described 
by three topologically different vertices shown 
in Fig.~\ref{fig:SUP_LADDERS}~\cite{SEDRAKIAN06}. 
We write out the explicit expression for the retarded components
of the $T$-matrix in operator form [their form in the momentum space
in identical to Eq. (\ref{T1})]
\bea \label{SUPT1}
T^{(1)}&=&  V \left[1 
            +   S_{GG}T^{(1)} +  S_{FG}T^{(2)}+ S_{FF} T^{(3)}\right],\\
\label{SUPT2}
T^{(2)}&=&  V \left[
              S_{FF}T^{(3)} +  S_{GG}T^{(1)}+ S_{FG} T^{(2)}
                 \right],\\
\label{SUPT3}
T^{(3)}&=&  V \left[S_{FG}T^{(2)} 
            +  S_{GG}T^{(3)}+ S_{FF} T^{(1)}\right],
\eea
where the two-particle retarded propagators are defined as 
\bea 
S_{GG} =  G^>G^>- G^<G^<, \quad 
S_{GF} = G^>F^>- F^<G^<, \quad 
S_{GF} = F^>F^> - F^< F^< ,
\eea
and their structure in the momentum space is given by Eq. (\ref{G2}).
To close the system of equations we define the normal and anomalous 
self-energies for the off-diagonal elements of the Schwinger-Keldysh 
structure
\bea\label{S1}
i\Sigma^{>,<} &=&  T^{(1)>,<} G^{<,>} +T^{(2)>,<} F^{<,>}, \\
\label{S2}
i\Delta^{>,<} &=&  T^{(3)>,<} F^{<,>} +T^{(2)>,<} G^{<,>}, 
\eea
which are shown in Fig.~\ref{fig:SUP_SELF}. 
\begin{figure}[tb] 
\begin{center}
\epsfig{figure=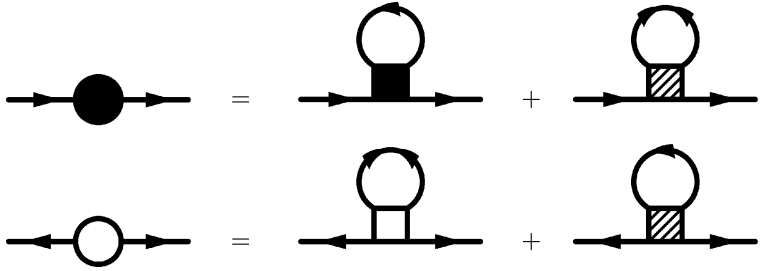,width=10.cm,angle=0}
\begin{minipage}[t]{16.5 cm}
\caption{ The normal (filled circle) and anomalous (empty circle) 
self-energies in the superfluid $T$-matrix theory. The 
self-energies couple through the $T$ matrices appearing in 
Fig.~\ref{fig:SUP_LADDERS}.
}
\label{fig:SUP_SELF}
\end{minipage}
\end{center}
\end{figure}
The retarded components of the 
self-energies which solve Dyson equations (\ref{eq:G1}) and (\ref{eq:G2})
can be constructed from Eqs. (\ref{S1}) and (\ref{S2}) via the dispersion 
relations, e.g., 
\be 
\Delta^R(\omega) = \int_{-\infty}^{\infty} \frac{d\omega'}{2\pi}
\frac{\Delta^>(\omega')-\Delta^<(\omega')}{\omega-\omega'+i\eta},
\ee
with an analogous relation for $\Sigma^R(\omega)$. The system of Eqs. 
(\ref{SUPT1})-(\ref{SUPT3}) can be used to derive the collective 
excitations of the superfluid nuclear matter in particle-particle 
channel. In the mean field approximation the self-energies decouple 
from the $T$-matrix equations (\ref{SUPT1})-(\ref{SUPT3}) and the
secular equation determining the frequencies of collective modes for 
vanishing center-of-mass momentum is
\be \label{SECULAR}
\left[A + (4\Delta^2-\omega^2) B\right]
\left[A -  (4\Delta^2 A+\omega^2) B + 2\Delta^2\omega^2B^2\right] = 0,
\ee
where, assuming that the pairing interaction can be approximated by 
a constant and the integrals regularized by a cut-off, one finds
\bea 
B(\omega) = \lambda \int_{\Delta}^{\infty}\frac{1-2f_F(\ep)}
{\sqrt{\ep^2-\Delta^2}}\frac{d\ep}{\omega^2-4\ep^2+i\omega\eta},\quad
A = 1 + \lambda \int_{\Delta}^{\Lambda} \frac{d\ep}{\sqrt{\ep^2-\Delta^2}}
[1-2f_F(\ep)],
\eea
where $\lambda$ is an effective coupling constant, $\Lambda$ is an ultraviolet
cut-off. The second equation is the stationary gap equation, i.e. $A = 0$;
the secular equation then leads to the solutions $B(\omega) = 0$,  
$B(\omega) = (2\Delta^2)^{-1}$, $\omega^2 = 0$ and $\omega = \pm 2\Delta$.
Among the first two non-trivial conditions the second one does not have 
a solution for weakly coupled systems $\lambda \ll 1$ and the collective 
modes are determined by the secular equation $B(\omega) = 0$, whereby the
real part of the solution determines the eigenmodes and the imaginary 
part their damping.

\subsubsection{\it Three-body $T$-matrix and bound states}

Up to now we were concerned with the correlations described
by the two-body $T$-matrix. The properties of dilute fermions 
or cold Fermi-liquids (the latter are characterized by a filled 
Fermi-sea) are well described in terms of two-body correlations 
between particles or quasi-particle excitations. 
However, the three-body correlations, which are next in 
the hierarchy, are important under certain circumstances. 
We turn now to the three-body problem in Fermi-systems within
the formalism developed in the previous sections. As is well known, 
the non-relativistic three-body problem admits exact free space
solutions both for contact and finite range 
potentials~\cite{SKORNYAKOV,FADDEEV}. 
Skorniakov--Ter-Martirosian--Faddeev equations sum-up the 
perturbation series to all orders with a driving term corresponding 
to the two-body scattering $T$-matrix embedded in the 
Hilbert space of three-body states.
The counterparts of these equations in the many-body theory
were first formulated by Bethe~\cite{BETHE} to access the 
three-hole-line contributions to the nucleon self-energy and the
binding of nuclear matter (Bethe's approach 
is discussed in Subsection~\ref{sec:BBG}). 
More recently, alternative forms of the three-body equations in 
a background medium have been developed that use either 
an alternative driving force (the particle-hole interaction or 
scattering $T$-matrix)~\cite{SEDRAKIAN_ROEPKE,SCHUCK,DUKELSKY,BLANKLEIDER} 
or/and  adopt an alternative version of the free-space 
three-body equations, known as the Alt-Grassberger-Sandhas 
form~\cite{AGS,BEYER1,BEYER2}.

The resummation series for three-body scattering amplitudes 
can be written down in terms of the three-body 
interaction ${\cal V}$ as~\cite{SEDRAKIAN_ROEPKE}
\bea\label{TFULL}
{\cal T}=  {\cal V} +   {\cal V}\,    {\cal G}\,   {\cal V}
=   {\cal V} +   {{\cal V}}\,   {{\cal G}}_0\,  {{\cal T}},
%=   {{\cal V}} +  {{\cal T}}\,   {{\cal G}}_0\,   {{\cal V}}
\eea 
where  ${\cal G}_0$  and ${\cal G}$ are the free and full three-particle 
Green's functions
(we use the operator form for notational simplicity; each operator, as in the 
two-particle case, is ordered on the contour).  If the three-body forces that 
act simultaneously between the three-particle are neglected, the interaction 
in Eq. (\ref{TFULL}) is simply the sum of pairwise interactions: 
${{\cal V}} =   {{\cal V}}_{12}+  {{\cal V}}_{23}
+  {{\cal V}}_{13}$, where ${{\cal V}}_{\alpha\beta}$ is the 
interaction potential between 
particles $\alpha$ and $\beta$.  The kernel of Eq. (\ref{TFULL})
is not square integrable: the potentials ${{\cal V}}_{\alpha\beta}$
introduce delta-functions due to momentum conservation for the spectator 
non-interacting particle and the iteration series contain  singular terms
(e.g., of type ${{\cal V}}_{\alpha\beta}{{\cal G}}_0 {{\cal V}}_{\alpha\beta}$ 
to  the lowest order in the interaction).
\begin{figure}[tb] 
\begin{center}
\epsfig{figure=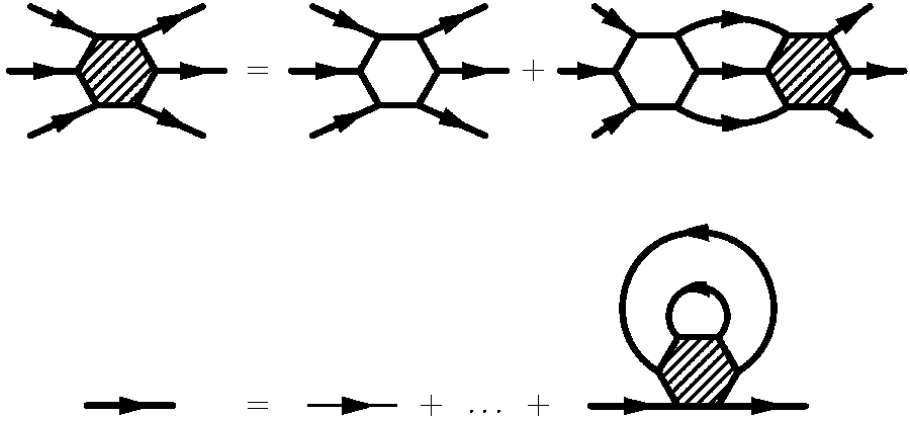,width=12cm,angle=0}
\begin{minipage}[t]{16.5 cm}
\caption{ 
Coupled integral equations for the three-body ${\cal T}$-matrix (first line)
and the single particle propagator (second line). The shaded
vertex stands for the amplitude $T^{(i)}$, the empty vertex stands for 
a channel $T_{kj}$ matrix. The second line shows the Dyson 
equation for the single particle Green's function including 
the contribution from three-body scattering. The dots stand for 
the contribution from the two-body scattering shown in Fig~\ref{fig:T2}. 
}
\label{fig:T3}
\end{minipage}
\end{center}
\end{figure}
The problem is resolved by summing up the ladder series in a 
particular channel (specified by the indices $\alpha,\beta$) 
to all orders~\cite{FADDEEV}. This summation
defines the channel $ {{\cal T}}_{\alpha\beta}$-matrix, 
which is essentially the two-body $T$-matrix embedded 
in the Hilbert space of three-particles states
\be\label{TDOWN}
  {{\cal T}}_{\alpha\beta}=    {{\cal V}}_{\alpha\beta} +  
 {{\cal V}}_{\alpha\beta}\,    {{\cal G}}_0\,    {{\cal T}}_{\alpha\beta}.
\ee
The three-body ${\cal T}$-matrix can be decomposed as 
$
  {{\cal T}} =   {{\cal T}}^{(1)}+   {{\cal T}}^{(2)} +   {{\cal T}}^{(3)}, 
$
where 
\bea\label{TUP}
  {{\cal T}}^{(\alpha)} &=&   {{\cal V}}_{\beta\gamma} +
  {{\cal V}}_{\beta\gamma}\,    {{\cal G}}_0\,    {{\cal T}},
\eea
and $\alpha\beta\gamma=123,\, 231,\, 312$. Now,  Eqs.
(\ref{TDOWN}) and (\ref{TUP}) are combined to eliminate the interaction 
terms ${\cal V}_{\alpha\beta}$ and one is left with three coupled
integral equations for ${{\cal T}}^{(\alpha)}$ ($\alpha = 1,2,3$)
\be\label{FTFA}
 {{\cal T}}^{(\alpha)}=   {{\cal T}}_{\beta\gamma} +
 {{\cal T}}_{\beta\gamma}\,   { {\cal G}}_0\, \left(
 {{\cal T}}^{(\beta)}+ { {\cal T}}^{(\gamma)}\right),
\ee
where the driving terms are the channel $T$-matrices.
The new equations are non-singular  Fredholm type-II integral equations.
Note that their formal structure is identical to the Faddeev equations in the  vacuum~\cite{FADDEEV}, however their physical meaning is different.
To see the physical content of Eqs. (\ref{FTFA}) we need to convert the contour
ordered equations into equations for the components (so that the KB ansatz 
(\ref{ANSATZ}) can be applied) and to 
transform them from the operator form into momentum representation. 
Proceeding as in Subsection~{\ref{sec:RTG}}, the retarded component of
Eq. (\ref{TUP}) reads
\bea\label{FTFA0} 
{\cal T}^{R\,(1)}(t, t')
= \ {\cal T}^{R}_{23}(t, t') 
+\int \left[ \ {\cal T}^{R\, (2)}(t, \bar t)  
+ {\cal  T}^{R\, (3)}(t, \bar t)\right] {\cal G}_0^{R}(\bar t, t'') 
{\cal T}^{R}_{23}(t'', t')  d\bar t  dt'',
\eea
were we used the time-locality of the interaction and omitted the momentum 
arguments of the functions (for the explicit expressions see ref.~\cite{SEDRAKIAN_ROEPKE}). Next, to apply the KB ansatz we need
to specify the particle hole content of the three-body $T$-matrix, 
i.~e. assign each incoming/outgoing state a particle or a hole.
Figure~\ref{fig:T3} shows the Feynman diagram for the three-body 
$T$-matrix where all the incoming (outgoing) states are particles (holes).
The remaining three-body $T$-matrices are obtained by reverting the 
direction of the arrows in the diagram. Depending on the particle-hole 
content of the three-body $\cal T$-matrix  (in the sense above) the 
intermediate state retarded Green's function is 
\bea\label{G3} 
{\cal G}_{0}^R(t_1,t_2) &=& 
\theta (t_1-t_2) \left\{
\begin{array}{cl}
G^{>}G^{>}G^{>}(t_1,t_2) - (> \, \leftrightarrow \, <) & ~(3p)\\
G^{>}G^{>}G^{<}(t_1,t_2) - (> \, \leftrightarrow \, <) & ~(2ph)\\
G^{>}G^{<}G^{<}(t_1,t_2) - (> \, \leftrightarrow \, <) & ~(p2h)\\
G^{<}G^{<}G^{<}(t_1,t_2) - (> \, \leftrightarrow \, <) & ~(3h)\\
\end{array}
\right.
\eea
where $p$ and $h$ refer to particle and hole states, the
brackets [e.~g. (2ph)] indicate the particle-hole content of 
Green's function;
for simplicity the time argument in a product is shown only once.
The short hand $> \, \leftrightarrow \, <$ stands for
a term where all the $G^<$ and $G^>$ functions are interchanged.
Upon applying the KB ansatz and Fourier transforming
Eq. (\ref{FTFA0}) one finds
\bea\label{FTFA1}
{\cal T}^{R\,(1)}(\Omega)
= \ {\cal T}^{R}_{23}(\Omega')
+\int \left[ \ {\cal T}^{R\, (2)}(\Omega')  
+ {\cal  T}^{R\, (3)}(\Omega')\right] 
\frac{Q_3(\Omega')}{\Omega-\Omega'+i\eta}
{\cal T}^{R}_{23}(\Omega') d\Omega',
\eea
where the the four-momentum space is  spanned in terms of Jacobi coordinates,
$K = p_{\alpha}+p_{\beta}+p_{\gamma},\quad k_{\alpha\beta} 
=(p_{\alpha} - p_{\beta})/2,\quad q_{\gamma} = (p_{\alpha}+p_{\beta})/3 
- 2p_{\gamma}/3,$ the center-of-mass energy $\Omega \equiv K_0$ and 
\be\label{Q3} 
Q_3(p_{\alpha},p_{\beta},p_{\gamma}) = a(p_{\alpha})a(p_{\beta})a(p_{\gamma})
\left\{[1-f_F(p_{\alpha})][1-f_F(p_{\beta})][1-f_F(p_{\gamma})] - f_F(p_{\alpha})f_F(p_{\beta})f_F(p_{\gamma})
\right\}.
\ee
This form of three-body equation incorporates off-mass-shell propagation
if the spectral function is taken in the form (\ref{SPEC}). The 
quasiparticle (on-mass-shell propagation) limit follows by 
using (\ref{SPEC_QP}) in Eq. (\ref{Q3}). Thus, the many-body environment 
modifies the three-body equation in a twofold way: 
first, the single-particle spectrum 
is renormalized in the resolvent of Eq. (\ref{FTFA1}) [which becomes
explicit after taking the quasiparticle limit], second, the intermediate
state propagation is statistically occupied according to Eq. (\ref{Q3}).
The limit $Q\to 1 $ and $a(\omega) = 
2 \pi \delta(\omega-\epsilon(\vecp))$ 
recovers the original Faddeev equations.
\begin{figure}[tb]
\begin{center}
\epsfig{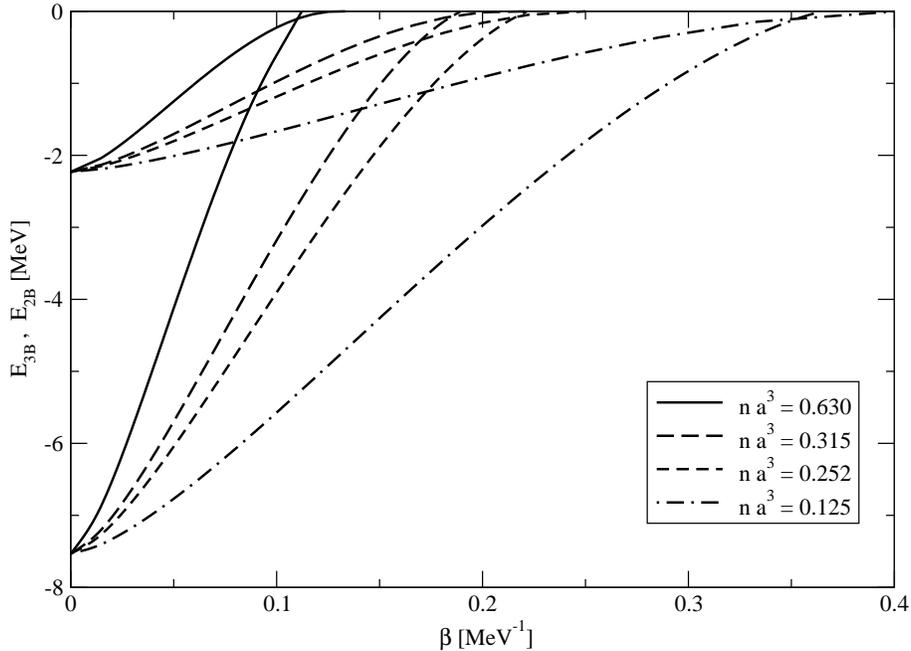}
\begin{minipage}[t]{16.5 cm}
\caption{
Dependence of the three-body and two-body bound state energies
on the inverse temperature for fixed parameter values $na^3$, 
where $n$ is the density and $a$ the neutron-proton triplet scattering
length. 
}
\label{fig:EB3}
\end{minipage}
\end{center}
\end{figure}
The single-particle self-energy obtains contributions from the three-body
$T$-matrix, which is shown in Fig.~\ref{fig:T3}. In the case of $(3p)$ 
scattering $T$-matrix the three-body self-energy is written as 
\bea\label{SIGMIN}
{\mathfrak S}^{>,<}(p_1) = \int_{p_2p_3}
\Bigl[ {\cal T}^{(1)}_{>,<}(\veck_{23}, \vecq_1, K)
+ {\cal T}^{(2)}_{>,<}(\veck_{13}, \vecq_2, K)
+ {\cal T}^{(3)}_{>,<}(\veck_{12}, \vecq_3, K)\Bigr] G^{>}(p_2)\,G^{>}(p_3),
\eea
where we use the notation $\int_{p} = d^4p/(2\pi)^4$.
The optical theorem relates the $T^{>,<}$ components of the
three-body $T$-matrix to the retarded component given by Eq.~(\ref{FTFA1}):
\bea\label{TMIN3}
&&
{\cal T}^{<}(\veck_{23}, \vecq_1,\veck_{23}', \vecq_1';K)
= \sum_{\alpha,\beta}\int_{p_4,p_5,p_6} 
{\cal T}^{(\alpha)\, R}(\veck_{23}, \vecq_1,\veck_{45}, \vecq_6;K) 
G^{<}(p_4)\,G^{<}(p_5)\,G^{<}(p_6)\nonumber \\
&&\hspace{2cm}\times 
{\cal T}^{(\beta)\, A}(\veck_{45}, \vecq_6,\veck_{23}', \vecq_1'; K) 
(2\pi)^4\, \delta^4(K-p_4-p_5-p_6) .
\eea 
To illustrate the usefulness of the three-body equations discussed above 
we show in Figure~\ref{fig:EB3} the binding energy of a three-body bound
state (triton) in nuclear matter as a function of the inverse temperature
for several values of density, $n$, measured in units of $a^{-3}$, where
$a = 5.4$ fm is the neutron-proton triplet scattering 
length. The asymptotic free-space value of the binding 
energy in this model is $E_{3B} = 7.53$ MeV. 
The binding energy was obtained from the solution 
of the homogeneous counterpart of Eq. (\ref{FTFA1}) in quasiparticle limit
assuming free single particle spectrum~\cite{SEDRAKIAN_CLARK}. 
In addition we show the temperature
dependence of the deuteron energy $E_{2B}(\beta)$ obtained within 
the same approximations from the homogeneous counterpart of Eq. 
(\ref{T1}). The continuum for the break-up process $3B \to 2B +N$, where
$N$ refers to nucleon, is temperature/density dependent as well and 
is found from the condition $E_{3B} (\beta) = E_{2B}(\beta)$.

\subsubsection{\it The quantum virial equation of state}\label{sec:VIRIAL}

The equation of the state of a  Fermi-system characterized by
small damping (long-lived, but finite life-time quasiparticles) 
can be written in the form of a virial expansion for 
density~\cite{ZIMMERMAN85,SCHMIDT90,SEDRAKIAN_ROEPKE}
\be\label{VIRIAL_ExP_DENS}
n(\beta,\mu) = \int\frac{d^4p}{(2\pi)^4}\left[a(E,\vecp)f_F(E) + 
               b(E,\vecp)g_B(E)+c(E,\vecp)f_F(E)\right],
\ee
where the virial coefficients $a(p)$, $b(p)$, and $c(p)$ are the one-, 
two- and three-particle spectral functions. Below we show that the virial 
coefficients $ b(E,\vecp)$ and $c(E,\vecp)$ can be written entirely
in terms of  the two- and three-body $T$-matrices and their derivatives. 
In the dilute limit the on-shell $T$-matrices are related to the 
scattering phase-shifts; since the damping in this limit is small 
a direct relation between the scattering observables in free-space
and the equation of state can be established.
We have seen that in the small damping limit the 
spectral function can be approximated by Eqs. (\ref{ASMALL})
and (\ref{ZFACTOR})  to leading order.   
Thus our starting point is the expression for the density of the system 
which we write as
\be\label{VIRIAL_ExP} 
n(\beta,\mu) =\int\frac{d^3p}{(2\pi)^3}f_F(\ep_p)+
\int\frac{d^4p}{(2\pi)^4} a(\omega)\left[f_F(\omega)-f_F(\ep)\right].
\ee
The first term is the contribution from the ``uncorrelated" quasiparticles
(note that the notion of quasiparticle already requires correlations which
renormalize the single particle spectrum, however the quasiparticles are
still characterized by a sharp relation between the energy and the momentum
as the ordinary particles).  The second term is the correlated density which,
upon using Eqs. (\ref{ASMALL})  and (\ref{ZFACTOR}), becomes
\be\label{NCORR}
n_{\rm corr}(\beta,\mu) = - \int\!\!\frac{d^4p}{(2\pi)^4}
\gamma(p)\partial_{\omega}\left(\frac{\cal P}{\omega-\ep_p}\right)
\left[f_F(\omega)-f_F(\ep_p)\right].
\ee
Let us first evaluate this expression neglecting the three-body correlations.
The damping can be written as $\gamma(p) = i[\Sigma^<(p)-\Sigma^>(p)] = -2\Im \Sigma$ where the self-energies are given by Eqs. (\ref{SIGMA1}). Using
the optical theorem for the two-body $T$-matrix we obtain
\be
\gamma(p_1) = 2 \int \frac{d^4p_{2}}{(2\pi)^4} a(\omega_2)\,
\left[g_B(\omega_1+\omega_2)+f_F(\omega_2)\right]
{\rm Im}T^{R}(\frac{\vecp_1-\vecp_2}{2},\frac{\vecp_1-\vecp_2}{2}; p_1+p_2),
\ee
where the spectral function can be taken in the quasiparticle approximation
at the order of interest. Substituting the damping in Eq. (\ref{NCORR})
and using the identity
$
\left[g_B(E)+f_F(\omega_2)\right]\left[f_F(E-\omega_2)-f_F(\omega_1)\right]
=\left[g_B(E)-g_B(\omega_1+\omega_2)\right]
\left[1-f_F(\omega_1)-f_F(\omega_2)\right],
$
we recover the second term of the expansion (\ref{VIRIAL_ExP})
with the second (quantum) virial coefficient~\cite{ZIMMERMAN85,SCHMIDT90}
\bea\label{NCORR2}
b(\ep_{p_1}, E)& =& 2\int \frac{d^3p_{2}}{(2\pi)^3}
Q_2(\ep_{p_1},\ep_{p_2})\Bigg[
{\rm Im}T^{R}(\frac{\vecp_1-\vecp_2}{2},\frac{\vecp_1-\vecp_2}{2};E)
\frac{d}{d E}{\rm Re}{R}^{R}(E)\nonumber \\
&+&{\rm Im}{R}^{R}(E)\, \frac{d}{d E} {\rm Re}T^{R}(\frac{\vecp_1-\vecp_2}{2},\frac{\vecp_1-\vecp_2}{2};E)
\Biggr],
\eea
where ${R}^{R}(E) = \left[E-\ep(p_1)-\ep(p_2)+i\eta\right]^{-1}$
is the two-particle resolvent. For systems which support bound states
in the free space  the second virial
coefficient obtains contributions both from the negative energy
bound states and the
continuum of scattering states. The bound states appear as simple poles
of the two-body $T$-matrix on the real axis. The scattering states can
be characterized by the phase-shifts in a given partial wave channel,
after the two-body scattering $T$-matrix is expanded into partial waves.
The phase shift is defined simply as the phase of the on shell
complex valued matrix $T_{\alpha}(p,p;E = \ep(p)+\ep(p)) 
=\vert T_{\alpha}(p,p,E)\vert {\rm exp}~({\delta_{\alpha}})$
where $\alpha = TSJLL'$ specifies the partial wave channel in terms of total
spin $S$, isospin $T$ and angular $L$ and total $J$ momenta.
The second virial coefficient can now be written in terms 
of the bound state energies $E_{\beta}$ (where $\beta = 1,2...$ 
enumerate the poles of the $T$-matrix) and the scattering phase shifts
\be
b(\ep_{p_1}, E) = 2\pi \sum_{\beta}\delta(\omega-E_{\beta})
+ 2\sum_{\alpha} ~c_{\alpha}~ 2\, {\rm sin}^2\delta_{\alpha}(E) 
\frac{d\delta_{\alpha}(E)}{dE},
\ee
where $c_{\alpha}$ are channel dependent constants. 
In the non-degenerate limit one recovers the classical 
Beth-Uhlenbeck formula~\cite{BETH_UHLENBECK}.

The third virial coefficient is obtained by including the contribution to
the damping from the three-body processes ${\mathfrak G} = i({\mathfrak S}^<-
{\mathfrak S}^>) = -2 \Im {\mathfrak S}$,
where the self-energies are defined by Eq. (\ref{SIGMIN}). Now the correlated
density is written as
\be\label{NCORR4}
n_{\rm corr}(\beta,\mu) = - \int\!\!\frac{d^4p}{(2\pi)^4}
[\gamma(p) + {\mathfrak G}(p)]\partial_{\omega}\left(\frac{\cal P}{\omega-\ep_p}\right)\left[f_F(\omega)-f_F(\ep_p)\right].
\ee
The damping ${\mathfrak G}(p)$ is expressed in terms of the ${\cal T^{<,>}}$
matrices which in turn can be related to the retarded component ${\cal T^{R}}$
if we use the optical theorem obeyed by the three-body matrices.
The off-shell form of the optical theorem reads
\bea
{\rm Im}{\cal T}^{R}(\veck_{23} \vecq_1,\veck_{45} \vecq_6;K)
&=& \frac{i}{2} \sum_{\alpha\beta}\int_{p_4,p_5,p_6}
{\cal T}^{R\,(\alpha)}( \veck_{23} \vecq_1, \veck_{45} \vecq_6; K)
\Bigl[G^<(p_4) G^<(p_5) G^<(p_6) \nonumber \\
&&\hspace{-2cm} - G^>(p_4) G^>(p_5) G^>(p_6) \Bigr]
{\cal T}^{A\,(\beta)}(\veck_{45} \vecq_6,\veck_{23} \vecq_1;K)
\delta^4\left(K-p_4-p_5-p_6\right),
\eea
where the proper momenta $p_i$ $(i = 1,2,3)$ and Jacobi momenta $k_{ij}$
$q_k$ transform into each other according to the rules given after Eq. (\ref{FTFA1}).
Since we seek corrections that are first order in damping, the ${\cal T}$-matrix
in  expression (\ref{NCORR4}) can be taken in the quasiparticle approximation;
the on-shell optical theorem implies then
\bea\label{OPT_THEOREM} 
{\cal T}^{< }(\veck_{23} \vecq_1,\veck_{45} \vecq_6;K)
&=& 2i f_F\left(\omega_1+\omega_2 + \omega_3\right)
{\rm Im}{\cal T}^{R}(\veck_{23} \vecq_1,\veck_{45} \vecq_6;K),
\eea
and an analogous expression for ${\cal T}^{> }$ with the replacement
$f_F(\omega) \to 1-f_F(\omega)$. Evaluating the three-body damping
${\mathfrak G}(p)$ with the help of on-shell optical theorem
one arrives at the third  quantum virial 
coefficient~\cite{SEDRAKIAN_ROEPKE}
\bea
c(\ep_{p_1}, E)& =& 2\int_{{\vecp}_2{\vecp}_3}
 Q_3 (\ep_{p_1},\ep_{p_2},\ep_{p_3}) 
\Bigl[ 
{\rm Im}{\cal T}^{R}( \veck_{23} \vecq_1,\veck_{23} \vecq_1; K)
\frac{d}{d E} {\rm Re} {\cal R}(E)\nonumber\\
&+& {\rm Im} {\cal R}(E) \frac{d}{d E} {\rm Re} {\cal T}
(\veck_{23} \vecq_1, \veck_{23} \vecq_1;K)\Bigr],
\eea
and where ${\cal R} = \left[E- \ep_{p_1}-\ep_{p_2}-\ep_{p_3}
+i\eta\right]^{-1}$is the three-particle resolvent.
The third virial coefficient can be decomposed into scattering
and bound-state contributions in analogy to the two-body case.
Complications arise in attractive systems where apart from the
three-body bound states one needs to take into account the
break-up, recombination and rearrangement channels which are
absent in the two-body case.
The knowledge of the virial expansion (\ref{VIRIAL_ExP_DENS}) completely
specifies the equation of state of the system; the pressure can be 
computed from the Gibbs equation 
\be 
p(\mu,\beta) = \int_{-\infty}^{\mu} d\mu' n(\mu',\beta).
\ee
The common form of the equation of state 
$p(n,\beta)$ is obtained upon eliminating the parametric 
dependence on the chemical potential $\mu$. The theories based on the 
second virial coefficients smoothly interpolate between the classical 
gas theory at low densities and high temperatures 
and the Bruckner-Bethe-Goldstone theory  at low temperatures 
and high densities~\cite{SCHMIDT90}. The effects of the third quantum
virial coefficient on the equation of state of nuclear matter
have not been studied to date.
\begin{figure}[tb] 
\begin{center}
\epsfig{figure=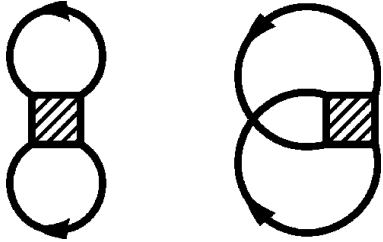,width=5.cm,angle=0}
\begin{minipage}[t]{16.5 cm}
\caption{ The lowest order diagrams of the BBG theory
for the direct ({\it left}) and the exchange ({\it right})
contributions to the energy. The square denotes the $K$-matrix,
the solid lines correspond to the hole propagators.
}
\label{fig:BBG1}
\end{minipage}
\end{center}
\end{figure}

\subsection{\it The Bruckner-Bethe-Goldstone theory}
\label{sec:BBG}

In a number of cases it is more convenient to evaluate the ground state 
energy, or at finite temperatures the thermodynamic potential, directly 
instead of first obtaining the Green's function from the Dyson
equations and then calculating the thermodynamic quantities. The
Brueckner-Bethe-Goldstone (BBG) theory evaluates the ground state 
energy of nuclear matter in terms of certain diagrammatic expansion 
of the energy, which has two important ingredients: (i) the effective
interaction is built up from the bare nucleon-nucleon force by summing 
the ladder diagrams into an effective interaction; (ii) the perturbation
expansion is organized according to the number of independent hole lines
in all topologically non-equivalent linked 
diagrams~\cite{BRUECKNER54,BRUECKNER55a,BRUECKNER55b,BETHE56,BETHE62,BETHE71}. 

The diagrams describing the perturbation series for the energy have 
the form of closed loops. Only the connected (linked) diagrams contribute, 
i.e. those diagrams which have the property that by starting at a vertex 
one can return to the same vertex moving along all the interaction and 
propagator lines. The diagrammatic rules are the same as those of
the ordinary Feynman perturbation theory
(except of the overall constant factor)~\cite{ABRIKOSOV75}. 
The lowest order diagrams of the BBG theory
are shown in Fig.~\ref{fig:BBG1}. The energy of nuclear matter including 
the contributions of the lowest order diagrams is written as
\be 
E^{(2)} = g\int \frac{d\vecp}{(2\pi)^3} \frac{p^2}{2m}f_F(\vecp) 
+ \frac{1}{2} \int \frac{d\vecp d\vecp'}{(2\pi)^6} 
K(\vecp,\vecp';\vecp,\vecp')_A f_F(\vecp) f_F(\vecp'),
\ee
where it is understood that at small temperatures 
the Fermi-functions are approximated by the 
step functions $f_F(p) =\theta(p_F-p)$; the superscript $(2)$ indicates that
the hole line expansion is carried out up to the terms of second order. 
The interaction is approximated
by the $K$-matrix (We use here the term $K$-matrix instead of the $G$-matrix
to avoid confusion with Green's functions). The $K$-matrix sums
the ladder diagrams, where the driving term is the bare nucleon-nucleon interaction
\be \label{K1}
K(\vecp,\vecp';\vecp,\vecp') = V(\vecp,\vecp';\vecp,\vecp') + 
\int\frac{d\vecq d\vecq'}{(2\pi)^6}V(\vecp,\vecp';\vecq,\vecq') 
\frac{   
[1-f_F(\vecq)][1-f_F(\vecq')]
}{\omega-\varepsilon(\vecq)-\varepsilon(\vecq')+i\eta}
K(\vecq',\vecq;\vecp,\vecp').
\ee 
In the intermediate state the $K$-matrix propagates two particles; the 
hole-hole propagation $\propto f_F(q)f_F(q')$ which appeared in the
$T$-matrix defined by Eq. (\ref{T1}) is absent here.
As a consequence the  $K$-matrix does not develop 
a singularity at the critical temperature of superfluid phase 
transition and is well defined at zero temperature.
It is clear that to obtain the equation for the $K$-matrix from Eq. (\ref{T1})
the quasiparticle limit (\ref{SPEC_QP}) must be taken with the wave-function 
renormalization $Z(\vecp) = 1$. Note that the effective interaction entering
the BBG expansion is real, therefore the $+i\eta$ term in Eq. (\ref{K1}) is commonly
dropped and the integration is treated as principal value integration. 
Since the perturbation expansion is now 
carried out for a macroscopic quantity, it is not obvious what the 
single particle energies $\varepsilon(\vecq)$ in Eq. (\ref{K1}) represent. 
This ambiguity leads to several choices of the single particle spectrum, 
one possible form being 
\be\label{SPS1} 
\varepsilon(\vecp) = \frac{p^2}{2m} +\int\frac{d^3p'}{(2\pi)^3} K(\vecp,\vecp';\vecp,\vecp')f_F(\vecp') = \frac{p^2}{2m} +U(p), 
\ee
where $U(p)$ is called the auxiliary potential. 
(The term arises from the rearrangement
of the original Hamiltonian 
$H = T+V = T+ U + \delta V$, where  $\delta V = V-U$
is small, $T$ and $V$ are the kinetic and potential energies).
The so-called `gap choice' keeps the auxiliary potential
for the states below the Fermi surface, which leads to a gap in the 
spectrum at the Fermi-energy; the `continuous choice' keeps 
this potential both for the particle and the hole states. Another definition 
arises upon using the Landau's Fermi-liquid theory, where the quasiparticle 
energy is defined as the functional derivative of the total energy 
with respect to the occupation
$\varepsilon(\vecp) = \delta E /\delta f_F(\vecp)$~\cite{BROWN71}
\bea
\varepsilon(\vecp)&=& \frac{p^2}{2m}
 + \int\frac{d^3p'}{(2\pi)^3} 
K(\vecp,\vecp';\vecp,\vecp')f_F(\vecp')\nonumber\\
&&\hspace{1cm}+\int\frac{d\vecp' d\vecq d\vecq'}{(2\pi)^6} 
\vert K(\vecp,\vecp';\vecq,\vecq')\vert^2
f_F(\vecq)f_F(\vecq')\frac{\delta(\vecp+\vecp'-\vecq-\vecq')}
{\varepsilon(\vecp)+\varepsilon(\vecp')-\varepsilon(\vecq)
-\varepsilon(\vecq')}.
\eea
The last term, known as the rearrangement term, guarantees that the quasiparticle 
energy is in fact the energy needed to extract a particle from the system.
Returning to the BBG theory, it should be noted that the choice of the single-particle 
spectrum (i.e. the self-energy) specifies the set of diagrams that are already
included in the Green's functions from which the closed diagrams for the energy
are constructed; and it is a matter of convenience which building blocks are chosen
as fundamental. The gap choice implies that the self-energy insertions for the particles are treated explicitly by grouping them into the higher order clusters. 
Thus, the particles and the holes are treated asymmetrically both in obtaining
the effective  $K$-matrix interaction and in defining the single particle energies. 
Note that in contrast to the $T$-matrix theory 
where the self-energies $\Sigma^{>,<}(p)$ are  defined
symmetrically, the BBG theory breaks this symmetry. 
\begin{figure}[tb] 
\begin{center}
\epsfig{figure=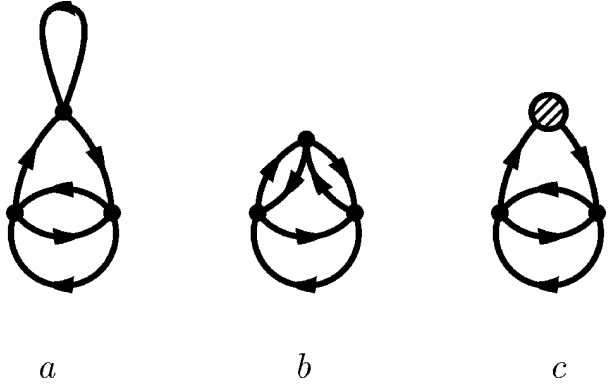,width=8.cm,angle=0}
\begin{minipage}[t]{16.5 cm}
\caption{ The lowest order three-hole line contributions to the energy
of nuclear matter: (a) the third-order bubble diagram,
(b) the third order ring diagram, (c) the bubble diagram with
an auxiliary potential insertion. The particles above the Fermi
sphere propagate from the left to the right, the holes - from the
right to the left. The dots denote the $K$-matrix, the shaded
vertex - the auxiliary potential $U$.
}\label{fig:BBG2}
\end{minipage}
\end{center}
\end{figure}

Given the form of the single particle spectrum, the next
natural question is the organization
of the diagrams in an expansion which has reasonable convergence properties. 
The BBG theory identifies such expansion parameter and 
organizes the diagrams order by order in this small parameter according to the
number of independent hole lines (i.~e. the number of hole lines that remain 
after the momentum conservation in a given diagram is taken into account).
A diagram with $i$ independent hole lines is of the order of $\kappa^{i-1}$
where the $\kappa$ parameter (pair excitation probability) is defined as 
\be
\kappa =  n \int \vert \psi(x)-\phi(x)\vert^2 d^3x,
\ee
where $\psi(x)$ is the perturbed pair-wave function satisfying the Schr\"odinger equation $K\phi = V\psi$, while $\phi(x)$ is the uncorrelated wave function and 
$n$ is the density. The integral extends to the surface where wave-function restores
its free-space form and defines an `interaction' volume $\sim 4\pi\lambda^3/3$, where
$\lambda$ is the hard-core radius. Thus, we see that $\kappa$ 
is essentially the ratio of the volumes occupied by the hard-core interaction 
and a particle. At the saturation density of 
nuclear matter $\kappa$ is of the order $15\%$, which allows one to estimate the 
error introduced by neglecting an $i$-hole line diagram to the potential energy 
$$
\Delta E_{\rm pot}(i) = \Delta E_{\rm pot}(2) \kappa^{i-1}.
$$
The expansion clearly breaks down when the interparticle distance is of the
order of the hard-core radius of the potential. 

The computation of the next-to-leading order three-hole line diagrams is 
complicated by the fact that the scattering problem of three particles
in the nuclear medium needs to be solved. The appropriate equations are due
to Bethe and are known as Bethe-Faddeev equations~\cite{BETHE}.
These equations are the counterparts of the three-body Faddeev equations 
in the free space, which take into account the influence of the background
medium. The lowest order three-hole line diagrams 
are shown in Fig.~\ref{fig:BBG2}~\cite{DAY67,DAY81}.
The particles above the Fermi sphere propagate from the left to
the right, the holes - from the right to the left. 
The $K$-matrix is represented by  a dot, since
the BBG theory assumes $K$ to be local in time. In the case where the 
spectrum is chosen to have a gap, one needs to evaluate  
only the diagrams a and b in Fig.~\ref{fig:BBG2}, 
while in the case of a continuous spectrum  the diagram c 
in Fig.~\ref{fig:BBG2} should be evaluated as well;
here the shaded vertex is an insertion of the auxiliary potential $U$. (The 
rationale behind the gap choice is the cancellation of this type of diagrams
in the BBG expansion, albeit, such a choice requires evaluation 
of the three-hole line diagrams, contrary to the continuous choice which is 
well converged at the two-hole line level)~\cite{SONG98,BALDO01,SARTOR03}.
\begin{figure}[tb]
\begin{center}
\epsfig{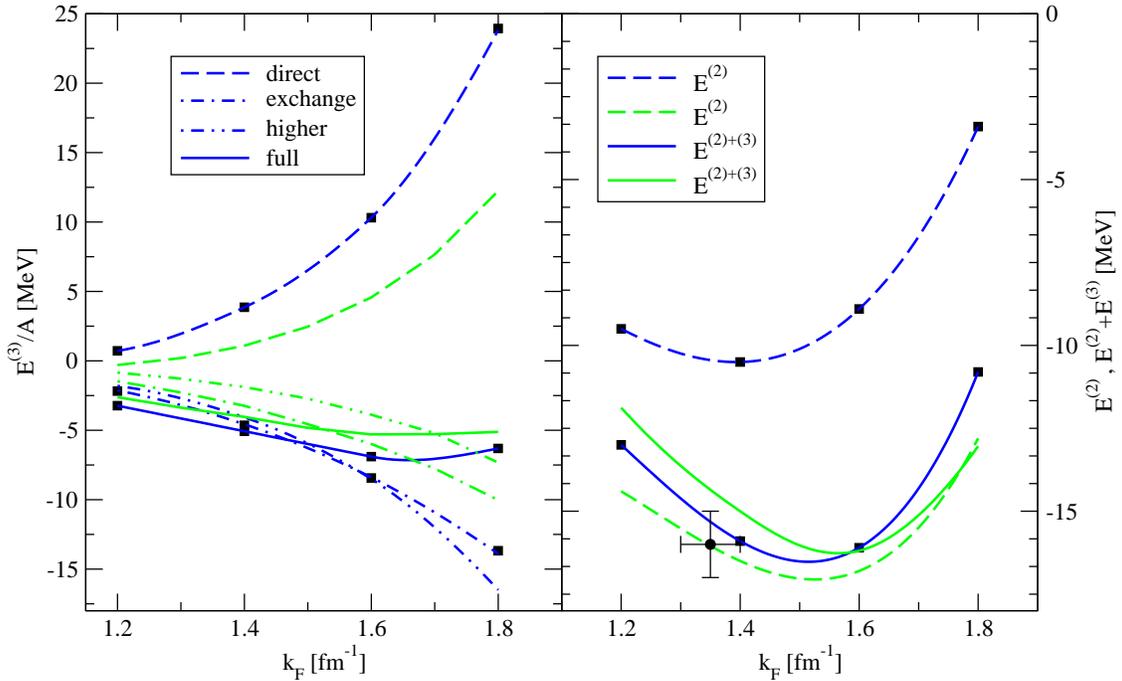}
\begin{minipage}[t]{16.5 cm}
\caption{ {\it Left panel}:
The three-hole line contributions to the energy
per particle of the isospin symmetrical nuclear matter as a function of
the Fermi wave vector. Displayed are the contributions from the
direct ({\it dashed line}), exchange ({\it dashed-dotted line}),
higher order ({\it dotted line}) diagrams and their sum $E^{(3)}/A$
({\it solid line}). The heavy lines marked with squares  show the
results obtained with the Reid potential~\cite{DAY81,SARTOR03}
the light lines -  with the  Argonne V14
potential~\cite{SONG98,BALDO01}. {\it Right panel}:
The energy per particle of isospin symmetrical nuclear
matter within the BBG theory.
$E^{(2)}$ ({\it dashed line}) is the contribution which includes
the two-hole line diagrams only;  $E^{(2)}+E^{(3)}$ include in addition
the three-hole line contributions; the labeling of curves and references
are the same as in the left panel, however the results for
refs.~\cite{SONG98,BALDO01} for the Argonne V14 potential are shown
for the continuous spectrum.  The empirical saturation 
point is shown with error bars.
}\label{fig:D3}
\end{minipage}
\end{center}
\end{figure}
The contribution to the energy from the three-hole line terms can be 
written, using for simplicity the operator notations, as 
\be 
E^{(3)} = \frac{1}{2} 
K_B \left(\frac{Q_3}{{\cal E}_B}\right) x {\cal K},
\ee
where$x = P_{123}+P_{132}$ and  
the three-body ${\cal K}$ matrix satisfies the Bethe-Faddeev equation
\be\label{BBGK}
{\cal K} = K_A x \left(\frac{Q_3}{{\cal E}_B}\right) K_B (1+x)
-K_Ax\left(\frac{Q_3}{{\cal E}_A}\right) {\cal K}.
\ee
The first term in Eq. (\ref{BBGK}) is the sum of the direct ($\propto 1$) and 
exchange ($\propto x$) diagrams shown in Fig.~\ref{fig:BBG2}; 
the second term  corresponds to the so-called higher order diagrams.
Here the permutation operator is defined as 
$P\vert \vecp_3\vecp_1\vecp_2\rangle = \vert \vecp_1\vecp_2\vecp_3
\rangle + \vert \vecp_2\vecp_3\vecp_1\rangle$, the index $A$ indicates
that the spectator third particle is above the Fermi-surface and the index 
$B$ indicates that there is non. This implies that the matrix elements
of the energy denominators in the three-body basis (here $q$ 
denotes the spectator particle) are defined as 
\bea
 {\cal E}_{A/B} = \left\{\begin{array}{cc}
E(\vecq)+ E(\veck)+E(\vecp)-\omega_3 & ({\rm if}~~q >q_F,~~{\rm case~A}),\\
 E(\veck)+E(\vecp)-\omega_2          & ({\rm if}~~q <q_F,~~{\rm case~B}),
\end{array}\right.
\eea
where $\omega_3 = E(\vecp_1)+ E(\vecp_2)+E(\vecp_3)$ and $\omega_2 = E(\vecp_1)+ E(\vecp_2)$, where $E(\vecp)$ is the single particle energy.
The action of the three-body Pauli operator is written as 
\bea\label{QBBG}
Q_3 \vert \vecq,\veck \vecp \rangle &=& [1-f_F(\veck)][1-f_F(\vecp)],
\quad \quad f_F(\vecp) \equiv \theta(\vert p_F-p \vert),
\eea
which implies that the particles in the two-body subspace must be outside 
the Fermi-sphere, while the propagation of the third, spectator particle,
is not restricted by the Pauli principle. The BBG Pauli
operator  should be compared to Eq. (\ref{Q3}) which is  the most general 
form of a three-particle Pauli operator for intermediate particle 
propagation which preserves the particle-hole symmetry.

Fig.~\ref{fig:D3} shows the various contributions to the three-hole line
energy $E^{(3)}$ per particle for several densities parameterized in terms
of the Fermi wave vector ($n = 2k_F^3/3\pi^2$ in symmetrical nuclear matter).
An important feature is the mutual cancellation of the positive contribution
from the direct term and the negative contributions from the exchange and higher
order terms. The differences between the results of refs.~\cite{DAY81,SARTOR03}
and ~\cite{SONG98,BALDO01} shown in the left panel Fig.~\ref{fig:D3} 
are due to the differences in the Reid and Argonne V14 potentials. 
The differences in the right panel are mainly due to 
the choice of the spectrum - gaped spectrum in the first case
and continuous spectrum in the second.
The three-hole line contribution to the energy leads to a
saturation curve of nuclear matter, which predicts a binding energy that is
consistent with the empirical saturation point (Fig.~\ref{fig:D3}, right
panel). The minimum of the saturation curve lies at densities that are larger
than the empirically deduced one - the missing ingredient 
is the three-body forces.
The convergence of the hole-line expansion in the case of continuous spectrum
is faster than in the case of the gaped spectrum; in the
first case the two-hole line expansion provides a satisfactory results within
the errors which are introduced by ignoring additional physics, such as
three-body forces and relativistic dynamics 
(these two aspects of many body problem cannot be disentangled in general).

\subsection{\it Relativistic $T$-matrix theory}

In describing the nuclear phenomenology within relativistic theory two 
distinct approaches are possible: the phenomenological approach starts with 
a meson-baryon Lagrangian whose parameters are fitted to reproduce the known
empirical properties. A typical set is the binding energy at saturation 
$E_B\simeq -16.0$ MeV, saturation density $\rho_0 = 0.16$ fm$^{-3}$,
compression modulus $K\sim 300$ MeV, symmetry energy $E_S\sim 30$ MeV
(see Subsec.~\ref{sec:SYM} below) and effective nucleon mass 
at saturation $m_N^*=0.8m_N$, where $m_N$ is the bare nucleon 
mass~\cite{GLEN_BOOK}. The microscopic approach constructs first
the free-space scattering $T$-matrix from a one-boson-exchange potential,  
which fits to the scattering phase-shifts and the deuteron binding energy; 
given the free-space interaction a many-body scheme is applied to describe 
the physics in matter. These models are then extrapolated to the 
large densities (and temperatures) to describe the properties of matter under
stellar conditions. This bottom to top approach 
(with respect to energy scales)  
should be contrasted to the top to bottom approaches that attempt to
constraint the form of the nucleon-meson Lagrangian and the couplings
by the symmetries of the underlying fundamental theory -  quantum 
chromodynamics (QCD). The models that incorporate the chiral symmetry - the 
dynamical symmetry of strong interactions - are based on low-momentum 
expansions of chiral Lagrangians; the usefulness of chiral models 
for treating dense hadronic matter, where momenta are generally 
not small  compared to other relevant scales (e.g. Fermi-energies) 
is unclear. However, chiral models are useful in treating the meson 
nucleon interactions in matter; for example, these have been used 
extensively in the studies of the kaon nucleon interactions in 
matter~\cite{KAPLAN86} (see Subsec.~\ref{meson_condensation} for 
a discussion and Subsec.~\ref{other_CHAP2} for further 
references).

\subsubsection{\it Dyson-Schwinger equations and mean field}

The elementary constituents of the relativistic models of nuclear matter 
are the meson and the baryons, whose interaction can be described by a
model Lagrangian
\be\label{LAGRANGIAN1}
{\cal L}_I = -g_{S}\bar\psi\psi\phi_{S}-ig_{PV}\bar\psi
\frac{\gamma^5\gamma^{\mu}}{2m_N}\psi(\partial_{\mu}\phi_{PV})
-g_{V}\bar\psi\gamma_{\mu}\psi\phi^{\mu}_{V} -if_{V}\bar\psi
\frac{\sigma^{\mu\nu}}{2m_N}\psi (\partial_{\mu} \phi_{\nu}),
\ee
where the $g_S$, $g_V, f_V$ and $g_{PV}$ are the coupling constants
of the nucleon fields $\psi$ to the meson fields $\phi$,  the indices
$S$, $V$, and $PV$ refer to scalar, vector and pseudovector couplings.
 Table~\ref{tab:MESONS} lists the (non-strange) mesons, 
their quantum numbers and typical values of meson-nucleon
couplings. The $\sigma$ meson is believed to represent the two-pion 
exchange contribution to the interaction within the one-boson-exchange 
models.  Chiral symmetry of strong interactions allows the presence 
of self-interacting meson terms  in Eq. (\ref{LAGRANGIAN1}) which 
we neglect for simplicity.
\begin{table}
\begin{tabular}{cccccccc}
\hline\hline
%\\
 meson & $\pi$ & $\omega$ & $\rho $& $\sigma$ & $\eta$ & $\delta$ & $\phi$\\
%\\ 
\hline
\\
Mass [MeV]   & 139 & 784 & 764 & 571 & 550 &  962 & 1020\\
\\
$I, J^{\pi}$ & 1,0$^{-}$ & 0,1$^{-}$ & 1,1$^{-}$ & 0,0$^{+}$ & 0,0$^{-}$ & 1, 0$^{+}$
& 1$^{-}$, 0\\ 
\\
$g^2/4\pi$ & 14.16 & 11.7 &  0.43 & 7.4 & 2.0 & 1.67  & - \\
\\
coupling & $PV$ & $V$ & $V$ & $S$ & $PV$ & $S$ & $V$\\
\\
\hline\hline
\end{tabular}
\caption{The masses $M$, the isospin $I$, spin and parity $J^{\pi}$, 
and couplings $g$ of non-strange mesons. The ratio of the tensor 
to vector coupling $f/g = 0$ for the $\omega$ meson and $ 5.1$ for 
the $\rho$ meson.}
\label{tab:MESONS}
\end{table}
The Euler-Lagrange equations for the baryon and meson fields lead to the 
following set of Schwinger-Dyson equations for nucleons
\bea\label{SCWINGER_DYSON2}
{\bm G}(p) &= &{\bm G}_0(p) + {\bm G}_0(p){\bm \Sigma}(p){\bm G}(p),\\
\label{REL_SELF}
{\bm \Sigma}(p) &=& -ig_{V}^2\int\frac{d^4q}{(2\pi)^4}\gamma_{\mu} 
{\bm D}_{V}^{\mu\nu}(q){\bm G}(p-q){\bm \Gamma}_{\nu}(p-q, p; q)\\
&-& ig_{S}^2\int\frac{d^4q}{(2\pi)^4} {\bm D}_{S}(q)
{\bm G}(p-q){\bm \Gamma}(p-q, p; q),\\
&-& ig_{PV}^2\int\frac{d^4q}{(2\pi)^4}\gamma^5\gamma_{\mu}\tau^i 
q^{\mu}q^{\nu}{\bm D}_{PV}(q){\bm G}(p-q)
{\bm \Gamma}^{5j}_{\nu}(p-q, p; q),
\eea
where ${\bm G}_0(p)$ and ${\bm G}(p)$ are the free and 
full nucleon propagators. The summation over the mesons  
with the same type of coupling is implicit. The nucleon 
self-energy ${\bm \Sigma}(p)$ contains the vector, scalar, and pseudo-vector
meson propagators and the associated three-point meson nucleon vertices 
${\bm \Gamma}(p)$. The meson propagators obey the  following Schwinger-Dyson 
equations, which we write explicitly for the case of vector coupling,
\bea\label{D_PROPAGATOR}
{\bm D}^{\mu\nu}(q) &=& {\bm D}_0^{\mu\nu}(q) + {\bm D}_0^{\mu\lambda}(q) 
                       {\bm \Pi}^{\lambda\rho}(q) {\bm D}^{\rho\nu}\\
\label{PI}
		       {\bm \Pi}^{\mu\nu} (q) &=& -ig_V^2\int\frac{d^4p}{(2\pi)^4}
		       {\rm Tr}[\gamma^{\mu}{\bm G}(p){\bm \gamma}^{\nu}(p,p+q;q)
		       {\bm G}(p+q)].
\eea
The vertices ${\bm \Gamma}(p)$ obey their own Schwinger-Dyson equations which
connect the three-point functions to four-point and higher order functions. The 
lowest order truncations of this hierarchy (i. e. replacing the vertices 
$\Gamma(p)$ by their bare counterparts) leads to the relativistic Hartree 
and Hartree-Fock theories. Another common approximation is to replace
the meson propagators by their free-space counterparts; 
the resulting nucleon self-energy is written as 
\bea\label{SIGMA_HF}
{\bm \Sigma}(p) &=& -ig_{V}^2\int\frac{d^4q}{(2\pi)^4}
\left[\gamma_{\mu} \gamma_{\nu}{\bm D}_{V}^{\mu\nu}(0){\bm G}(q)
-\gamma_{\mu} \gamma_{\nu}{\bm D}_{V}^{\mu\nu}(q){\bm G}(p-q)\right]\nonumber\\
&-& ig_{S}^2\int\frac{d^4q}{(2\pi)^4} 
\left[{\bm D}_{S}(0){\bm G}(q)-{\bm D}_{S}(q){\bm G}(p-q)\right]\nonumber\\
&-& ig_{PV}^2\int\frac{d^4q}{(2\pi)^4}
\gamma^5\gamma_{\mu}\tau^i \gamma^5\gamma_{\nu}\tau^j
q^{\mu}q^{\nu}{\bm D}_{PV}(q){\bm G}(p-q).
\eea
Note that pions (which couple by the pseudo-vector coupling) contribute
to the self-energy only via the Fock exchange 
term in last line of Eq.~(\ref{SIGMA_HF}). 
One recovers the conventional relativistic mean field models upon
dropping the Fock exchange terms. (It should be noted that the meson 
self-interactions, which we neglected from the outset, play an important role
in the relativistic mean-fields models. The self-interaction coupling 
provides a further tool for adjusting the models to the phenomenology). 
The phenomenological 
models that are based on the Hartree (or Hartree-Fock) description of nuclear 
matter (the theory is known also as quantum hadrodynamcis) have 
been used extensively to study the properties of nuclear matter; we will not
discuss these models here (see the monographs~\cite{GLEN_BOOK,WEBER_BOOK,
WALECKA95,SEROT_WALECKA}).  

\subsubsection{\it Covariant $T$-matrix}

The theories which are based on the covariant treatment of the $T$-matrix
and self-energy in nuclear matter are know as the Dirac-Bruckner-Hartree-Fock 
(DBHF) theories. These theories were developed during the last two decades
mostly in the zero-temperature and quasiparticle 
limits~\cite{AMORIM_92,ANASTASIO_83,HOROWITZ_87,MACHLEIDT_87,DE_JONG_91,ENGVIK,HUBER_94,HUBER_95,DE_JONG_97,GROSS_B,SCHILLER_00,ALONSO_03,DALEN}. 
This section gives a brief overview of the ideas underlying this theory.

We start our discussion of the relativistic $T$-matrix theory by writing down 
the four-dimension Bethe-Salpeter equation (BSE) in the free space
\be
T(p_1p_2; p_1'p_2') = V (p_1p_2; p_1'p_2') + i \int\frac{d^4q}{(2\pi)^4}
V(p_1p_2;p_+'',p_-'')G_D(p_+'')G_D(p_-'')T(p_+'',p_-'';p_1'p_2'), 
\ee
where $P = p_1+p_2$ is the center of mass momentum and 
$p_{\pm} ={P}/{2}\pm p$, and $G_D$ is the Dirac propagator
\be 
iG_D(p) =\left[\gamma\cdot p- m +i\eta\right]^{-1} 
\ee
The reduction of the four-dimensional BSE to 
the three-dimensional form requires
certain constraints on the zero-components 
of the four-momenta of the in- and outgoing
particles. These constraints, the first one due to Gross~\cite{GROSS} and 
the second one due to Logunov-Tavkhelidze~\cite{LT}, 
Blankenbecler-Sugar~\cite{BS} and Thompson~\cite{TH}, require that 
\bea
p_1^0 = \sqrt{\vecp_1^2+m^2}, \quad  p_2^0 = \sqrt{s} - p_1^0,\\
p_1^0=p_2^0 = \frac{\sqrt{s}}{2},
\eea
where $s  = (p_1 + p_2)^2$ is one of 
the Mandelstam invariants. The three-dimensional
reduction of the BSE within the Thompson prescription is written as
\be \label{BSE1}
T(\vecp_1\vecp_2; \vecp_1'\vecp_2';s) 
= V (\vecp_1\vecp_2; \vecp_1'\vecp_2') 
+ i \int\frac{d^4q}{(2\pi)^4}
V(\vecp_1\vecp_2;\vecp_+'',\vecp_-'')
G_2(\vecp'',s)T(\vecp_+'',\vecp_-'';\vecp_1'\vecp_2';s), 
\ee
where for $\vecP =0$ the two-particle propagator is
\be 
G_2(\vecp,s) = -\frac{m^2}{E_p^2} \frac{\Lambda^+(\vecp)\Lambda^
+(-\vecp)}{\sqrt{s} -2 E_p+i\eta},
\ee
where $E_p$ is the on-shell particle energy and $\Lambda^{\pm}(\vecp)  $ 
are the projectors on the positive energy states (the 
negative energy states are commonly neglected, although a complete
analysis of the covariant form of the nucleon-nucleon amplitude 
requires information for both positive- and negative-energy Dirac 
spinors~\cite{AMORIM_92}). 

After the Thompson reduction the interaction is instantaneous, i.e. the 
retardation effects intrinsic to the full BSE are removed.
The reduced relativistic scattering two-body problem is thus 
described by BSE (\ref{BSE1}) which permits
one to adjust the parameters of the interaction to the experimental 
phase shifts; the bound state spectrum is described by the homogeneous
counterpart of Eq. (\ref{BSE1}) and can be used to constrain the 
interactions to reproduce the deuteron binding energy.  

Now we turn to the scattering problem in nuclear matter and 
write the formal solution of the Schwinger-Dyson equation for  
nucleons as
\be\label{PROP2} 
i{\bm G}(p) =\left[\gamma\cdot p- m -{\bm \Sigma}(p)\right]^{-1}.
\ee
The self-energy has a decomposition in terms of the Lorentz invariants
\bea 
{\bm \Sigma}(p)&=& {\bm \Sigma}_S(p) 
                  +\gamma_{\mu}{\bm \Sigma}_V^{\mu}(p)
                  +\frac{\sigma_{\mu\nu}}{2}{\bm \Sigma}_T^{\mu\nu}(p)  
                  +\gamma_5{\bm \Sigma}_{PS}(p)
                  +\gamma_5\gamma_{\mu}{\bm\Sigma}_{PV}^{\mu}(p)\nonumber\\
&=& {\bm \Sigma}_S(p_0,p^2) 
       +\gamma_{\mu}p^{\mu}{\bm \Sigma}_V(p_0,p^2)
      +\frac{\sigma_{\mu\nu}}{2} p^{\mu}p^{\nu}{\bm \Sigma}_T(p_0,p^2).
\eea
The assumption that the theory is invariant under 
parity transformations requires
that the terms involving $\gamma_5$ vanish;
the last term in the second line
vanishes by the anti-symmetry of the tensor $\sigma_{\mu\nu}$.
Since we are interested in the equilibrium properties of matter, 
we shall not carry along the Schwinger-Keldysh structure and 
will specify the discussion to the retarded propagators.
Upon separating the zero-component of the vector self-energy, 
$\gamma_{\mu}p^{\mu}{\Sigma}_V(p_0,p^2) = -\gamma_0\Sigma_0(p)+
{\bm\gamma}\cdot \vecp \Sigma_V(p)$, the propagator (\ref{PROP2}) 
can be written as a quasi-free (retarded) propagator
\be \label{DP}
iG(p) =\left[\sla p^{\star} - m^{\star}(p)\right]^{-1},
\ee
where the effective momenta and masses are defined as 
\bea 
m^{\star}(p) = m + { \Sigma_S}(p),\quad
p_0^{\star} = p_0 +  {\Sigma}_{0}(p),\quad
\vecp^{\star} = \vecp\left[1 + {\Sigma}_{V}(p)\right];
\eea
the analogy to the Dirac propagator 
is formal because the new quantities are coupled via 
the self-energies and are complex in general.
The form of the new propagator (\ref{DP}) suggests defining effective spinors
which are the on-shell positive energy solutions of the medium modified 
Dirac equation where the imaginary part is set to zero, i.~e.
\be 
{u}_r (\vecp) = \left(\frac{E_p^{\star}+m^{\star}}{2m^{\star}}\right)^{1/2}
\left(\begin{array}{c}
1\\
{\vecsigma \cdot \vecp}{(E_p^{\star}+m^{\star})^{-1}}
\end{array}\right)\chi_r ,
\ee
where $\chi_r$ is a  state-vector in the spin-space and $E_p^{\star} = 
\sqrt{\vecp^{\star 2}+m^{\star 2}}$ is the energy eigenvalue. The effective 
spinors are normalized according to  ${u}_r (\vecp){u}_s (\vecp) = 
\delta_{sr}$. For further purposes it is useful to  define the effective
quantities  according to $\tilde F(p) = (m^{\star}/E_p^{\star}) \bar u(\vecp) 
F(p) u(\vecp)$. Acting on equation (\ref{DP}) by the unity operator 
$\Lambda^{+\star}+\Lambda^{-\star} = 1$, where the 
positive and negative energy 
projectors are defined as $\Lambda^{+\star} ={u}_r\otimes \bar {u}_r$ and 
upon neglecting the negative energy part one finds 
\be
\tilde G(p) = 
\left[p_0^{\star}-E_p^{\star}+i\zeta^{\star}(p)\right]^{-1}.
\ee
where the damping is defined as 
\be 
\zeta^{\star}(p) = \Im\left[\Sigma_{0}(p)  
-\frac{m^{\star}}{E_p^{\star}}\Sigma_S(p)
- \frac{p^{\star}}{E_p^{\star}}p\Sigma_V(p)\right].
\ee
The spectral function can be constructed in full analogy 
to the non-relativistic case
\be\label{SPECFUNC_REL}
\tilde A(p) = 
\frac{\zeta^{\star}(p)}{(p^{\star}-E_p^{\star})^2
+\left[\zeta^{\star}(p)/2\right]^2},
\quad {\rm lim}_{\zeta^{\star}\to 0}\tilde A(p) 
= 2\pi\delta(p^{\star}-E_p^{\star}).
\ee
The second relation, which corresponds to the quasiparticle limit of the 
spectral function, defines the single particle energies
\be \label{QP_REL}
\varepsilon_p = \sqrt{\vert p^{\star}\vert^2+m^{\star}}-\Sigma_{0}(p).
\ee
As in the non-relativistic case, the form of the spectral function
is Lorentzian and the spectral sum rule (\ref{SUM_RULE}) is 
fulfilled for  the general and quasiparticle forms
of the spectral function in Eq. (\ref{SPECFUNC_REL}).

The Bethe-Salpeter equation in the background medium and in the reference
frame of the center-of-mass of two particles (suppressing spin 
indices) is written as
\be\label{BSE_REL}
T(p,p';s^{\star},P) = V(p,p')+\int\frac{d^3q}{(2\pi)^3}
V(p,q) \frac{m_+^{\star}}{E_{+}^{\star}} \frac{m_-^{\star}}
{E_{-}^{\star}} \frac{Q_2(q,s^{\star},P)}
{\sqrt{s^{\star}}-E^{\star}_{+}-E_{-}^{\star}+i\eta} T(q,p';s^{\star},P)
\ee
where $Q_2(q,s^{\star},P) = 1-f_F(\ep_{P/2+q})-f_F(\ep_{P/2-q})$ 
is the Pauli blocking and $\pm$ is a short hand for $P/2
\pm p$. The dependence of the Pauli blocking on $s^{\star}$ and $P$
is due to the function evaluated in the two-particle center-of-mass frame, 
where the Fermi-sphere is deformed because of Lorentz transformation from 
the lab to the center of mass frame. A closed set of equations is obtained
upon introducing the retarded self-energy in terms of the $T$-matrix 
(\ref{BSE_REL})
\be\label{SIG_REL}
\Sigma(p) = \int \frac{d^3q}{(2\pi)^3} \frac{1}{2E_q^{\star}} 
f_{(1)}^{\alpha}[ {\rm Tr}(\sla q^{\star}+m^{\star}) f_{(2)}^{\alpha}]
T^{\alpha}- (\sla q^{\star}+m^{\star})f_{(2)}^{\alpha} T_{\rm ex}^{\alpha}],
\ee
where the subscript ex stands for exchange,
$T^{\alpha}$ are coefficients of the expansion of the full $T$-matrix
in Lorentz invariants 
\be\label{F_ExPANSION} 
T = \sum_{\alpha} f_{(1)}^{\alpha} f_{(2)}^{\alpha} T^{\alpha},\quad\quad 
f_{(i)}^{\alpha} \in \{1, \gamma^{\mu}_i, \sigma^{\mu\nu},
        \gamma_5\gamma^{\mu}_i,  \gamma_5\sla q_i\}.
\ee
\begin{figure}[tb]
\begin{center}
\epsfig{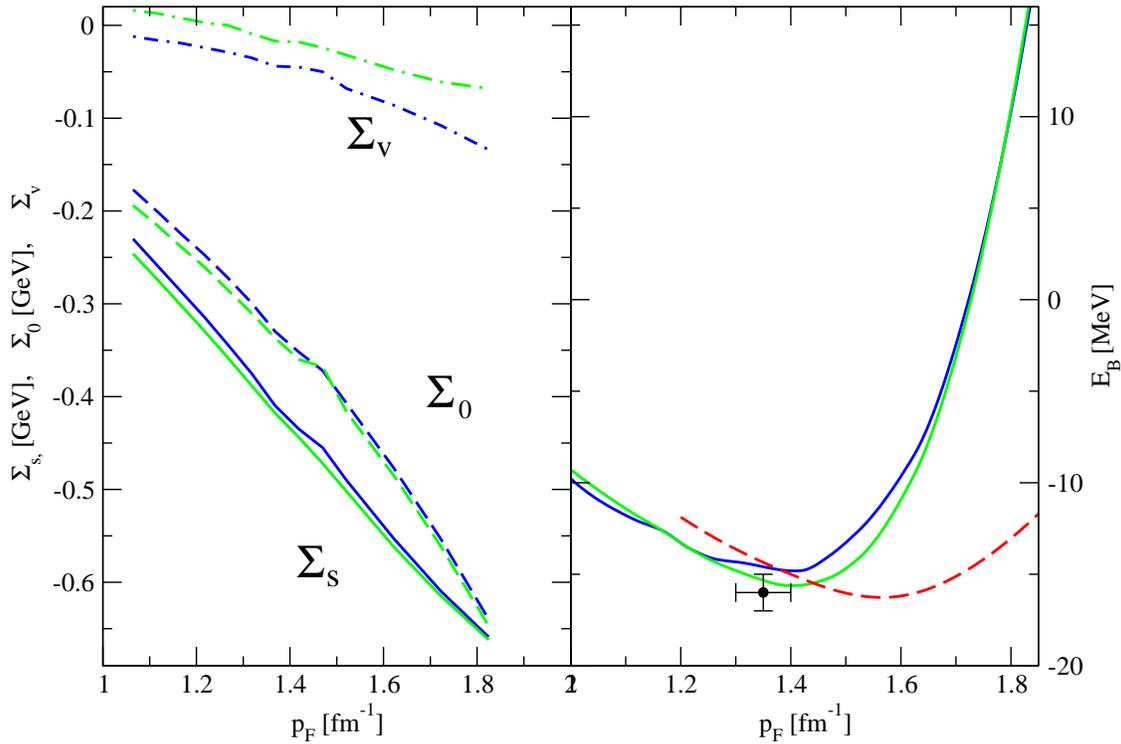}
\begin{minipage}[t]{16.5 cm}
\caption{ {\it Left panel}:
The non-vanishing components of the 
self-energy $\Sigma_S$, $\Sigma_0$ and $\Sigma_V$ 
as a function of the Fermi-momentum within the Dirac-Brueckner (heavy lines)
and the relativistic $T$-matrix (light lines) approximations.
{\it Right panel}:
The  binding energy per particle of isospin symmetrical nuclear
matter within the Dirac-Brueckner and $T$-matrix theories~\cite{DE_JONG_91}. 
The dashed line shows binding energy
in a non-relativistic Brueckner theory where the three-body forces 
are neglected. The empirical saturation point is shown with error bars.
}\label{fig:selfe}
\end{minipage}
\end{center}
\end{figure}
\noindent
The chemical potential appearing in the Fermi-functions is adjusted 
to reproduce the density of the  system. The solutions of the 
self-consistent, finite-temperature relativistic
$T$-matrix theory allow one to compute the energy density as
\be\label{ENERGY_REL}
E = \int\frac{d^3p}{(2\pi)^3}\left[
\langle \bar u^{\star}(\vecp)\vert {\bm \gamma}\cdot \vecp+ m 
+ \frac{1}{2}\Sigma(\vecp)\vert  u^{\star}(\vecp)\rangle
\right] f_F(\ep_{p}),
\ee
and the thermodynamic quantities introduced in Eqs. (\ref{FREE_ENERGY})
and (\ref{ENTROPY}). The binding energy at zero-temperature 
[$f_F(\ep_{p}) \equiv \theta (p_F-p)$] is obtained from 
Eq.~(\ref{ENERGY_REL}) 
\be 
E_B(p_F) = \rho^{-1} E(p_F) - m_N.
\ee
While we have kept only the positive energy states in our
discussion, an unambiguous treatment of the nucleon self-energy in 
matter requires keeping the negative energy states as 
well~\cite{AMORIM_92,HUBER_95,DE_JONG_97}. 
An example of such an ambiguity is the pion
exchange part of the Lagrangian  which can be described by a pseudo-scalar
or a pseudo-vector coupling. Both couplings produce the same free space
matrix elements 
for the on-shell nucleons when the coupling constants $f_{PV}$ and $g_{PV}$ 
are related as $f_{PV}/g_{PS} = m_{\pi}/2m$, where $m_{\pi}$ is the pion mass. 
If only the positive energy states are kept, a recipe to overcome 
this problem is to divide the $T$-matrix into the Born 
term plus a correlation term~\cite{GROSS_B,SCHILLER_00}. 
The ansatz (\ref{F_ExPANSION}) is applied only to the correlation term, 
since the structure of the Born term is dictated by the interaction 
$V$,  which is fixed.

Another often used approximation is the neglect of the momentum 
dependence of the self-energies, which are approximated by their 
value at the Fermi-momentum $p_F$. 
The form of Pauli-operator in Eq. (\ref{BSE_REL}) which keeps only 
the particle-particle propagation in the intermediate state is the 
counterpart of the non-relativistic Bruckner theory (as it relies 
on the ideas of the hole-line expansion)~\cite{HOROWITZ_87,MACHLEIDT_87}. 
Including the hole-hole propagation leads to the relativistic counterpart
of the original $T$-matrix theory~\cite{DE_JONG_91}.

Fig.~\ref{fig:selfe}, left panel,   
shows various components of the nucleon self-energy 
from a simplest type calculation which ignores the momentum dependence 
of the self-energies, the negative energy sea and works at 
zero temperature~\cite{DE_JONG_91}.
The contribution of $\Sigma_0$ component is negligible in the case where 
the negative energy contributions are neglected and the nucleon-nucleon 
amplitude is expanded according to the ansatz (\ref{F_ExPANSION}).
The components $\Sigma_S$ and $\Sigma_0$ are large on the nuclear scale, 
but of the same order of magnitude, so that their contributions mutually 
cancel. The binding energy of isospin symmetrical matter with the 
Bruckner and $T$-matrix approaches is shown in Fig~\ref{fig:selfe}, 
right panel. 
The additional density dependence of the Dirac spinors is an important 
ingredient of the relativistic $T$-matrix theories which leads to 
a new saturation mechanism. Compared to the non-relativistic theory 
the saturation density is correctly reproduced by the relativistic 
theories. The role of the three-body forces within the relativistic 
theories, in particular the role played by the $\Delta$ isobar, 
has not been addressed in the literature.

\subsection{\it Isospin asymmetric matter}
\label{sec:SYM}

The proton fraction $Y_p$ in neutron star interiors is constrained by the 
condition of equilibrium with respect to the weak processes. 
The disparity between the neutron and proton numbers 
(breaking the $SU(2)$ symmetry in matter) motivates the study of the 
nuclear matter under isospin asymmetry, which  conventionally is
described  by the asymmetry (or neutron excess) 
parameter $\alpha = (\rho_n -\rho_p)/(\rho_n+\rho_p)$, 
where $\rho_n$ and $\rho_p$ are the number densities 
of neutrons and protons, or alternatively by  the proton fraction 
$Y_p = (1-\alpha)/2.$ The isospin asymmetry is accommodated
in the $T$-matrix and related theories by working with two-point functions
(self-energies, $T$-matrices etc.) which are $2\times 2$ matrices in the 
isospin space. The fundamental quantity characterizing the asymmetric
nuclear matter is the symmetry energy, $E_S$ (i.~e. the energy cost of
converting a proton into a neutron). 
\begin{figure}[tb]
\begin{center}
\epsfig{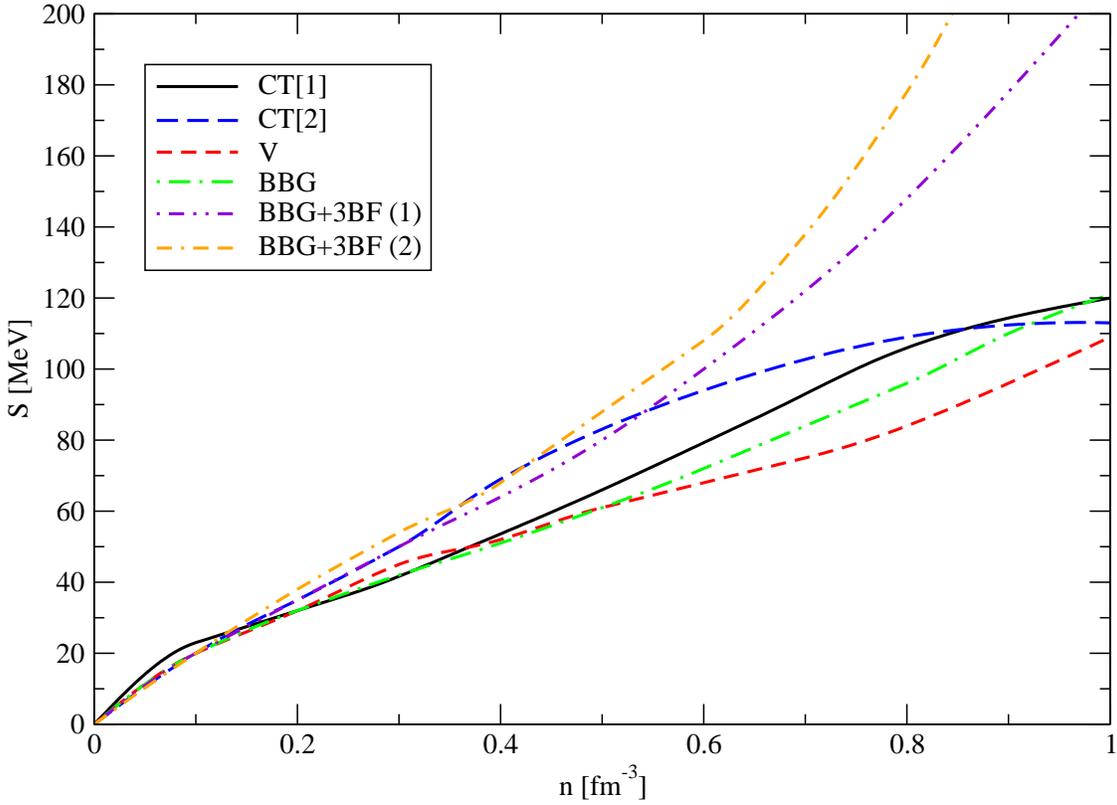}
\begin{minipage}[t]{16.5 cm}
\caption{ Dependence of symmetry energy of nuclear matter on density.
CT[1] and CT[2] refer to the results of refs.~\cite{MPA} and~\cite{LI}
obtained within the covariant $T$-matrix theory. V refers to 
the variational result of ref.~\cite{AKMAL} which includes a three-body force; 
BBG refers to the results based on the BBG theory with two-body 
forces~\cite{ZHOU};  BBG(1) includes in addition a microscopically 
derived three-body force~\cite{MICRO_TBF1,MICRO_TBF2,MICRO_TBF3}; BBG(2)
is the same as BBG(1) but with a phenomenological three-body 
force~\cite{PHENO_TBF1,PHENO_TBF2}.
}\label{fig:sym_energy}
\end{minipage}
\end{center}
\end{figure}
For small values of $\alpha$ the symmetry energy can be expanded in series
\be\label{SYM_ENERGY} 
E_S = S_2  \alpha^2 + S_4\alpha^4 +O(\alpha^6),
\ee
where $S_2$ is the coefficient in the symmetry energy term of the 
Bethe-Weizs\"acker formula. At zero temperature the contribution 
from the kinetic energy to $E_S$ can be evaluated explicitly
in the theories where the matter effects are included in interaction
energy alone
\bea
E_K(\rho,\alpha ) - E_K(\rho,0) = \frac{E_K(\rho,0)}{2}\Bigl\{
\left[(1-\alpha)^{5/3}+(1+\alpha)^{5/3}\right] -1\Bigr\}
\simeq \frac{5}{9}E_K(\rho,0) \alpha^2 + O(\alpha^4);
\eea
the coefficient of $\alpha^2$ identifies the contribution
of the kinetic energy to $S_2$. The interaction energy 
contribution is clearly model dependent, in particular 
it depends on the many-body theory and the nuclear interaction 
adopted. The importance of studying the symmetry energy 
arises from the importance of  neutron $\beta$-decay reactions
in high density matter, whose density threshold depends 
on the proton concentration (see Subsec~\ref{sec:CLASS}). Some 
constraints on the symmetry energy can be obtained around the 
saturation density $\rho_0$. An expansion of $S_2$ with respect to 
small deviations  from $\rho_0$ gives
\be
S_2(n) =  S_2(\rho_0) 
+ \frac{P_0}{\rho_0^2} (\rho-\rho_0) + O[(\rho-\rho_0)^2]. 
\ee
The first derivative of $S_2$ determines the change in the 
pressure at the saturation point due to the asymmetry of the system.
Upon using the expansion (\ref{SYM_ENERGY}) one finds~\cite{DIEPERINK03}
\be 
P(\rho)  = \rho^2 \frac{\partial E(\rho, \alpha)}{\partial \rho} \simeq P_0.
\ee    
The symmetry energy as a function of density is shown in 
Fig.~\ref{fig:sym_energy} for several models which are based
on the relativistic Dirac-Bruckner approach~\cite{MPA,LI}, variational 
approach~\cite{AKMAL} and BBG approach~\cite{ZHOU}; the latter 
two approaches include the three-body forces. 

The values of the symmetry energy and its derivative
at the saturation density vary in a narrow range: 
$S_2(\rho_0) \simeq 29\pm 2 $ MeV and $P_0 = 3 \pm 1$ MeV fm$^{3}$. 
The predictions of various models of the high density 
behavour  of the symmetry energy differ substantially.
The  relativistic, variational and BBG  theories (the latter without 
three-body force) vary within $10\%$ of an ``average'' value.
The BBG theories supplemented by either 
microscopic~\cite{MICRO_TBF1,MICRO_TBF2,MICRO_TBF3} 
or phenomenological~\cite{PHENO_TBF1,PHENO_TBF2} 
three-body forces predict symmetry energy that is by 
a factor of two larger than predictions of other models. 
However, the discrepancies in the magnitude of the 
symmetry energy at asymptotically large densities are not essential, 
since other degrees of freedom such as hyperons, mesonic condensates,
or other states of matter are likely to occupy the stable ground state.
\begin{figure}[tb]
\begin{center}
\epsfig{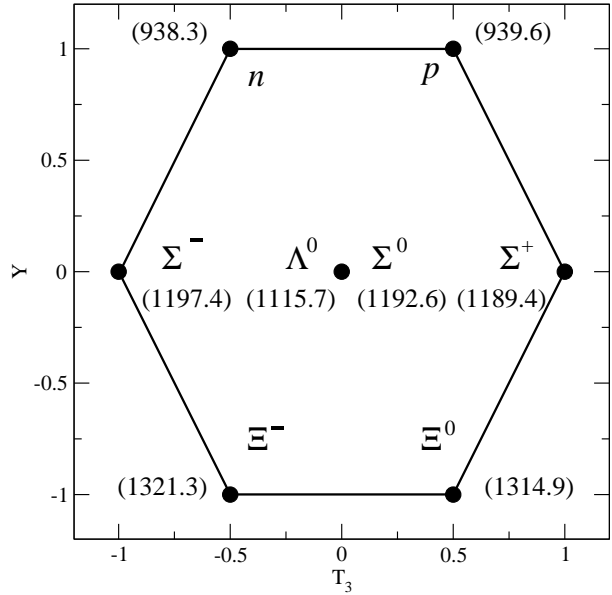}
\begin{minipage}[t]{16.5 cm}
\caption{ The $SU(3)$ baryon octet. The axis are 
the isospin $T_3$ and the hypercharge $Y$. 
Baryon masses are given in  brackets.}
\label{fig:bar_octet}
\end{minipage}
\end{center}
\end{figure}

\subsection{\it Hyperons}
\label{sec:Hyperons}

At densities around the saturation density the only baryonic 
degrees of freedom are protons and neutrons, which form 
an iso-duplet whose approximate free-space $SU(2)$ symmetry
is largely broken in matter.  At larger densities the number 
of stable baryons increases. These include the isospin 3/2 nucleon 
resonances $\Delta^{\pm}$, $\Delta^0$ and the strangeness carrying baryons 
(hyperons). The hyperonic states can be classified according 
to the irreducible representation of the $SU(3)$ group. The two diagonal 
generators of the group are linear combinations of the isospin $T_3$ and 
hypercharge $Y$, which is equal to the sum of the baryonic number
and strangeness, $Y = B+S$. The charges of baryons are determined by the 
Gell-Mann--Nishijima formula $Q = Y/2 + T_3$. Fig.~\ref{fig:bar_octet}
shows the octet of the baryons whose strangeness carrying members 
can appear in neutron star matter. If we neglect 
the interactions between the hyperons and nucleons, the threshold 
for hyperons to become stable is determined by comparison of 
the hyperon mass to the largest available energy scale - the neutron 
Fermi energy~\cite{AMBARTSUMYAN}. 
The $\Sigma^-$ hyperons can appear in matter 
through the weak hyperonic (inverse) beta-decay 
reactions $e^- + n \to \Sigma^- + \nu_e$ and hadronic weak decay 
$n + \pi^-\to \Sigma^-$. The energy balance in the first reaction implies 
$2\mu_n \simeq M_{\Sigma^-}=1197$ MeV, where $\mu_n$ is the 
chemical potentials of neutrons (we used the fact that the 
chemical potential of neutrons and electrons are almost equal in matter under 
$\beta$ equilibrium, see subsection~\ref{sec:CNWE}). The r. h. side of 
the second reaction is $O(\mu_n)$, therefore it is negligible
compared to the first reaction. Similar arguments apply to other 
hyperons which are stabilized either through the 
hyperonic $\beta$ decays or hadronic weak decays. 
For example for the lightest hyperon $\Lambda^0$ one finds 
\bea
\left.
\begin{array}{cc}
p + e^- \to \Lambda^0 + \nu_e, \quad  p + \pi^- \to \Lambda^0,
& O(\mu_p)\\
n + \pi^0 \to \Lambda^0 & O(\mu_n)
\end{array}
\right\} = M_{\Lambda} = 1116\,\, {\rm MeV}
\eea
The reactions in the first line being $O(\mu_p)$, where $\mu_p\ll\mu_n$ 
is the proton chemical potential, can be neglected and 
$\Lambda^0$ appear primarily through the weak hadronic 
process in the second line. Since the r.~h. side of this reaction is 
$O(\mu_n)$ and the mass difference $M_{\Sigma^-} - M_{\Lambda} < \mu_n$ 
at relevant densities,  $\Sigma^-$ hyperons appear first.

Interacting hypernuclear matter was initially studied
within variational approaches by Pandharipande~\cite{PANDA71} 
and Bethe and Johnson~\cite{BETHE75}. With the advent of 
the powerful phenomenology of relativistic mean-field models
these were extended to the hyperonic 
sector~\cite{GLEN85,WEBER89,GLEN91,GLEN91b}.
The extension of the $T$-matrix and related theories to include hyperons
requires the interactions between the hyperons and nucleons ($YN$)
and hyperons and hyperons 
($YY$)~\cite{JUL1,NIJM_HYP1,NIJM_HYP2,NIJM_HYP3}. 
The experimental information about the interaction involving hyperons 
is rather scarce. The $YN$ potentials are fitted to the $\Lambda N$ 
and $\Sigma N$ scattering data. The information on $YY$ interactions 
is limited to the ground state of double-$\Lambda$ hypernuclei~\cite{GIBSON}. 
Additional constraints come from the $SU(3)$ symmetry arguments.
The main difference between the $YN$ and the ordinary $NN$ interactions 
is that the direct $YN$ interaction does not contain the one-pion-exchange 
(hereafter OPE) part of the $NN$ interaction, therefore the short range 
part of the nuclear force is not hidden under the dominant OPE interaction. 
The $\Lambda$ hyperon couples to the neutral pion due to 
$\Lambda-\Sigma^0$ mixing. 

The extension of the $T$-matrix theory to include the hyperonic degrees
of freedom requires a treatment of the coupled-channel problem. 
The possible interaction channels in the isospin basis are given in
Table~\ref{tab:isospin_basis}. 
\begin{table}
\caption{The interaction channels within the isospin basis,
         for combinations of the total strangeness $S$ and
         total isospin $I$.}
\begin{tabular}{lccccc}
\hline
\\
         & $I=0$ & $I={\textstyle\frac{1}{2}}$ & $I=1$
                 & $I={\textstyle\frac{3}{2}}$ & $I=2$ \\
\\
\hline
  \\
  $S= 0$ & $N\!N$ & & $N\!N$ & & \\
  $S=-1$ & & $(\Lambda N,\Sigma N)$ & & $\Sigma N$ & \\
  $S=-2$ & $(\Lambda\Lambda,\Xi N,\Sigma\Sigma)$ &
         & $(\Xi N,\Sigma\Lambda,\Sigma\Sigma)$ & & $\Sigma\Sigma$ \\
  $S=-3$ & & $(\Xi\Lambda,\Xi\Sigma)$ & & $\Xi\Sigma$ & \\
  $S=-4$ & $\Xi\Xi$ & & $\Xi\Xi$ & & \\
  \\
  \hline
  \hline
\end{tabular}
\label{tab:isospin_basis}
\end{table}
In the strangeness $S=0, -4$ sectors there is a single channel.  
In the $S = -1$ sector the channels  $\Lambda N, \Sigma N$ are
coupled and the $T$-matrix equation reads
\bea \label{TBB}
\left(\begin{array}{cc}
T_{\Lambda N;\Lambda N} & T_{\Lambda N;\Sigma N}\\
T_{\Sigma N;\Lambda N} & T_{\Sigma N;\Sigma N}
\end{array}\right) &=& 
\left(\begin{array}{cc}
V_{\Lambda N;\Lambda N} & V_{\Lambda N;\Sigma N}\\
V_{\Sigma N;\Lambda N} & V_{\Sigma N;\Sigma N}
\end{array}\right)  \nonumber\\ 
&&\hspace{-2cm}+\left(\begin{array}{cc}
V_{\Lambda N;\Lambda N} & V_{\Lambda N;\Sigma N}\\
V_{\Sigma N;\Lambda N} & V_{\Sigma N;\Sigma N}
\end{array}\right)
\left(\begin{array}{cc}
G_{\Lambda N}T_{\Lambda N;\Lambda N} & G_{\Lambda N}T_{\Lambda N;\Sigma N}\\
G_{\Sigma N}T_{\Sigma N;\Lambda N} & G_{\Sigma N}T_{\Sigma N;\Sigma N}
\end{array}\right),
\eea  
where the intermediate state propagator, which generalizes the 
single species result (\ref{G2}) to a multi-component system, is   
\bea\label{G2BB}
G_{2BB'}(\vecp;P) &=& 
	\int\!\!\frac{d^4P'}{(2\pi)^4}\int\!\!\frac{d\omega}{(2\pi)}
                       A_{B}(p_+)A_{B'}(p_-) 
    \left[1 - f_B(p_+) - f_{B'}(p_-)
   \right] \frac{\delta(\vecP-\vecP')}{\Omega-\Omega'\pm i\eta},	       
\eea
where $B$ stands for any baryon of the $SU(3)$ octet. The spectral functions 
in Eq. (\ref{G2BB}) are related to the self-energies 
\bea \label{SIGMAR_BB} 
\left(\begin{array}{cc}
\Sigma^R_{\Lambda N} & \Sigma^R_{\Sigma N}\\
\Sigma^R_{\Sigma N} & \Sigma^R_{\Sigma N}
\end{array}\right)(p)  &=& \sum_{p'}\left(\begin{array}{cc}
T^R_{\Lambda N;\Lambda N} & T^R_{\Lambda N;\Sigma N}\\
T^R_{\Sigma N;\Lambda N} & T^R_{\Sigma N;\Sigma N}
\end{array}\right) (p+p') \left(\begin{array}{cc}
G^<_{\Lambda N} & G^<_{\Lambda N}\\
G^<_{\Sigma N} & G^<_{\Sigma N}
\end{array}\right)(p')\nonumber\\
&&\hspace{-4cm}+\sum_{p'} 2g(\omega+\omega') \left(\begin{array}{cc}
T^R_{\Lambda N;\Lambda N} & T^R_{\Lambda N;\Sigma N}\\
T^R_{\Sigma N;\Lambda N} & T^R_{\Sigma N;\Sigma N}
\end{array}\right) (p+p'){\rm Im}\left(\begin{array}{cc}
G^R_{\Lambda N} & G^R_{\Lambda N}\\
G^R_{\Sigma N} & G^R_{\Sigma N}
\end{array}\right)(p').
\eea
Equations (\ref{TBB}) and (\ref{SIGMAR_BB}) are the generalization 
of the equations of (\ref{T1}) and (\ref{SIGMAR}) to the case of coupled
$\Lambda N, \Sigma N$ channels. The example above is sufficiently general 
to illustrate the treatment of other coupled channels shown in Table~\ref{tab:isospin_basis}. The Brueckner-Bethe-Goldstone theory for
hyperonic matter is recovered from the above equations by 
(i)~taking the quasiparticle and zero temperature limits and 
(ii)~dropping the hole-hole propagation from the intermediate 
state propagators. 

The BBG calculations of hypernuclear matter were carried
out over several decades in parallel to the development of the theory
for ordinary nuclear matter with a special attention to the problem of 
binding of $\Lambda$ particle in nuclear matter 
(see e.g. \cite{BODMER,NAGATA,YAMAMOTO1,YAMAMOTO2}). 
Recent work on Bruckner theory for hypernuclear matter shifted the 
interest towards understanding the $\beta$-equilibrated matter 
in neutron stars~\cite{BALDO_00,VIDANA_00,RIOS_HYP,NICOTRA} 
(see Subsec.~\ref{sec:stellar_models} below).

\subsection{\it Charge neutrality and weak equilibrium}
\label{sec:CNWE}

Neutron stars evolve towards equilibrium with respect to the weak 
interactions on long time scales. However, in many cases, the time-scales 
of interest are much shorter than these equilibration time-scale 
and we can assume, to a good approximation, that the NS interiors 
are in approximate weak equilibrium between hadrons and leptons.
If the temperatures in the interiors of NS are below several MeV, which 
corresponds to time scales of the order of months or less
after star's formation, 
the neutrinos propagate through the star without interactions. 
 Their chemical potentials can be set to zero. 
In a matter composed of neutrons ($n$), 
protons $(p)$ and electrons $(e)$ the weak equilibrium is established 
by the $\beta$-decay and electron capture reactions 
\be 
n \to p + e^- + \anu_e, \quad \quad p + e^-\to n +\nu_e.
\ee
\begin{figure}[tb]
\begin{center}
\epsfig{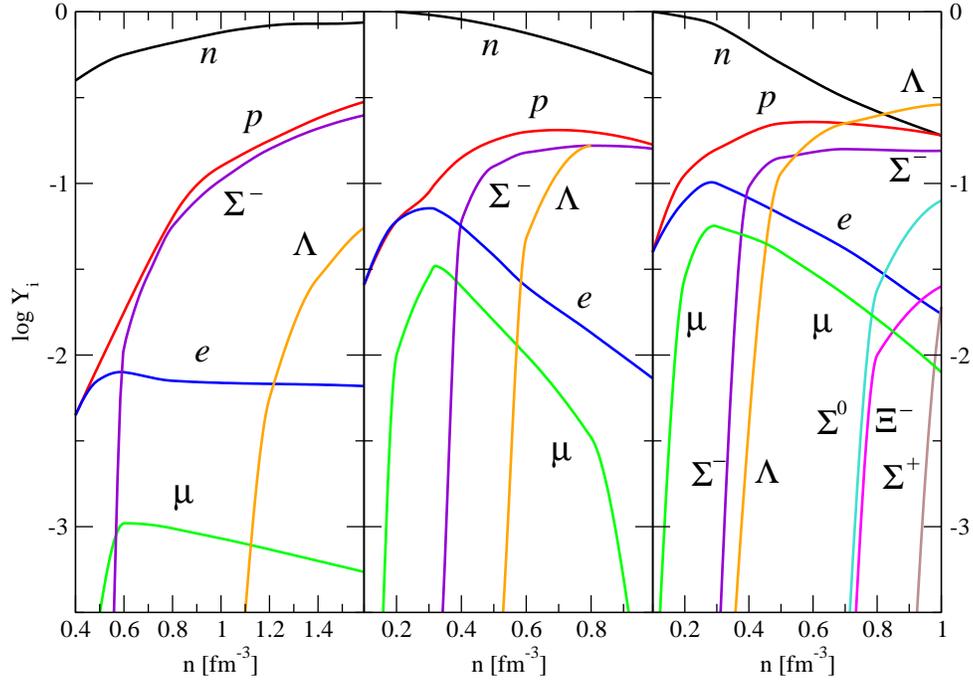}
\begin{minipage}[t]{16.5 cm}
\caption{ Matter composition in $\beta$-equilibrated 
hypernuclear matter. {\it Left panel:} Noninteracting 
matter~\cite{AMBARTSUMYAN};
{\it Central panel:}
Interacting mixture according to the BBG theory~\cite{NICOTRA}; 
{\it Right panel:} Interacting mixture within a relativistic 
mean-field theory~\cite{GLEN85}.
}\label{fig:amba}
\end{minipage}
\end{center}\
\end{figure}
The equilibrium condition requires that the chemical potentials obey the 
equality $\mu_e = \mu_n-\mu_p.$ Given the energy density of asymmetric 
nuclear matter $\varepsilon$ the chemical potentials can be expressed as 
\be 
\mu_{n/p} = \frac{\partial\varepsilon}{\partial n_{n/p}} 
= \frac{\partial\varepsilon}{\partial n} + \frac{1\mp\alpha}{n}
\frac{\partial \varepsilon}{\partial\alpha}, \quad \delta\mu =\frac{2}{n}
\frac{\partial\varepsilon}{\partial\alpha},
\ee
where $\delta\mu = \mu_n-\mu_p = \mu_e$. The charge neutrality 
requires that the number densities $n_p = n_e$; treating electrons as 
a non interacting and massless $n_e = \mu_e^3/3\pi^2$,  which 
in turn implies that 
\be\label{FRACTION}
Y_p= Y_e = \frac{8}{3\pi^2n}
\left(\frac{\partial\varepsilon}{\partial n}\right)^3 
= \frac{8}{3\pi^2n} S_2^3 \left(\frac{1}{2}-Y_p\right)^3  \quad 
(\mu_e \ge m_{\mu}),
\ee
where in the last step the expansion (\ref{SYM_ENERGY}) has been used.
Eq. (\ref{FRACTION}) allows one to compute the proton fraction at a given 
density if the symmetry coefficient $S_2$ is known.
When the electron chemical potential exceeds the 
muon rest mass $m_{\mu} =  105.7$ MeV the 
electrons decay into muons via the reaction
\be 
e^-\to \mu^- + \anu_{\mu} + \anu_{e}.
\ee
Equilibrium with respect to this process and its inverse requires 
$\mu_e = \mu_{\mu}$, where $\mu_{\mu}$ is the chemical potential of muons.
The proton fraction in this case is given by the parametric equation
\be\label{FRACTION2} 
Y_p = \frac{1}{3\pi^2n}\left[(\delta\mu^2
-m_{\mu}^2)^{3/2}+\delta\mu^3\right].
\ee
The conditions for $\beta$-equilibrium and charge neutrality are readily 
generalized to the case of arbitrary numbers of baryons and leptons; the 
key observation is that there are two conserved charges available - the 
total baryonic charge, which is related to the conservation of the 
baryonic density and the total electrical charge, which is related 
to the charge neutrality of matter. The thermodynamic potential of matter
is a functional of the baryon $n(Y_i)$ and lepton  $n(Y_j)$ densities,
where $i$ and $j$ enumerate the baryon and lepton species respectively:
\be 
\Omega(Y_i,Y_j) = E - \mu_B \left[n_B-\sum_i B_i Y_i n\right] 
-  \mu_L n\left[ \sum_i Q_i Y_i +\sum_j Q_j Y_j \right],
\ee
where $B_i$ and $Q_i$ are the baryon and electric charges and 
$\mu_B$ and $\mu_L$ are the associated Lagrange multipliers, $n_B$ is 
the baryon density.  The 
equilibrium conditions require 
\be
\frac{\partial\Omega}{\partial Y_i } = 0, \quad \frac{\partial\Omega}{\partial Y_j } = 0.
\ee
The baryon and lepton chemical potentials are obtained as
\bea \label{CHARGES}
\mu_i =\frac{\partial E}{\partial Y_i} = -\mu_B B_i+\mu_LQ_i,  \quad\quad
\mu_j =\frac{\partial E}{\partial Y_j} = \mu_L Q_j.
\eea
If the neutron and electron chemical potentials are chosen as 
independent parameters, Eqs. (\ref{CHARGES}) identify 
$\mu_B = -\mu_n$  and $\mu_e = -\mu_L$ and the chemical potentials 
of arbitrary baryonic species are written as $\mu_i = B_i\mu_n 
-Q_i \mu_e$. For example for a mixture consisting of
$n$, $p$, $\Sigma^{\pm}$ and $\Lambda^0$ baryons the 
equilibrium conditions 
give 
\be 
\mu_{\Sigma^-} = \mu_{n} + \mu_{e},\quad 
\mu_{\Lambda} = \mu_{n},\quad
\mu_{\Sigma^+} = \mu_{n} - \mu_{e} = \mu_p, \quad \mu_{\mu} =\mu_e,
\ee
\be
Y_p+Y_{\Sigma^+} - (Y_e+Y_{\mu}+Y_{\Sigma^-}) = 0.
\ee
Fig.~\ref{fig:amba} shows the abundances of various species in a 
baryon-lepton mixture within the non-interacting hyperonic
matter theory~\cite{AMBARTSUMYAN},  BBG theory~\cite{NICOTRA} and 
the relativistic mean-field theory~\cite{GLEN85}. The interacting models
agree qualitatively at low densities, however at larger densities
the relativistic theory allows for the $\Sigma^{0,+}$ hyperons
and the cascade $\Xi^-$.

\subsection{\it Meson condensation}
\label{meson_condensation}
Bose-Einstein condensation (BEC) of mesons has an important impact 
on the properties of hadronic matter in NS. First, it softens 
the equation of state and, second, it leads to an 
enhanced neutrino emission. Contrary to the ordinary Bose-Einstein 
condensation, where above the critical temperature one deals with 
a normal Bose gas, the pion condensation is associated with an 
unstable mode in nuclear matter which has the {\it quantum numbers
of pions}~(ref.~\cite{MIGDAL} and references therein). 
The threshold of pion condensation is derived by 
considering the pion retarded propagator in nuclear medium
\be\label{PION_PROPAGATOR} 
D_{\pi}(\omega,\vecq) = \left[\omega^2-\vecq^2 -m_{\pi}^2 
-\Pi(\omega,\vecq)\right]^{-1},
\ee
where the pion self-energy, which is represented by the 
polarization tensor of nuclear matter $\Pi(\omega,\vecq)$,  
sums the particle-hole ($ph$) and 
$\Delta$-resonance -- hole ($\Delta h$) states to all orders. 
[The equations determining the pion propagator and the self-energy 
are identical to Eqs. (\ref{D_PROPAGATOR}) and 
(\ref{PI}) with the vector coupling replaced by the pseudo-vector coupling.]
The resummation of these channels in nuclear matter is a complex problem
in general~\cite{DICKHOFF1,DICKHOFF2}, but can be performed analytically 
if one approximates the driving term in the $ph$ and $\Delta h$ 
series by Landau parameters
\bea \label{LANDAU_PAR}
&&\Gamma_{ph} = f + g~\vecsigma_1\cdot \vecsigma_2 +
                 (f' + g'~\vecsigma_1\cdot \vecsigma_2) 
                   (\vectau_1\cdot \vectau_2),
\eea
where $\vecsigma$ and $\vectau$ are the vectors of Pauli matrices in
spin and isospin spaces. In analogy to (\ref{LANDAU_PAR}) one may define
the interaction in the $\Delta h$ channel as
$\Gamma_{\Delta h} = 
g'_{N\Delta} (\vecsigma_1 \cdot {\bf S}_2)(\vectau_1 \cdot {\bf T}_2)$ 
and 
$\Gamma_{\Delta \Delta} = g'_{\Delta\Delta} ({\bm S}_1 
\cdot {\bm S}_2 )({\bm T}_1 \cdot {\bm T}_2), $ where ${\bm S}$ 
and ${\bm T}$ are the spin and isospin operators for the $\Delta$ resonance. 
The net polarization tensor is the sum of nucleon and 
$\Delta$ contributions, $\Pi  = 
(\omega^2-\vecq^2)\tilde\chi_N + \vecq^2 f \tilde\chi_{\Delta}$, where 
$f = \left[(m_N+m_{\Delta})^2 - (\omega^2-\vecq^2)\right]/4m_{\Delta}^2$, with
$m_N$ and $m_{\Delta}$ being the nucleon and $\Delta$-resonance mass.
The ``susceptibilities'' which include the effects of short-range correlations 
are given in terms of one-loop Lindhard functions $\chi_N$ and  $\chi_{\Delta}$ 
as~\cite{DMITRIEV}
\be
\tilde\chi_N = 
[1+(g'_{\Delta\Delta}-g'_{N\Delta})\chi_{\Delta}] {\cal D}^{-1},
\quad 
\tilde\chi_N = 
[1+(g'_{NN}-g'_{N\Delta})\chi_{\Delta}] {\cal D}^{-1},
\ee
where ${\cal D} = 1+ g'_{NN}\chi_N + g'_{\Delta\Delta}\chi_{\Delta} + 
(g'_{NN}g'_{\Delta\Delta}+g_{N\Delta}^{'2})\chi_{N}\chi_{\Delta}$. Due to 
the isospin symmetry  of nuclear matter the polarization tensors of 
neutral and charged pions are equivalent. Their spectral function can
be written as
\be 
B_{\pi}(\omega,\vecq) = \frac{-2{\rm Im}\Pi^R(\omega,\vecq)}
{[\omega^2-\vecq^2-m_{\pi}^2-{\rm Re}\Pi^R(\omega,\vecq)]^2
+[{\rm Im}\Pi^R(\omega,\vecq)]^2}.
\ee
The pion condensation in symmetric nuclear matter is characterized
by the condition $B(0,\vecq_c)\to \infty$ at ${\rm Im}\Pi^R(0,\vecq_c)\to 0$. 
In the presence of $\pi$ condensate the uniform nuclear matter acquires 
a periodic nucleonic spin wave structure with wavenumber $\vecq_c$.
These qualitative features remain intact for neutral pion condensation in
neutron matter.  Charged pion condensation in neutron matter is 
characterized by additional instabilities. Apart from 
 the collective modes mentioned 
above, a mode appears which carries the quantum numbers of $\pi^+$ above 
some critical density $\rho_+$, which depends on the details of the 
repulsive interaction in the $S = 1$ and $T =1$ channel.
In terms of nucleon excitations it represents a bound state of a 
proton and a neutron hole. At higher densities the sum of the poles of the 
$\pi^+$ and  $\pi^-$ propagators [c.~f. Eq.~(\ref{PION_PROPAGATOR})] vanishes,
which signals the instability of matter towards formation 
of $\pi^-\pi^+$ meson pairs.

Whether the pion condensation occurs in compact stars is an open issue. 
The answer depends crucially on the values of the $g'$ parameters; the 
currently accepted range of $g'_{NN} \in 0.6-0.8$ and the universality 
ansatz, which sets all the $g'$ parameters equal to $g_{NN}'$,
precludes pion condensation
in finite nuclei and in compact stars within the density 
range where the nucleonic and mesonic components retain their identity.
Recent analysis of Gamow-Teller resonance in the
$^{90}$Zr(p,n)$^{90}$Nb reaction~\cite{WAKASA}
suggests the following values of parameters: $g'_{NN} = 0.59$
and $g'_{N\Delta} = 0.18 + 0.05g'_{\Delta\Delta}$, with $g'_{\Delta\Delta}$ being
undetermined.  With this new information 
the critical density of pion condensation turns out to be 
lower than that with the universality ansatz~\cite{SUZUKI}. An independent
evidence for low-density neutral pion condensation was obtained in variational
calculations of ref.~\cite{AKMAL}.

The mechanism by which kaons may from a Bose condensate 
in neutron star matter was developed by Kaplan 
and Nelson~\cite{KAPLAN86}. Since the anti-kaon interactions in
nuclear matter are attractive, their effective mass could be 
substantially lower than their mass in the vacuum. Thus, instead 
of neutralizing the positive charge by the negative charge
of energetic electrons, this can be done by stabilizing $K^-$
in matter. Most of the effects of $K^-$ condensation on the properties 
of compact stars are similar to those discussed for pion condensation.
These include a substantial softening of the equation of state 
which reduces the maximum mass of a compact star to  $1.5 M_{\odot}$.
This reduction may have important implications for the low-mass black 
hole population in our Galaxy~\cite{BROWN_BETHE}.

\subsection{\it Stellar models}
\label{sec:stellar_models}
The equilibrium configurations of compact stars are described by the 
Einstein equations of General Theory of Relativity (GTR)
\be\label{EINSTEIN} 
R_{ik} -\frac{1}{2} Rg_{ik} = \frac{8\pi k}{c^4} T_{ik}\quad  i,k = 0,1,2,3
\ee
where $R_{ik}$ is the Ricci  tensor, $R$ - the scalar curvature, $g_{ik}$ 
 - the metric tensor, $T_{ik}$ - the energy-momentum
tensor and $k$ - the Newton's constant; 
the cosmological constant is omitted from the Einstein's equations.
The gravitational field created by a spherically symmetric mass 
distribution is itself spherically symmetric. For such fields
the components of the metric tensor are functions of the radial coordinate 
$r$ and time $t$. The metric can be written
\be 
ds^2 = e^{\nu}c^2dt^2 -e^{\lambda} dr^2 -r^2 
(d\theta^2+\Sin^2\theta d\phi^2).
\ee
The energy and momentum tensor can be expressed through the 
mass density $\rho$ and pressure $P$ as
\be 
T_{ik} = (\rho + c^{-2}P) u_iu_k - Pg_{ik},
\ee
where $u_i$ is the matter four-velocity. For a static mass distribution 
$T_0^0 = \rho c^2$ and  $T_1^1 =T_2^2 = T_3^3 = -P$. In the static 
(time-independent) limit and for spherically symmetric gravity   
 Eq. (\ref{EINSTEIN}) reduces to 
\bea\label{SS_GR1}
e^{-\lambda}\left(\frac{\lambda'}{r}-\frac{1}{r^2}\right)+\frac{1}{r^2}
 &=& \frac{8\pi k}{c^2}\rho\\
\label{SS_GR2}
e^{-\lambda}\left(\frac{\nu'}{r}+\frac{1}{r^2}\right)-\frac{1}{r^2}
 &=& \frac{8\pi k}{c^4}P\\
\label{SS_GR3}
\frac{1}{2}e^{-\lambda}
\left(\nu''+\frac{\nu^{'2}}{2}+
\frac{\nu' - \lambda'}{r}-\frac{\nu'\lambda'}{2}\right)
 &=& \frac{8\pi k}{c^4}P,
\eea
where primes denote radial derivatives.
The last equation can be replaced by the equation for hydrodynamic
equilibrium 
\be \label{HYDRO}
P' +\frac{1}{2} (P+c^2\rho)\nu' = 0,
\ee
which is nothing else than the explicit form of the covariant 
hydrodynamic equations $T^n_{m;n} = 0.$ 
Eqs. (\ref{SS_GR1}), (\ref{SS_GR2}), and 
(\ref{HYDRO}) should be supplemented with the equation of state 
$P(\rho)$ and boundary conditions $P(0)$, $\nu(0)$ and $\lambda(0)$.
The inner solution within the configuration should be matched with 
the external Schwarzschild solution 
\be\label{SCHWARTZ} 
e^{\nu} = e^{-\lambda} = 1-\frac{2kM}{c^2r},
\ee
where $M$ is the mass of the configuration. The quantity 
$R_S = {2kM}/{c^2}$ is the Schwarzschild radius; the importance of 
general relativity is determined by the ratio $R_S/R$ where $R$ is 
a characteristic length (e.g. the star radius).

It is evident that at the center of the configuration $\lambda (0) = 0$, 
$\lambda' (0) =\nu'(0)= 0$ to avoid singularities as $r\to 0$.
The problem of finding the internal solutions of stellar configurations 
is simplified upon introducing a new variable via the relation
$ e^{-\lambda} = 1-{2km(r)}/{c^2r}$. One finds~\cite{OPPENHEIMER}
\bea \label{TOV1}
\frac{dm(r)}{dr} &=& 4\pi \rho r^2,\\
\label{TOV2}
\frac{dP}{dr}&=& -\frac{k(P+c^2\rho)}{[c^2r-2km(r)]r}\left[m(r)+
\frac{4\pi}{c^2}Pr^3\right].
\eea
Thus we have two equations for three variable $p$, $\rho$, and $m$.
The system of equations is closed by specifying the equation of 
state $P(\rho)$. We implicitly assumed that the pressure is independent
of temperature, as is the case for a system at zero temperature. However
in those cases where the temperature is important, the complete 
set of equations includes  apart from the equation of state also two 
new differential equations which describe the energy balance and 
thermal transport equations (see Chapt.~\ref{Sec:cooling}).

To gain insight in the equations of hydrostatic equilibrium (\ref{TOV1}) 
and (\ref{TOV2}) consider a self-gravitating star with constant density
$\rho_0$. Integrating the first equation gives
$
m(r) = 4\pi\rho_0r^3/3.
$
The integration of the second equation gives
\be 
r = R_0 \left[1- \left(\frac{3P+c^2\rho_0}{P+c^2\rho_0}
\frac{P_0+c^2\rho_0}{3P_0+c^2\rho_0}\right)^2
\right]^{1/2}, \quad R_0 = \left(\frac{3c^2}{8\pi k\rho_0}\right)^{1/2}.
\ee
\begin{figure}[tb]
\begin{center}
\epsfig{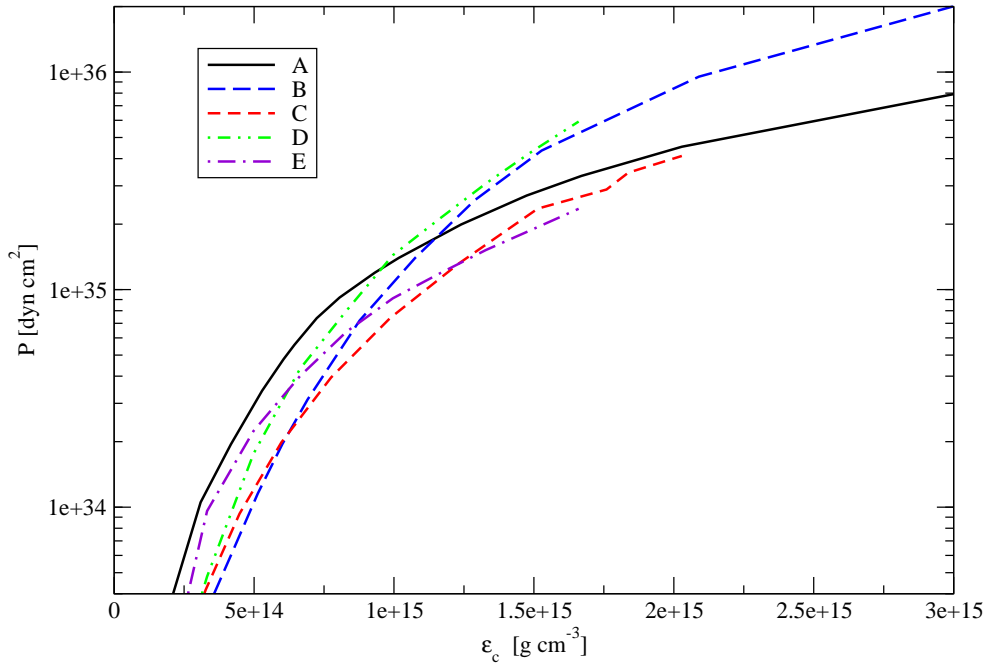}
\begin{minipage}[t]{16.5 cm}
\caption{Pressure vs central energy density for
representative equations of state (EoS). Model A ({\it solid line}) 
 - relativistic mean field model, features hyperons 
and pion condensate~\cite{GLEN85}. Model B ({\it dashed line}) 
-  non-relativistic variational EoS based on 
AV14 potential supplemented by  UVII three-body force~\cite{WFF}. 
Model C ({\it short dashed line}) - relativistic model based 
on covariant $T$-matrix~\cite{MPA}. 
Model D ({\it dashed double-dotted}) -  non-relativistic EoS 
based on the BBG theory, includes a three-body force. Model E 
({\it dashed dotted})  - same as model D, but includes 
interacting hyperons~\cite{NICOTRA}.
}\label{fig:pressure}
\end{minipage}
\end{center}
\end{figure}
Substituting $P= 0$ one finds the radius $R$ of the configuration. 
The metric within the sphere is written as 
\be 
e^{-\lambda} = 1- \frac{r^2}{R_0^2}, \quad e^{\nu} = 
\left(1-\frac{2kM}{c^2R}\right)\left( \frac{c^2\rho_0}{c^2\rho_0
+P(r)}\right)^{2},\quad r\le R.
\ee
These solutions match the Schwarzschild solution (\ref{SCHWARTZ})
at the surface of the configuration. For the constant density 
$\rho_0 = 2.85 \times 10^{14}$ g cm$^{-3}$ and maximal pressure
$P = 5 \times 10^{33}$ erg cm$^{-3}$ one finds the maximal radius 
and the mass of the configuration: $R = 6.4$ km, $M = 0.16 ~M_{\odot}$.
In the case of an incompressible fluid $P(r)/c^2 \rho_0 = 1/3$ and one
finds $R = 17.74$ km, $M = 3.3 ~M_{\odot}$. The numbers above give 
approximate lower and upper limits on the mass and radius of a 
compact object. For a more realistic estimate we now turn to the 
many-body equations of state of hadronic matter.

\begin{table}
\begin{tabular}{cccl}
\hline
EoS & method & composition & forces\\
\hline
A   & RMF    & $npeH\pi$  & contact 2B\\
B   & variational    & $npe $        &  realistic 2B + 3B\\
C   & DBHF    & $npe$        &  realistic 2B\\
D   & BBG    & $npe$         &  realistic 2B + 3B\\
E   & BBG    &$ npeH $       &  realistic 2B + 3B\\
\hline
\hline
\end{tabular}
\caption{Summary of EoS A-E introduced in the text;
 $H$ refers to hyperons, $\pi$ -- pion condensate, 
$2B$ and $3B$ -- two and three-body forces, RMF -- relativistic 
mean field. The remaining abbreviation are introduced in the 
text.}
\label{EOS_COLLECTION}
\end{table}
\begin{figure}[tb]
\begin{center}
\epsfig{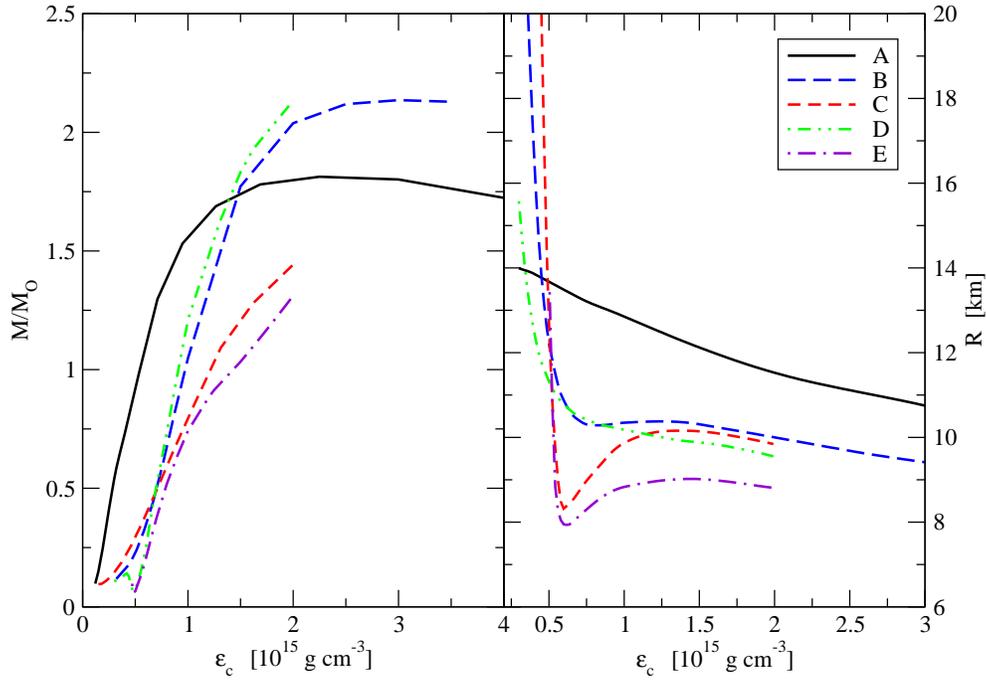}
\begin{minipage}[t]{16.5 cm}
\caption{ {\it Left panel:} Dependence of the
gravitational mass (in units of solar mass $M_{\odot}$) on the 
central energy density for configurations 
constructed from EoS A-E specified in the text.  {\it Right panel:}
Dependence of the star's radius on its central density for 
EoS A-E. 
\label{fig:models}
}
\end{minipage}
\end{center}
\end{figure}

\begin{figure}[tb]
\begin{center}
\epsfig{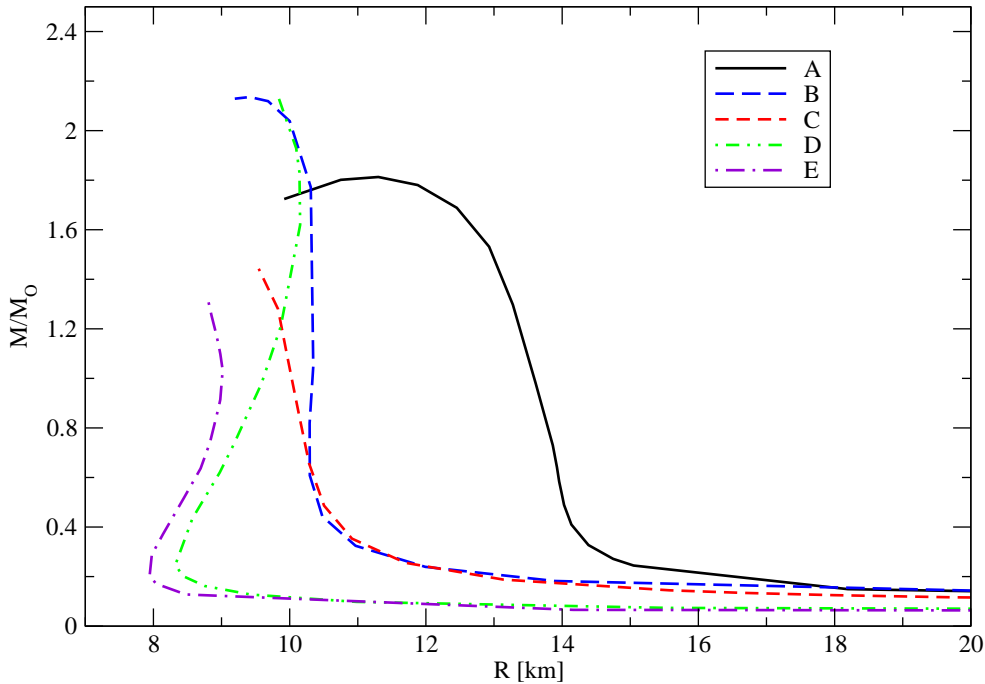}
\begin{minipage}[t]{16.5 cm}
\caption{ Dependence of configuration masses on their 
radii for EoS A-E specified in the text.
}\label{fig:mass_radius}
\end{minipage}
\end{center}
\end{figure}

For illustrative purposes we have chosen several equations
of states which are either based on the theories discussed in 
previous sections or are alternatives listed in Subsec.~\ref{other_CHAP2}; 
their properties are summarized in Table~\ref{EOS_COLLECTION}. 
Dependence of pressure on central density 
for models labeled A-E are shown in Fig.~\ref{fig:pressure}.
Model A is a relativistic mean field model which 
allows for hyperons and onset of pion condensation~\cite{GLEN85}. 
Model B is a non-relativistic variational model based 
on the Argonne AV14 two-body potential which is supplemented by  
the Urbana UVII three-body force~\cite{WFF}. 
Model C is a covariant model based on the Dirac-Bruckner (covariant 
$T$-matrix) theory~\cite{MPA}.  Finally, models D and E are based
on the BBG theory which employs  two-body and  
three body forces~\cite{NICOTRA}.  Model E includes hyperons  
along with the nucleonic component, whose interactions are 
taken into account within the BBG theory.

Equations of state are commonly characterized
by their stiffness, which is expressed through the adiabatic index
$\gamma = {\rm log}P/{\rm log}\rho$ (the larger the adiabatic the stiffer
the equation of state). The non-relativistic models B 
and D which include only nucleonic component interacting with 
two- and three-body forces are the stiffest equations of state at large 
densities. The relativistic mean field model A is stiff at low densities 
where the hyperons are still not present but becomes softer at higher 
densities as the result of hyperonization of matter and pion condensation. 
As a general trend  new degrees of freedom  soften 
the equation of state since the new constituents of matter share 
the stress due to the pressure with other constituents. Such a softening is 
apparent from a comparison of non-relativistic models  D and E which 
differ by the presence of hyperons in the latter model.
Note that the models are shown in their published density range. They
were supplemented  by the equations of state for low-density matter 
from refs.~\cite{Negele73} and \cite{Baym71} to construct sequences 
of stellar models.

The masses and radii for sequences of configurations with different central 
energy densities $\varepsilon_c$ are shown in Fig.~\ref{fig:models} for 
equations of state (hereafter EoS) A to E. The configurations were computed
with the RNS code written by Stergioulas~\cite{STERGIOULAS}.
As a common trend one finds that the stiffer is the 
EoS at the central density of a configuration, the larger is its  mass. 
Each configuration features a maximal mass, but because of the limited
density range across which our models are defined
this mass is apparent only for models A, B and D . 
Mass measurements of compact stars obtained from 
the timing of binary pulsars are broadly consistent with a canonical 
mass of 1.4 $M_{\odot}$. Pulsars that have undergone long periods of accretion
have larger masses. The most massive measured pulsar to date
PSR J0751+1807 
is a millisecond pulsar in a 6hr binary system with a helium white dwarf
secondary. The mass of the pulsar is measured through the orbital decay 
which is interpreted as due to emission of gravity waves. Combined with 
measurements of the Shapiro delay, this implies a pulsar mass $2.1\pm 0.2$
$M_{\odot}$~\cite{CORDES}.
While models A-D feature stars with  canonical masses of 1.4~$M_{\odot}$ 
(an exception is the model E), the latter observation places a severe 
limit on the EoS of nuclear matter and excludes all the models except B and D.
Recent observations of binary pulsars that have undergone extended
periods of accretion without losing their stability 
indicate that the EoS of compact stars ought
to be rather stiff. All except the lightest
members of the sequences shown in  Fig.~\ref{fig:mass_radius}
are characterized by small radii which are confined within the range of 8-11
km, their value being  almost independent  of the central energy 
density~(see Fig.~\ref{fig:models}). This leads
to degeneracy in the mass-radius relationship in 
Fig.~\ref{fig:mass_radius}:  for models B, D and E 
there exist configurations with different masses but same radii.
\begin{table}
\begin{tabular}{ccccccccc}
\hline
EoS & $\varepsilon_c$ & $M$ &  $R$ & $\Omega_K$ & $P_K$ & $T/W$ &$cJ/kM_{\odot}^2$ & $I$ \\
& [10$^{15}$ {\rm g/cm}$^3$] & $M_{\odot}$ 
& [km] & [$10^4$ {\rm s}$^{-1}$]& [$10^{-3}$ {\rm s}]
&  &     & $[10^{45}$ {\rm g cm}$^2]$ \\
\hline
A & 4.00  & 1.98  & 13.1  & 1.05  & 0.60 &  0.088 & 2.32  & 1.94\\
B & 2.00  & 2.50  & 12.9  & 1.20  & 0.52 & 0.145  &  4.53  &   3.29\\
C & 2.00  & 1.77  & 12.9  & 1.03  & 0.61 &  0.118 & 2.08  &  1.77\\
D & 1.60  & 2.49  & 13.5  & 1.14  & 0.55 & 0.181  &  4.67   &   3.62\\
E & 1.60  & 1.47  & 12.8  & 0.97  & 0.64 & 0.200 & 1.55  & 1.41\\
\hline  
\end{tabular}
\caption{Parameters of compact stars rotating at their Keplerian frequency
constructed from EoS A-E specified in the text. The models listed 
are the largest mass (stable) configuration for the EoS A and B and 
the largest central density configuration for EoS C-D.  The entries 
are the central energy density, the gravitational mass, the equatorial 
radius,  the Keplerian rotation frequency and period $P_K = 2\pi/\Omega_K$, 
the ratio of the kinetic (rotational) to the potential (gravitational) energy,
the angular momentum $J$, and the moment of inertia~$I$.
 }
\label{tab:eos_models}
\end{table}
Apart from the stationary global observable like the mass and radius, 
stellar configurations are characterized by the dynamical quantities
such as the moment of inertia $I$, rotational angular momentum $J$ 
and limiting rotation frequency $\Omega_K$, know as the Kepler 
frequency. The rotation of the star modifies the metric and thus 
the rotation frequency in the local inertial frame, which 
we denote by $\omega_L$. Slowly rotating stars admit perturbative 
approach, where the small parameter is the ratio of the rotational
kinetic energy to the gravitational binding 
energy~\cite{HARTLE1,HARTLE2,HARTLE3,SEDRAKIAN_CHUBARYAN1,SEDRAKIAN_CHUBARYAN2,WEBER92}. 
The angular velocity of a slowly rotating star 
obeys the following equation
\be \label{Diff_omega}
(r^4 j\omega_L')' + 4r^3 j' \omega_L = 0, \quad
\quad j = e^{-(\nu+\lambda)/2}.
\ee
The external solution of Eq. (\ref{Diff_omega}) is given by 
\be 
\omega_L(r) = \Omega - \frac{2kJ}{c^2r^3}
\ee
where $\Omega$ is the rotation frequency. The internal solution can be
obtained through integration of Eq. (\ref{Diff_omega})  with the 
boundary conditions $\omega_L' = 0$, when $r = 0$ and $j(r) = 1$ when 
$r = R$. The moment of inertia for slowly rotating objects
can be written as 
\be 
I = \frac{J}{\Omega} = \frac{8\pi}{3}
\int_0^R dr ~r^4 e^{(\lambda-\nu)/2}\left(\rho + \frac{P}{c^2} \right)
\frac{\omega_L}{\Omega},
\ee
where the volume integration assumes approximate spherical symmetry. 
The Keplerian frequency $\Omega_K$ is 
defined as the frequency at which the centrifugal
force on a test particle at the equator equals the gravitational 
binding force. For rotational frequencies $\Omega > \Omega_K$ mass 
shedding from the equator leads to instability of the configuration.
Table \ref{tab:eos_models} lists the parameters of the 
general relativistic configurations rotating at the Kepler 
frequency. The values of the central density correspond either
to the maximum mass stable (non-rotating) configuration (models A, B)
or the largest central density available for a given EoS (models C-E).
N. B. The central densities of these configurations exceed the 
nuclear saturation density by factors of the order of 10. It is likely 
that the baryonic and mesonic degrees of freedom are not appropriate 
at such large densities; nevertheless the EoS of quark matter can 
not differ dramatically from those listed above, and the conclusions 
drawn from the limiting frequencies should hold, at least qualitatively.
The fastest millisecond pulsar measured to date is IGR J00291+5934 
with rotation period $P = 1.7$ ms. 
According to Table \ref{tab:eos_models} and the studies 
of limiting frequencies of rotating superdense stars based on 
alternative EoS (ref.~\cite{WEBER_BOOK}, Chapter~16), stars that 
are gravitationally bound can not rotate faster 
than half a millisecond. The situation is different if a configuration 
is {\it self-bound} due to strong interactions, as is the case for the 
strange stars. Thus, if an object is observed in the future with rotation 
periods smaller than half a millisecond it must be  an exotic object, 
e.g. a compact star made up of strange matter (ref.~\cite{WEBER_BOOK}, 
Chapter~18).

\subsection{\it A guide to alternative methods}
\label{other_CHAP2}

Here we complement the discussion of the nuclear many-body problem above
by a brief  summary of some of the alternative methods that are used  
in the studies of the nuclear many-body problem. 

\underline{\it Qunatum Monte Carlo (QMC) methods} 
have been used extensively to study the 
properties of infinite nuclear matter, 
neutron matter and light nuclei with various 
versions of the Argonne two-body and the Urbana three-body 
interactions~(ref.~\cite{QMC} and references therein). Recent developments 
include calculations of the ground state energy of nuclear and neutron matter 
using the Argonne AV18 and Urbana UIX potentials~\cite{AKMAL}, Green's 
functions Monte-Carlo (GFMC) calculations of light nuclei~\cite{GMC2002}, 
and QMC calculations of neutron matter~\cite{CARLSON03,SCHMIDT,SARSA}. 
These calculations are variational in nature (i.~e. provide an upper bound 
on the energy of the system) and are  based on a non-relativistic description 
of nucleonic matter in terms of the Schr\"odinger equation.

\underline{\it Chiral Perturbation theory} seeks to establish a 
connection between  non-perturbative QCD and low-energy 
phenomenology of nuclear systems. This top to bottom 
(with regard to the energy scales) approach is anchored in the symmetries 
of the QCD Lagrangian (notably 
the approximate chiral symmetry of strong interactions) 
and in the QCD sum rules. In the nuclear 
and neutron matter problems the chiral 
perturbation theory offers a method of treating  the long-range  
pion-nucleon dynamics explicitly, the short-range correlations
being incorporated in contact terms~\cite{LUTZ,KEISER1,KEISER2,FINELLI}. 
As mentioned in Subsec.~\ref{meson_condensation}, chiral Lagranians are 
useful in deriving the properties of the kaon condensate in dense matter. 
Furthermore, chiral Lagrangians have been used to derive
free-space nucleon-nucleon interactions that are input in 
many-body and few-body calculations~(ref.~\cite{EPPELBAUM} 
and references therein).

\underline{\it Relativistic density functional theories} 
are based on the ideas of the 
mean-field theory of nucleons and mesons. These models incorporate a limited
number of phenomenological constants, that are fitted to the properties of 
bulk nuclear matter and finite nuclei, and provide a powerful tool to 
study many aspects of nuclear phenomenology at an elementary (Hartree
or Hartree-Fock) level~\cite{RING}. Some recent 
models incorporate the chiral symmetry
in the Lagrangian of the theory~\cite{TSANG,FURN,SEROT}.

\underline{\it Lattice field theory methods} 
have gained attention in recent years. 
The nuclear and neutron matter problems were studied in close 
analogy to the numerical simulations of the Hubbard model in condensed 
matter systems~\cite{MULLER,LEE1,LEE2}. Lattice methods were also 
applied to dilute, spin 1/2  non-relativistic 
fermions~\cite{KAPLAN2}. A lattice realization of the scalar 
$\phi^6$ field theory was applied to study $\alpha$ clusters
in  nuclear matter~\cite{SMS96}.

During the past decade the 
\underline{\it Effective field theory methods} were developed for 
nuclear systems and applied to the few-body nuclear 
physics; see refs.~\cite{HAXTON,VANKOLCK,KAPLAN3} and references
therein. Some aspects of the many-body problem are discusses
within this method in refs.~\cite{HAMMER,SCHAEFFER}.

\underline{\it Coupled cluster method}  is one of the most powerful 
and universally applicable techniques in quantum many-body theory, 
that is well suited for treating non-perturbative problems.
Its application to the nuclear many-body problem and finite nuclei
has a long history~(see ref.~\cite{MUTHER_POLLS} and references therein).
Recent developments include the quantum chemistry inspired coupled cluster 
calculations of ground and excited states of nuclei~\cite{KOW,DEAN2}.

\section{Neutrino interactions in dense matter}\label{sec:sec3}

The temperature of a NS born in a supernova explosion
is of the order of several tens of MeV. During the first seconds
neutrinos are trapped inside the star (i.e. the neutrino mean free
path $\lambda_{\nu} \ll R$, where  $R$ is the star radius). The energy 
that is lost in neutrinos is radiated from the surface of the 
neutrino-sphere - the surface where the optical depth for neutrinos drops
to zero. Once the star cools down to temperatures of the order of several 
MeV, the matter becomes transparent to neutrinos ($\lambda_{\nu} \gg R$).
The subsequent thermal evolution of NS is controlled by neutrino radiation
from its interiors for the following $10^3$-$10^4$ years. This long term 
evolution of NS is independent of the cooling history during the first 
several hours when the interior cools down to temperatures of  the
order of 0.1 MeV. The thermal history of NS does strongly 
depend on the neutrino  emission rates from dense matter
during the neutrino radiation epoch $t \le  10^4-10^5$ years. The neutrino 
emission rates in turn depend crucially on matter composition, 
elementary particle content, and condensed matter properties, 
such as superfluidity and superconductivity. 
Thus the studies of thermal evolution
of neutron stars offer a unique tool to test the physics of neutron 
star interiors. The measurements of the surface temperatures of young 
NS by $X$-ray satellites have the potential of constraining 
the properties of dense matter. The late-time $t \ge 10^5$ thermal 
evolution of NS is dominated by cooling via the photoemission 
from the surface and  heating due to conversion of rotational 
and magnetic energy into heat. This chapter reviews the neutrino radiation 
processes that are relevant for the neutrino radiation era. 

\subsection{\it Classification of weak processes}
\label{sec:CLASS}

We start with a classification of the weak reactions and concentrate first 
on the nucleonic matter. It is useful to classify the processes by 
the number of baryons in the initial (or final) state. 
The simplest neutrino emission process that involves single baryon
in the initial (final) state are 
\bea \label{URCA}
&&n \to p + e + \anu , \quad p + e \to n + \nu , \\ 
\label{BREMS}
&&n \to n + \nu +\anu   \hspace{5cm} ({\rm forbidden}).
\eea
The first reaction is the charge current $\beta$-decay (and its inverse).
It is  known in astrophysics as the Urca processes~\cite{GAMOW}. 
The Urca reaction  is kinematically allowed in matter under 
$\beta$-equilibrium  if the proton fraction is sufficiently 
large~\cite{Boguta81a,LATTIMER_PRL}. The threshold for the Urca process
arises from the kinematical requirement of simultaneous conservation
of momentum and energy in the reaction; a simple estimate for cold
matter in $\beta$-equilibrium shows that the  proton fraction 
$Y_p \ge 11-14\%$ for the 
Urca process to work~\cite{LATTIMER_PRL,PETHICK_RMP}.

The second process - the neutral 
current neutrino pair bremsstrahlung is forbidden by the energy and 
momentum conservation.  This statement 
is true if $n$ refers to (quasi)particles whose 
spectral function is a delta function [cf. Eq. (\ref{SPEC_QP})]. 
If one chooses to work with excitations that are characterized 
by finite widths, the reaction (\ref{BREMS}) is allowed~\cite{SD99}. 
The point is that
the finite width incorporates multi-particle processes that we are going 
to include explicitly in the next to leading order of expansion. 
The process with two baryons in the initial (and final) states 
are the modified Urca and its inverse~\cite{CHIU_SALPETER,BAHCALL_WOLF}
\bea
\label{MOD_URCA}
&&n +n \to n+ p + e + \anu , \quad p +n \to p+ p + e + \anu, \\ 
\label{MOD_URCA_INV}
&&n + p + e \to n+ n+ \nu , \quad p + p + e \to p + n + \nu.
\eea
and the modified bremsstrahlung processes
\bea
\label{MOD_BREMS}
&&N+N \to N + N + \nu +\anu , \quad N\in n,p   .
\eea
The modified processes are characterized by a spectator baryon that 
guarantees the energy and momentum conservation in the reaction.
We can continue adding further spectator baryons on both sides of the 
reactions above, however the power radiated by higher order reactions
drops dramatically for two reasons. First, if we consider the 
three-body processes, which are next in the hierarchy, 
the probability of scattering of three quasiparticles
is suppressed compared to the two-body counterpart  
(Pauli principle). Second, adding an extra fermion 
in the initial and final state introduces a small factor $T/E_F\ll 1$ 
for each fermion, where $E_F$  is the Fermi  energy. Thus,
going one step higher in the hierarchy suppresses the reaction 
rate by a small parameter  $(T/E_F)^2$. The relevant quantity 
for numerical simulations of neutron star
cooling is the neutrino emissivity, which is defined as 
the power of energy radiated per unit volume.
The emissivities of the 
processes above are $\varepsilon_{\beta} \sim  10^{27}\times T_9^6$ 
for the reaction (\ref{URCA}), $\varepsilon_{mod.~\beta} \sim  10^{21} 
\times T_9^8$  for the reactions (\ref{MOD_URCA}) and (\ref{MOD_URCA_INV}), 
and  $\varepsilon_{\nu\anu} \sim  10^{19}\times T_9^8$ for the 
reactions (\ref{MOD_BREMS});
here $T_9$ is the temperature in units $10^9$ K.

The general arguments above apply to the reactions in the hypernuclear 
matter under $\beta$-equilibrium (Subsection~\ref{sec:Hyperons}). 
The charge current processes on hyperons (hyperon Urca processes) are 
\be\label{URCA_HYP}
\Sigma^- \to n + e+\anu ,\quad\quad
\left(
\begin{array}{c}
\Lambda^0\\ \Sigma^0
\end{array}
\right) \to \left(
\begin{array}{c}
p\\ \Sigma^+
\end{array}\right) + e + \bar \nu , \quad\quad 
\left(
\begin{array}{c}
\Sigma^-\\ \Xi^-
\end{array}
\right) \to \left(
\begin{array}{c}
\Lambda^0\\ \Sigma^0 \\ \Xi^0
\end{array}\right) + e + \anu  
\ee
and the reactions inverse to these. Note that some of the 
reaction involve violation of strangeness conservation by weak 
interactions with a change of strangeness by amount  
$\Delta S = 1$. The neutrino emissivity 
of the hyperon Urca process has the same  parametric dependence 
on the density of states and temperature as the nucleonic 
Urca process, however it is smaller than the nucleonic Urca process
because the weak matrix elements are suppressed by factors
of 0.01 - 0.6. Another important difference 
is that the threshold for the $\Lambda^0\to pe\anu$ 
reaction, corresponding to the $\Lambda^0$ hyperon fraction
$Y_{\Lambda}\simeq 0.0013$, is much smaller than 
the threshold for the process (\ref{URCA})~\cite{PRAKASH92}.
The remainder reactions have similar thresholds as the ordinary 
Urca and, being less effective, contribute a fraction of the 
total emissivity. The pair bremsstrahlung processes (\ref{BREMS})
on hyperons need to conserve the strangeness and are forbidden 
at the one-body level.  The modified Urca processes 
involving hyperons can be written as~\cite{MAXWELL87}
\bea\label{HYP_URCA}
B_1+B_2 \to B_3+B_2 +e +\anu,
\eea
with the baryons $B_i$ $(i = 1,2,3)$ chosen consistent
with charge conservation; the strangeness is either 
conserved or changed by  amount $\Delta S = 1$. The inverse 
of (\ref{HYP_URCA}) yields the same result as the direct reaction. 
The modified bremsstrahlung process can be written as~\cite{MAXWELL87}
\be \label{HYP_BREMS}
B_1+B_2 \to B_3+B_4 +\nu +\anu,
\ee
where the weak interaction vertex, written symbolically as 
$B\to B+\nu+\anu$, involves arbitrary octet baryons. (Note that 
contrary to ref.~\cite{MAXWELL87}, current work finds non-vanishing 
matrix elements for the $\Lambda \to \Lambda+\nu+\anu$ transition 
within the $SU(6)$ quark model for the baryon octet~\cite{TATSUMI}). 
The estimates of the reactions (\ref{HYP_URCA}) and (\ref{HYP_BREMS})
which are based on the one-pion-exchange (OPE) potential show 
that the emissivities of hyperonic processes are small compared 
to their nucleonic counterparts, contributing at most 2/3 of the
total nucleonic emissivity. The OPE interaction adequately represents
only the long-range part of the interaction. The short range repulsive 
components of the nuclear force reduces the rate of neutrino emission 
by modified processes by factors 4-5; since the reduction  
applies to the entire baryon octet the relative ratio of the 
nucleonic to hyperonic emissivities should not be affected. The exchange
of strangeness carrying mesons, such as the $K$-meson, opens new channels
for hyperonic modified processes and this will enhance the 
contribution of hyperons to the cooling rate. 

We mentioned above that the reaction (\ref{BREMS}) is forbidden for 
quasiparticles with $\delta$-function type spectral functions (\ref{SPEC_QP}) 
by the energy and momentum conservation. This constraint is lifted 
in the case where the baryons pair~\cite{RUDERMAN78,VOSK87,KAMINKER}. 
The binding energy of Cooper pairs makes it
energetically possible to create $\nu\anu$ pairs in the inelastic processes 
of pair creation and annihilation:
\be \label{CPBF}
 B(\Delta) \to B(\Delta) + \nu + \anu ,
\ee
where $B(\Delta)$ is a quasiparticle excitation of the superfluid state.
The rate of the Cooper pair breaking and formation (CPBF) 
processes is of the order of ($10^{19}-10^{21})\times T_9^7$
depending on the baryon pairing patterns~\cite{RUDERMAN78,VOSK87,KAMINKER}.
At extreme low temperature these processes are suppressed exponentially
as ${\rm exp}(-\Delta(0)/T)$, where $\Delta(0)$ is the zero-temperature 
pairing gap. 

The neutral current one-body process (\ref{CPBF}) 
induced by the superfluidity  have their charge current 
counterparts~\cite{SEDRAKIAN05}. 
While the former vanish when the 
temperature approaches the critical temperature of superfluid phase 
transition, the emissivity of the latter process approaches the value 
of the corresponding Urca process. Put another way, the suppression 
of the Urca processes in the superfluid state is not restricted to the 
reduction of the phase space; the pair breaking can take place for 
these processes as well. In the extreme low temperature limit the 
Urca processes are suppressed exponentially as 
${\rm exp}(-\Delta_{\rm max}(0)/T)$, where $\Delta_{\rm max}(0)$ is the 
largest gap for fermions involved in the reaction. The modified processes
are suppressed by factors ${\rm exp}\{-[\Delta_B(0)+\Delta_{B'}(0)]/T\}$
where $B$ and $B'$ label the pair of the initial or final baryons. 
As in the case of the Urca process the largest gaps enter the 
suppression factor.

If pions or kaons form a Bose-Einstein condensate (BEC) the  
meson decay reactions contribute via the reactions~\cite{VOSKRESENSKY86}
\bea\label{PION_DECAY}
&&n + \langle \pi^-\rangle \to n +  e^- + \anu_e,\\
\label{KAON_DECAY}
&&n + \langle K^-\rangle \to n +  e^- + \anu_e.
\eea
The emissivities of these reactions are large compared to 
those of the baryonic processes above; for pions 
$\varepsilon_{\pi} \sim 10^{26} T_9^6$ and for kaons $\varepsilon_{K} 
\sim 10^{25} T_9^6$; thus, a distinctive property of the models 
that feature a meson condensate is the rapid cooling. Note the
kinematical differences in these reactions, since pions condense at 
finite momentum and in the $P$-wave, while kaons form a zero momentum 
condensate in the $S$-wave.

\subsection{\it Transport equations for neutrinos}

This section introduces the real-time formalism for neutrino 
transport~\cite{SEDRAKIAN99b}. 
We shall treat neutrinos as massless particles, since
on the energy scales relevant for neutron star physics the masses
of neutrinos are small. For massless neutrinos it is irrelevant whether 
neutrinos are Dirac or Majorana particles and their 
free particle Lagrangians are identical. The interaction Lagrangian 
for the charge current is 
\be
 {\cal L}_W = \frac{g_{W}}{\sqrt{2}}
\sum_a \left( L^-_{a\mu} W^{-\mu} +  L^+_{a\mu} W^{+\mu} \right), 
\quad L^-_{a\mu} = \bar\psi_{La}^{\nu}\gamma_{\mu} \psi_{La}^l, \quad 
 L^+_{\mu a } = \bar\psi_{La}^l\gamma_{\mu}\psi_{La}^{\nu},
\ee
where $g_W$ is charge current coupling constant, $W^{\pm}$ are the 
gauge vector boson fields, $L^{\pm}_{a\mu}$ is the lepton current written 
in terms of left-handed chiral spinors of neutrino $\psi_{aL}^{\nu}$ 
and lepton $\psi_{aL}^l$, $a$ is the flavor index. The interaction Lagrangian
for the charge neutral interaction is 
\be
 {\cal L}_Z = \frac{g_{Z}}{{2}}
\sum_a \left(J_{a\mu} Z^{\mu} + J^f_{a\mu} Z^{\mu}\right), 
\quad J_{a\mu} = \bar\psi_{La}^{\nu}\gamma_{\mu} \psi_{La}^{\nu}, \quad 
 J^f_{\mu a} = \bar\psi_{a}^f\gamma_{\mu}(c_v-c_A\gamma^5)\psi_{a}^{f},
\ee
where $g_Z = g_W/\Cos \theta_W$, where $\theta_W = 28.7^o$, 
$\Sin^2\theta_W = 0.23$. These coupling constants are related to the
Fermi coupling constant by the relation $G_F = (1/\sqrt{2})(g_W/2M_W)^2 
= 1.166 \times 10^{-5}$ GeV$^{-2}$, where $M_W  \simeq 82$ GeV is the 
$W$-boson mass (the $Z$-boson mass $M_Z = M_W/\Cos \theta_W$).

The theory of neutrino radiation can be conveniently formulated in terms
of the real-time quantum neutrino transport, as discussed in 
Section~\ref{sec:sec2}.  The neutrino Greens functions are
written in the matrix form
  \be\label{MATRIX_GF}
   i {\bm S}(1,2) =i\left( \begin{array}{cc}
                              S^{c}(1,2) & S^{<}(1,2) \\
                              S^{>}(1,2) & S^{a}(1,2)
                           \end{array} \right) =
                    \left( \begin{array}{cc}
\left <T\psi(1) \bar \psi(2)\right > & -\left <\bar\psi(2)\psi(1)\right > \\
\left <\psi(1)\bar\psi(2)\right > & \left <\tilde T\psi(1)\bar\psi(2)\right >
                               \end{array} \right),
  \ee
where $\psi(1)$ are the neutrino field operators, 
$\bar\psi = \gamma^0\psi^*$, $T$ is the chronological 
time ordering operator, and $\tilde T$
is the anti-chronological time ordering operator; the indexes
$1, 2, \dots$ denote the space-time arguments.
The neutrino matrix propagator  obeys the Dyson equation
\bea\label{NU_DYSON1}
{\bm S}(1,2) & = & {\bm S}_0(1,2)+{\bm S}_0(1,3)
{\bm \Omega}(3,2){\bm S}(2,1) ,
\eea
where ${\bm S}_0(1,2)$ is the free neutrino
propagator and  ${\bm S}_0^{-1}(1,2) {\bm S}_0(1,2) = 
\sigma_z \delta(1-2)$, $\sigma_z$ is the third component of the 
(vector) Pauli matrix, ${\bm \Omega}$  is the neutrino proper 
self-energy and we assume  integration (summation) over 
the repeated variables. The self-energy ${\bm \Omega}$ is 
a $2\times 2$ matrix with  elements defined on the contour. 
 
The set of the four Green's functions above can be supplemented 
by the retarded and advanced Green's functions which are defined, 
in analogy to (\ref{GR}) and  (\ref{GA}), as
\bea
i S^R(1,2)=\theta(t_1-t_2)
     \langle \left\{ \psi(1),
       \overline{\psi}(2) \right\}\rangle ,\quad
   i S^A(1,2)=-\theta(t_2-t_1)
     \langle \left\{ \psi(1),
       \overline{\psi}(2) \right\} \rangle ,
\eea
where $\{\, ,\, \}$ stands for an anti-commutator.
The retarded and advanced Green's functions obey integral
equations in the quasiclassical limit.
By applying  the Langreth-Wilkins rules (\ref{LW1})
and (\ref{LW2}) to the Dyson equation (\ref{NU_DYSON1})
we find the transport equation for the off-diagonal 
elements of the matrix Green's function
 \bea
&&  \left[ \not\!\partial_{3} -\Re\, \Omega^R(1,3),S^{>,<}(3,2)\right]
  -\left[\Re\, S^R(1,3),\Omega^{>,<}(3,2)\right]\nonumber \\
 &&\hspace{3cm}  = \frac{1}{2}\left\{S^{>,<}(1,3),\Omega^{>,<}(3,2)\right\}
   +\frac{1}{2}\left\{\Omega^{>,<}(1,3),S^{>,<}(3,2)\right\},
\label{DYSON_OFF}
\eea
where  $[\, ,\,]$ stands for commutator. In arriving at Eq.~(\ref{DYSON_OFF}) 
we assumed the existence of the Lehmann representation for the neutrino 
propagators; as a result we find $\Re~ S^R = \Re~ S^A\equiv \Re~ S$ 
and $\Re~ \Omega^R = \Re~ \Omega^A\equiv \Re~ \Omega$.

The neutrino dynamics can be treated semiclassically, by separating 
the slowly varying center-of-mass coordinates from the rapidly 
varying relative coordinates. Carrying out a Fourier transform with
respect to the relative coordinates and keeping the first-order
gradients in the slow variable we arrive at a quasiclassical neutrino
transport equation
\bea
 &&  i\left\{\Re S^{-1}(q,x),S^{>,<}(q,x)\right\}_{P.B.}
  +i\left\{\Re\, S(q,x),\Omega^{>,<}(q,x)\right\}_{P.B.}  \nonumber\\
  &&\hspace{3cm}  =  S^{>,<}(q,x)\Omega^{>,<}(q,x)
   +\Omega^{>,<}(q,x)S^{>,<}(q,x),
\label{TRANS_EQ}
\eea
where $q\equiv (\vecq , q_0)$ and  $x$ are the neutrino four momentum 
and the center-of-mass space-time coordinate, respectively,
$\{\dots\}_{P.B.}$ is the four-dimensional Poisson bracket 
[cf. Eq. (\ref{KIN4})]. To eliminate the second Poisson bracket on the 
l. h. side of Eq. (\ref{TRANS_EQ}) we carry out a decomposition analogous to 
(\ref{ASMALL}) with respect to the small neutrino damping:
$S^{>,<}(q,x)=S_0^{>,<}(q,x)+S_{1}^{>,<}(q,x)$, where $S_0^{>,<}(q,x)$
is the  leading (quasi-particle) and $S_{1}^{>,<}(q,x)$ is the 
next-to-leading order term.
The quasiparticle part of the transport equation is then written as
~\cite{SEDRAKIAN99b}
\bea\label{QPA_TRANS}
  i\left\{\Re S^{-1}(q,x),S_0^{>,<}(q,x)\right\}_{P.B.}
   =S^{>,<}(q,x)\Omega^{>,<}(q,x)+\Omega^{>,<}(q,x)S^{>,<}(q,x)
\eea
and describes the evolution of the distribution
function (Wigner function) of  on-shell
excitations. The l. h. side of Eq.~(\ref{QPA_TRANS}) corresponds 
to the drift term of the Boltzmann equation, 
while the r. h. side corresponds to
the collision integral, where the self-energies
 $\Omega^{>,<}(q,x)$ are interpreted as 
the collision rates. The advantage of this form
of the (generalized) collision integral is that it admits systematic
approximations in terms of the Feynman perturbation theory. The remainder
part of the transport equation
\bea
 i\left\{\Re S^{-1}(q,x),S_{1}^{>,<}(q,x)\right\}_{P.B.}
  +i\left\{\Re\, S(q,x),\Omega^{>,<}(q,x)\right\}_{P.B.}  =  0,
\eea
relates the off-mass-shell part of the
neutrino propagator to the self-energies in a form of a local
functional which depends on the local (anti-)neutrino particle distribution
function and their coupling to matter.

\subsubsection{\it On-shell neutrino approximation}

The on-mass-shell neutrino propagator is related to the single-time
distribution functions (Wigner functions) of neutrinos and anti-neutrinos,
$f_{\nu}(q,x)$ and $f_{\bar\nu}(q,x)$, via the ansatz
\bea
S_0^<(q,x)
&=& \frac{i\pi\sla q}{\omnu(\vecq)}
    \Big[ \delta\left(q_0-\omnu(\vecq)\right)f_{\nu}(q, x)
-\delta\left(q_0+\omnu(\vecq)\right) 
\left(1-f_{\bar \nu}(-q,x)\right) \Big],
\eea
where
$\omnu(\vecq)=\vert q\vert$ is the on-mass-shell
neutrino/anti-neutrino energy.  Note that the ansatz
includes {\it simultaneously} the neutrino particle states and
anti-neutrino hole states, which propagate in, say, positive time
direction. Similarly, the on-shell propagator
\bea
S_0^>(q,x)
&=& - \frac{i\pi\sla q}{\omnu(\vecq)}
    \Big[ \delta\left(q_0-\omnu(\vecq)\right)
    \left(1-f_{\nu}(q,x)\right)
-\delta\left(q_0+\omnu(\vecq)\right)f_{\bar\nu} (-q,x)\Big],
\eea
corresponds to the states propagating in the reversed time
direction and, hence,
includes the anti-neutrino particle states and
neutrino hole states.

To recover the Boltzmann drift term, we take the trace on both
sides of the transport equation  (\ref{TRANS_EQ})
and integrate over the (anti-)neutrino energy $q_0$.
The single time Boltzmann equation (hereafter BE)
for neutrinos is obtained after integrating over the
positive energy range:
\bea\label{BE_NU}
& & \left[\partial_t + \vec \partial_q\,\omnu (\vecq) \vec\partial_x
\right] f_{\nu}(\vecq,x) =
\int_{0}^\infty \frac{dq_0}{2\pi} {\rm Tr} \left[\Omega^<(q,x)S_0^>(q,x)
-\Omega^>(q,x)S_0^<(q,x)\right];
\eea
a similar equation follows for the  anti-neutrinos if one integrates
in Eq. (\ref{TRANS_EQ})
over the range $[-\infty , 0]$.

The different energy integration limits select from the r. h. side of
the transport equations the  processes leading to
modifications of the distribution functions of (anti-)neutrinos.
The separation of the transport equation into neutrino and anti-neutrino
parts is arbitrary. It is motivated by the observation
that the  fundamental quantities of neutrino radiative transport, as the
energy densities or neutrino fluxes, can be obtained by taking the
appropriate moments of BEs and these quantities are not symmetric with
respect to the neutrino/anti-neutrino populations in general.

\subsubsection{\it Collision integrals}
\label{COLLISION}

\begin{figure}[tb]
\begin{center}
\epsfig{figure=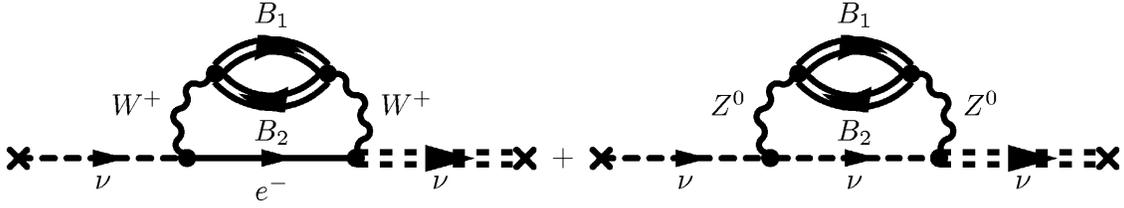,width=14.8cm,angle=0}
\begin{minipage}[t]{16.5 cm}
\caption{ The neutrino self-energies for 
charge and neutral current processes.  
The dashed and double-dashed curves correspond to the free and 
full neutrino propagators (reverting the time-direction one finds the
propagators for anti-neutrinos). The solid line is the electron propagator.
The loop is the baryon polarization tensor involving baryons $B_1$ and 
$B_2$. The wavy lines correspond to the $W^{+}$ and $Z^0$ boson 
propagators. The incoming and outgoing $\nu$ propagators are shown for
clarity and are not included in the self-energy.
}\label{fig:dyson_nu}
\end{minipage}
\end{center}
\end{figure}
The diagrams contributing to the neutrino emission rates
can be arranged in a perturbation expansion with respect
to the weak interaction. The lowest order in the weak interaction
Feynman diagrams which contribute to scattering,
emission, and absorption processes are shown in the 
Fig.~\ref{fig:dyson_nu}. The corresponding transport 
self-energies are read-off from the diagram
\bea
-i\Omega^{>,<}(q_1,x) &=& \int \frac{d^4 q}{(2\pi)^4}
\frac{d^4 q_2}{(2\pi)^4}(2\pi)^4 \delta^4(q_1 - q_2 - q)
i\Gamma_{L\, q}^{\mu}\, iS_0^{<}(q_2,x) i\Gamma_{L\, q}^{\dagger\, \lambda}
i \Pi^{>,<}_{\mu\lambda}(q,x),
\eea
where $\Pi^{>,<}_{\mu\lambda}(q)$ are the off-diagonal elements
of the matrix of the baryon polarization tensor,
$\Gamma_{L\, q}^{\mu}$ is the weak interaction vertex.
The contact interaction  can be used for the energy-momentum
transfers much smaller than the vector boson mass, $q\ll M_Z, M_W$, 
in which case the gauge boson propagators are approximated as
\be 
iD_{Z,W} = \frac{g_{\mu\nu} - q_{\mu}q_{\nu}/M_{Z,W}^2}{q^2-M_{Z,W}^2}
 \simeq - \frac{g_{\mu\nu}}{M_{Z,W}^2},
\ee
where $g_{\mu\nu}$ is the metric tensor. Let us first concentrate 
on the BE for neutrinos. Define the loss and gain terms of the 
collision integral as:
\bea
I_{\nu}^{>,<}(\vecq,x)=\int_{0}^\infty \frac{dq_0}{2\pi} {\rm Tr}
\left[\Omega^{>,<}(q,x)S_0^{>,<}(q,x)\right].
\eea
Substituting the self-energies and the propagators in the collision
integrals we find for, e.g., the gain part:
\bea\label{GAIN}
I_{\nu}^{\rm <}(\vecq_1,x)&=& -i\int_{0}^\infty \frac{dq_{10}}{2\pi}
 {\rm Tr}
\Biggl\{ \int_{-\infty}^{\infty} \frac{d^4 q}{(2\pi)^4}
\frac{d^4 q_2}{(2\pi)^4}(2\pi)^4\delta^4(q_1 - q_2 - q)
\Gamma^{\mu}_L\frac{\pi\sla q_2}{\omnu(\vecq_2)}
\Big[\delta\left(q_{02}-\omnu( \vecq_2)\right)f_{\nu}(q_2, x)\nonumber\\
&-&\delta\left(q_{02}+\omnu(\vecq_2)\right) \left(1-f_{\bar \nu}
(-q_2,x)\right) \Big]\Gamma^{\dagger\,\lambda}_L
\frac{\pi\sla q_1}{\omnu(\vecq_1)}
\delta\left(q_{10}-\omnu(\vecq_1)\right)\left(1-f_{\nu}(q_1,x)\right)
\Pi_{\mu\lambda}^{>}(q,x)\Biggr\}. \nonumber \\
\eea
The loss term is obtained by replacing in Eq. (\ref{GAIN}) the
neutrino Wigner functions by the neutrino-hole functions
$f_{\nu}(q, x) \to (1-f_{\nu}(q, x))$ and the anti-neutrino-hole
Wigner functions by the anti-neutrino  functions
$ \left(1-f_{\bar \nu}(-q,x)\right) \to f_{\bar \nu}(q,x)$.
The terms proportional $(1-f_{\nu}) f_{\nu}$ and
$(1- f_{\nu})(1-f_{\anu}) $ in the gain part of
the collision integral, $I_{\nu}^<(\vecq)$, correspond to the
neutrino scattering-in and emission  contributions, respectively.
The terms  proportional  $f_{\nu} (1-f_{\nu})$ and $f_{\nu}f_{\anu}$
in the loss part of the collision integral, $I_{\nu}^>(\vecq)$, are the
neutrino scattering-out and absorption contributions.

The loss and gain collision integrals for the anti-neutrinos can be defined
in a manner, similar to the case of neutrinos, with the energy integration
spanning the negative energy range
\bea
I_{\anu}^{>,<}(\vecq,x)=\int^{0}_{-\infty} \frac{dq_0}{2\pi} {\rm Tr}
\left[\Omega^{>,<}(q,x)S_0^{>,<}(q,x)\right].
\eea
Using the above expressions for the
self-energy and the propagators, we find, e.g.,
for the gain term:
\bea\label{GAIN2}
I_{\anu}^{<}(\vecq_1,x)&=& i\int^{0}_{-\infty}
\frac{dq_{10}}{2\pi} {\rm Tr}\Biggl\{ \int_{-\infty}^{\infty}
\frac{d^4 q}{(2\pi)^4}\frac{d^4 q_2}{(2\pi)^4}(2\pi)^4\delta^4(q_1- q_2- q)
\Gamma^{\mu}_L\frac{\pi\sla q_2}{\omnu(\vecq_2)}\Big[\delta\left(q_{02}
-\omnu(\vecq_2)\right)f_{\nu}(q_2, x)
\nonumber \\
&-&\delta\left(q_{02}
+\omnu(\vecq_2)\right)\left(1-f_{\bar \nu}(-q_2,x)\right)
\Big]\,\Gamma^{\dagger\lambda}_L\frac{\pi\sla q_1}{\omnu(\vecq_1)}
\delta\left(q_{10}+\omnu(\vecq_1)\right)f_{\anu}(-q_1,x)
\Pi^{>}_{\mu\lambda}(q,x)\Biggr\}.
\eea
The loss term is obtained by making replacements in Eq. (\ref{GAIN2})
analogous to those applied to Eq. (\ref{GAIN}).
The terms proportional $ f_{\nu}f_{\anu}$ and
$f_{\anu}(1-f_{\anu}) $ in the gain part of
the collision integral, $I_{\anu}^<(\vecq)$, then correspond to the
neutrino absorption and scattering-out  contributions.
The terms  proportional $(1-f_{\anu}) (1-f_{\nu})$ and
$(1-f_{\anu})f_{\anu}$ in the loss part of the collision integral,
$I_{\anu}^>(\vecq)$, are the neutrino 
emission and scattering-in contributions,
respectively. Note that, when the neutrinos are in thermal
equilibrium with the baryons, the collision integrals for the
scattering-in/scattering-out and for the absorption/emission cancel.
Under the conditions of detailed balance the (anti-)neutrino distribution
function reduces to the Fermi-Dirac form.

\subsubsection{\it Neutral current  processes (bremsstrahlung)}

The neutrino-pair emissivity (the power of the
energy radiated per volume unit) is obtained by multiplying the 
left-hand-sides of the neutrino and anti-neutrino
by their energies, respectively, summing the BEs, 
and integrating over a phase space element:
\bea
\epsilon_{\nu\bar\nu}&=&\frac{d}{dt}\int\!\frac{d^3q}{(2\pi)^3}
\left[f_{\nu}(\vecq) +f_{\bar\nu}(\vecq)\right]\omnu(\vecq)=
\int\!\frac{d^3q}{(2\pi)^3}\left[I_{\nu}^{<, {\rm em}}(\vecq)
-I_{\anu}^{>, {\rm em}}(\vecq)\right]\omnu(\vecq),
\eea
where in the collision integrals we kept only the terms
which correspond to the processes with the
neutrino and anti-neutrino in the final state (bremsstrahlung)
\bea
&&\int\!\frac{d^3q_1}{(2\pi)^3}
I_{\nu}^{>,<, {\rm em}}(\vecq_1)\omnu(\vecq_1)
= i \int\!\frac{d^3q_1}{(2\pi)^32 \omnu(\vecq_1)}
\frac{d^3 q_2}{(2\pi)^3 2\omnu(\vecq_2)}
\frac{d^4 q}{(2\pi)^4}(2\pi)^4 \delta^3(\vecq_1+\vecq_2-\vecq) \nonumber\\
\label{COLL_INT1}
&&\hspace{1cm}
\delta(\omnu(\vecq_1)+\omnu(\vecq_2)-q_{0})\omnu(\vecq_1)
\left[1-f_{\nu}(\omnu(\vecq_1))\right]
\left[1-f_{\anu}(\omnu(\vecq_2))\right]\Lambda^{\mu\lambda}(q_1,q_2)
\Pi_{\mu\lambda}^{>,<}(q,x),
\label{COLL_INT2}
\eea
and $\Lambda^{\mu\lambda} = {\rm Tr}\left[\gamma^{\mu}
(1 - \gamma^5)\sla q_1\gamma^{\nu}(1-\gamma^5)\sla q_2\right]$.
The collision integrals for neutrinos and anti-neutrinos can be
combined if  one uses the identities  $\Pi_{\mu\lambda}^{<}(q)
=\Pi_{\lambda\mu}^{>}(-q) = 2i g_B(q_0) {\Im}\,
\Pi_{\mu\lambda}^R(q)$; here $g_B(q_0)$ is the Bose distribution
function and $\Pi^R_{\mu\lambda}(q)$ is the retarded component
of the polarization tensor. With these modifications
the neutrino-pair bremsstrahlung emissivity 
becomes~\cite{SD99,VOSK87,SEDRAKIAN99b}
\bea\label{EMISSIVITY1}
\epsilon_{\nu\anu}&=& - 2\left( \frac{G}{2\sqrt{2}}\right)^2
\sum_f\int\!\frac{d^3q_2}{(2\pi)^32 \omnu(\vecq_2)}
\int\!\frac{d^3 q_1}{(2\pi)^3 2\omnu(\vecq_1)}
\int\!
\frac{d^4 q}{(2\pi)^4}
\nonumber\\
&&\hspace{1cm}
(2\pi)^4 \delta^3(\vecq_1 + \vecq_2 -  \vecq)
\delta(\omnu(\vecq_1)+\omnu(\vecq_2)-q_{0})\, 
\left[\omnu(\vecq_1)+\omnu(\vecq_2)\right]
\nonumber\\
&&\hspace{2cm}
 g_B(q_0)\left[1-f_{\nu}(\omnu(\vecq_1))\right]
\left[1-f_{\anu}(\omnu(\vecq_2))\right]
 \Lambda^{\mu\lambda}(q_1,q_2){\Im}\,\Pi_{\mu\lambda}^R(q).
\eea
The symbol $\Im$ refers to the imaginary part of the polarization 
tensor's resolvent and the $f$-sum is over the three neutrino 
flavors. We note that  Eq. (\ref{EMISSIVITY1}) is applicable for arbitrary
deviation from equilibrium. Therefore Eq.~(\ref{EMISSIVITY1}) 
is applicable beyond the boundaries
of the linear response theory or the $S$-matrix theory which explicitly
resort to the equilibrium properties of the system as a reference
point.

\subsubsection{\it Charged current processes ($\beta$-decay)}

In the case of charge current processes there is a single neutrino
or anti-neutrino in the initial and final states. It is sufficient
to compute, say, the direct $\beta$-decay and multiply the result by 
a factor 2 to account for the inverse process.  The anti-neutrino 
emissivity is obtained from the counterpart of Eq. (\ref{BE_NU}) for 
anti-neutrinos~\cite{SEDRAKIAN05}: 
\bea
\epsilon_{\anu}&=&\frac{d}{dt}\int\!\frac{d^3q}{(2\pi)^3}
f_{\anu}(\vecq)\omnu(\vecq).
\eea
In full analogy to the charge neutral current interactions we obtain
\bea\label{EMISSIVITY2}
\epsilon_{\anu}&=& - 2\left( \frac{\tilde G}{\sqrt{2}}\right)^2
\int\!\frac{d^3q_1}{(2\pi)^32 \ome(\vecq_1)}
\int\!\frac{d^3 q_2}{(2\pi)^3 2\omnu(\vecq_2)}
\int\! d^4 q \,\delta (\vecq_1 + \vecq_2 -  \vecq)
\nonumber\\
&&\delta(\ome+\omnu-q_{0})\, \omnu(\vecq_2)
 g_B(q_0)\left[1-f_{\anu}(\ome)\right]\left[1-f_{e}(\anu)\right]
 \Lambda^{\mu\lambda}(q_1,q_2)\Im\,\Pi^R_{\mu\lambda}(q),
\eea
where $\omega_e$ is the electron energy, $\tilde G = G_F\Cos \theta_C$, 
and $\theta_C$ is the Cabibbo angle ($\Cos \theta_C = 0.973$). 
Note that in cold matter (temperatures $T \le 5$ MeV) 
neutrinos propagate without  interactions and to a good 
approximation $f_{\nu},~f_{\anu}\ll 1$. The properties of 
matter to which neutrinos couple are encoded in the
polarization tensors according to 
Eqs. (\ref{EMISSIVITY1}) and (\ref{EMISSIVITY2}).

\subsection{\it Polarization tensors of hadronic matter}

The neutrino emission rates depend on the response of hadronic 
matter to the weak probes in the time-like domain.
The polarization tensors appearing in Eqs.~(\ref{EMISSIVITY1})
and (\ref{EMISSIVITY2}) can be viewed as self-energies of 
$W^+$ and $Z^0$ bosons. The neutrino self-energies 
in Fig.~\ref{fig:dyson_nu}  are then re-interpreted 
as the Fock contributions due to the exchange of renormalized gauge 
bosons. The picture adopted in Subsec.~\ref{COLLISION}
interprets the same diagrams as second order in weak interaction Born 
self-energies of neutrinos. Both interpretations are 
equivalent of course.

The classification of the reactions by the number of baryons participating 
in the reaction (Subsec.~\ref{sec:CLASS}) translates into expansion of the
polarization tensor in particle-hole loops. The one-body processes 
(\ref{URCA}) and (\ref{BREMS}) are described by the one-loop 
polarization tensor, 
the two-baryon processes, e.g. Eqs. (\ref{MOD_URCA}) and (\ref{MOD_BREMS})
are described by the two-loop polarization tensor, etc. This is the case
if one works with well-defined  quasiparticles described 
by the $\delta$-function spectral functions. For general forms of 
spectral functions with finite width the situation is more complex: 
the loop expansion in the particle-hole channel can still be carried 
out, however the width of the spectral function should not 
contain resummation in this channel to avoid double counting.

\begin{figure}[tb]
\begin{center}
\epsfig{figure=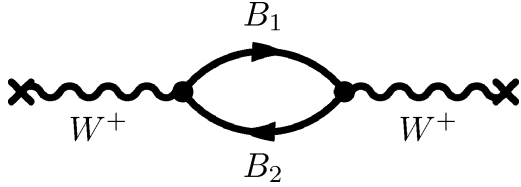,width=6.8cm,angle=0}
\begin{minipage}[t]{16.5 cm}
\caption{ The one-loop polarization tensor for charged current 
process. The wavy lines correspond to the $W^{+}$ propagators, 
the solid line to the baryonic propagators.
}\label{fig:one_loop}
\end{minipage}
\end{center}
\end{figure}

\subsubsection{\it One-loop processes}

The one-loop polarization tensor is shown in Fig.~\ref{fig:one_loop}; 
the corresponding analytical expression is 
\be 
i\Pi_{\mu\nu}^<(q) = \int\frac{d^4p}{(2\pi)^4}\frac{d^4p'}{(2\pi)^4}
\Tr \left[\Gamma_{\mu} G^<(p)\Gamma_{\nu} G^>(p')\right] (2\pi)^4
\delta^4(q+p'-p) = T_{\mu\nu} {\cal L}(q),
\ee
where the charged current weak interaction vertices are 
$\Gamma_{\mu} = \tilde G_F \gamma_{\mu}(1- g_A\gamma^5)$ with 
$g_A = 1.26$ being the axial coupling constant,  
the tensor $T_{\mu\nu} = -\tilde G^2 g_x^2$ where $g_X = 1$ 
for $\mu =\nu =0$ and  $g_X = g_A$ for $\mu =\nu =1,2,3$;  
the loop integral can be computed analytically in the 
quasiparticle limit
\bea\label{LOG}
{\cal L}(\vecq,\omega) &=&\frac{m^*_B m^*_{B'}}{2\pi\beta\vert q\vert}
{\rm ln}\Bigg\vert\frac{1+{\rm exp}\left[-\beta
\left(\xi-\mu_B\right)\right]}{1+{\rm exp}\left[-\beta\left(\xi-\mu_B\right)-\beta\omega\right]}
\Bigg\vert \equiv \frac{m^*_B m^*_{B'}}{2\pi\beta\vert q\vert} L(\omega,q),
\eea
where $m^*_B$ is the effective mass of a quasiparticle of baryon
type $B$, $\xi = \tilde p^2/2m -\mu_B$ and $\tilde p = (m^*_B/q) 
(\omega -\mu_B+\mu_{B'} -q^2/2m^*_B)$, where we assume that $m^*_B 
\simeq m^*_{B'}$. This result can be substituted in Eq. (\ref{EMISSIVITY2})
to obtain, for example, the emissivity 
of the Urca process $n \to p + e + \anu$~\cite{SEDRAKIAN05}
\bea\label{URCA_EMISSIVITY} 
\epsilon_{\anu} &=&  \epsilon_0 \int_{-\infty}^{\infty} 
dy ~g(y)~ L(y,p_{Fe})\int_{0}^{\infty} dz z^3 f_e(z-y), \quad 
\epsilon_0 =(1+3g_A^2)\frac{3\tilde G^2 m_{n}^*m_{p}^*p_{Fe}}{2\pi^5\beta^6},
\eea
where $p_{Fe}$ and $f_e$ are the Fermi-momentum and distribution function
of electrons and $y = \beta\omega$. In the zero temperature limit 
$ L(y,p_{Fe}) = y\theta (-\beta\xi)$, the integrals in 
Eq.~(\ref{URCA_EMISSIVITY}) can be performed analytically and one 
recovers the zero-temperature result of Lattimer et al.~\cite{LATTIMER_PRL}.
The zero temperature $\theta$-function can be rewritten as 
$\theta (p_{Fe} + p_{Fp} -p_{Fn})$~\cite{LATTIMER_PRL} which tells 
us that the ``triangle inequality" 
$p_{Fe} + p_{Fp} \ge p_{Fn} $ must be obeyed 
by the Fermi-momenta of the particles for the Urca process to operate. 

The emissivity of the Urca processes (\ref{URCA_EMISSIVITY}) scales as 
$\tilde G^2$, since the process is second order in the weak interactions;
its linear dependence on the effective masses of participating baryons
arises through their density of states $\nu \sim m^* p_F/\pi^2$, where 
$p_F$ stands for the Fermi momentum; the density of states of massless
electrons contributes the factor $p_{Fe}^2$ and a factor $1/p_{Fe}$ 
arises from the momentum conservation condition. The temperature dependence 
can be understood from the dimensional analysis of the reaction 
rate and arises as follows~\cite{PETHICK_RMP}: each degenerate fermion
being confined to a narrow band $\sim T$ around
the Fermi surface contributes a factor $T$,  the final state 
anti-neutrino contributes a factor $T^3$; 
an additional factor $T$ is due to the fact that we are interested in the 
energy rate and a compensating factor of $1/T$ arises due to 
the energy conservation constraint. 

Eq. (\ref{URCA_EMISSIVITY}) can be adapted to describe the Urca 
processes (\ref{HYP_URCA}) which involve hyperons~\cite{PRAKASH92}.
Among these the process $\Lambda^0\to p + e^- +\anu$ is potentially
important. The corresponding triangle inequality follows from the 
theta function $\theta (p_{Fe} + p_{Fp} -p_{F\Lambda})$. 
Since typically $p_{Fe} \sim p_{Fp} \ll p_{Fn}$  a small fraction 
of $\Lambda$'s is sufficient for the reaction to operate. Although 
it is less effective than the nucleonic Urca process, since 
it involves a change of strangeness and its matrix element 
is proportional to $\Sin^2\theta_C$,
it is still as efficient as other exotic cooling channels e.g.
the pion decay (\ref{PION_DECAY}) or kaon decay (\ref{KAON_DECAY}). 

For identical baryons the one-body bremsstrahlung process (\ref{BREMS})
vanishes, as can be seen from Eq.~(\ref{LOG}) in the limit of equal chemical 
potentials (note that in the non-relativistic kinematics spurious 
terms remain, which vanish exactly if the proper relativistic kinematics is used). However there are cases where the one-loop bremsstrahlung is possible 
because the baryons are embedded in a mean-field; 
an example are the CPBF processes which arise 
due to the pairing mean field (see 
subsection~\ref{sec:CPBF}). Yet another possibility arises
when the baryons are coupled to  external 
gauge fields. Canonical neutron stars
support magnetic fields, of the order of $10^{12}-10^{13}$ G. 
A separate class of neutron stars, know as magnetars, are believed to 
support fields that are much larger, of the order of $B\sim 10^{15}-10^{17}$
G~\cite{LAI}. The interaction of the baryon magnetic moment 
with the applied field induces a splitting in the energy of spin-up and 
spin-down baryons of the order of $\mu^{(B)} B$, where $B$ is the magnetic 
induction, $\mu^{(B)}$ is the fermion magnetic moment. For neutrons 
$\mu^{(n)} = g_n\mu_N$, where the gyromagnetic ratio $g_n = -1.913$;
for protons $\mu^{(p)} = g_p\mu_N$, $g_p = 2.793$, where
$\mu_N = e\hbar/2m_p = 3.152\times 10^{-18} $ G$^{-1}$ MeV is the nuclear
magneton. Thus, we can expect non-vanishing neutrino bremsstrahlung as
a result of the Pauli spin-paramagnetic splitting whenever 
the $2\mu^{(B)} B\sim T$~\cite{VANDALEN00}. 
For neutrons the Pauli paramagnetism is 
the only effect that affects the quasiparticle spectrum, which is written 
as (to the leading order in  $B^2/m_N$)
\begin{eqnarray}
\varepsilon_n (s)=(m_n^2+p^2-2 s m_n \mu^{(n)} B)^{1/2},
\end{eqnarray}
where the spin projections on the magnetic axis are $s = \pm 1$.
The quasiparticle spectrum of charged particles includes in addition 
to the Zeeman splitting the Landau quantization of orbits
in the plane transverse to the direction of the applied field
\be
\varepsilon_{pN}(s)=[m^2+p^2_z+(2N+1-s g_p) eB]^{1/2}, 
\ee
where $N$ labels the Landau levels. The emissivity due to the 
$n \to  n +\nu + \anu$ process is obtained as~\cite{VANDALEN00}
\bea\label{BBREMS}
\epsilon_{\nu \overline{\nu}}= \frac{G_F^2 c^2_A m_N^2}{2 (2
\pi)^5} T^7\int_{0}^{\infty} dy \frac{y^4}{e^{y}-1} 
\int_{0}^{y} dx\Bigg( \ln  \frac{e^{-\beta E_{-}}+1}
{e^{-\beta E_{-} -y}+1}- \ln  \frac{e^{-\beta E_{+}}+1}
{e^{-\beta E_{+} -y}+1} \Bigg),
\eea
where $E_{\pm} = (m_N^2+p^2_{\pm}-sg_neB)^{1/2}$ and $p_{\pm}$ are 
the upper and lower bounds on the momenta imposed by the integration 
over the angle between the particle momentum and the momentum transfer
in the process. 
\begin{figure}[tb]
\begin{center}
\epsfig{figure=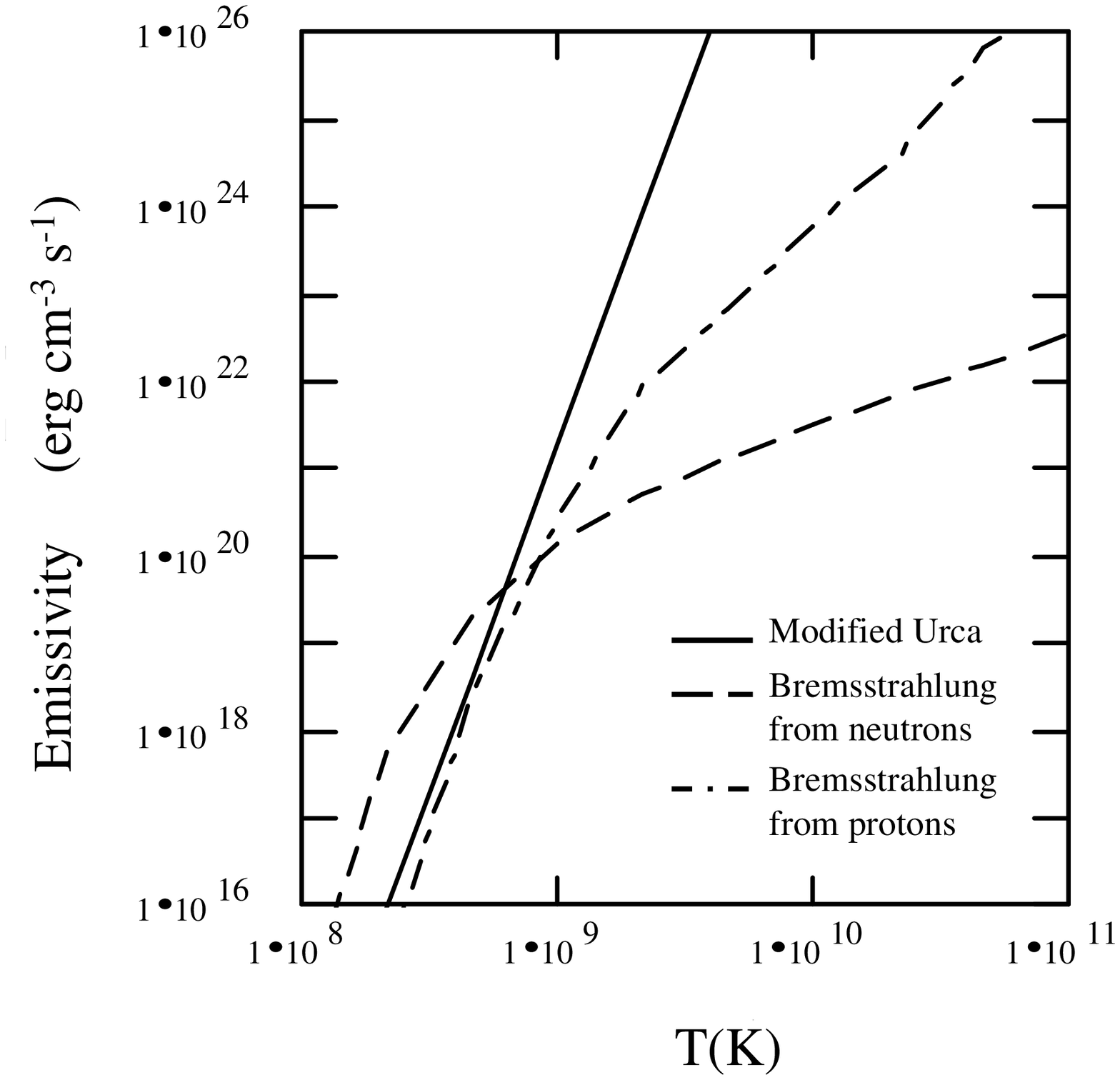,width=8.cm,angle=0}
\epsfig{figure=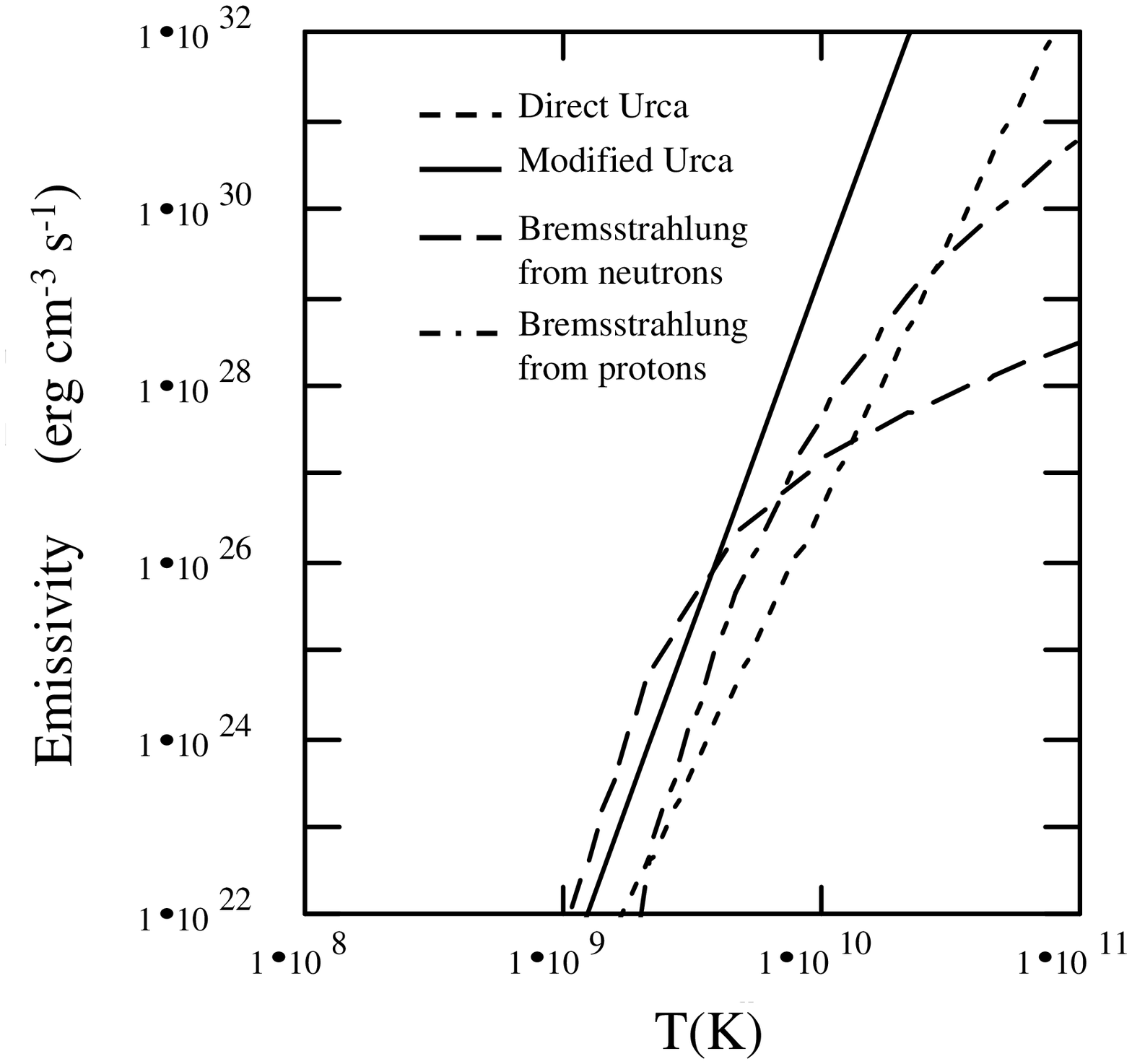,width=8.cm,angle=0}
\begin{minipage}[t]{16.5 cm}
\caption{The emissivities of various processes
vs the temperature for fields $B = 10^{16}$ ({\it left panel})  
and  $B = 10^{17}$ G ({\it right panel}).
}\label{fig:Bemission}
\end{minipage}
\end{center}
\end{figure}
The emissivity involves only axial currents because the 
process requires a spin-flip whereby a quasiparticle is transfered from 
one Fermi surface (of the spin-up population) to the other surface (spin-down
population). The dependence on the effective masses arises from the 
phase space integration which introduces a density of states per particle. 
The analysis of the temperature dependence is similar to the case of the
Urca process, with the only difference that the transfered momentum 
is of the order of $T$ (rather than $p_{Fe}$) therefore an additional 
factor $T$ appears.
Note that this bremsstrahlung process vanishes in 
the zero field limit. Several authors
considered the Urca process in strong magnetic fields, where the
effects of Landau quantization change the qualitative picture 
by removing the ``triangle inequality" 
constraint~\cite{VANDALEN00,LEINSON98,BAIKO}.  

The temperature region where the one-body neutrino-pair bremsstrahlung
is important increases with increasing magnetic field
(see Fig.~\ref{fig:Bemission}).  The pair bremsstrahlung
from neutrons  is efficient whenever $|\mu_n| B \sim T$, since 
then the energy involved in the spin-flip is of the same order 
of magnitude as the thermal smearing of the Fermi surface.
The temperature at which neutrino-pair bremsstrahlung from neutrons
becomes comparable to the competing processes roughly coincides with 
this condition. For lower temperatures the emissivity drops 
exponentially, because the energy transfer becomes larger than the 
thermal smearing. Neutrino-pair bremsstrahlung from protons
is important when  $\mu_p B \sim T$. The emissivity due to the 
protons increases faster than the emissivity due to the neutrons 
with the temperature, since the smearing of the proton transverse 
momenta provides an additional relaxation on the kinematical
constrains.
According to Fig.~\ref{fig:Bemission} the emissivity
of the modified Urca process is
larger than that of neutrino-pair bremsstrahlung
from neutrons and protons at high temperatures, mainly
due to the different temperature  dependencies of these processes
($\propto T^7$ for the one-body bremsstrahlung as compared to
$\propto T^8$ for the modified Urca).
However, for temperatures that are smaller than $\mu_p B \sim T$ the
emissivity drops just as for neutrons -  exponentially.
In the case of a superstrong magnetic field $B\ge 10^{17}$ G 
the large uncertainty in the transverse momenta of the protons 
and electrons allows the direct Urca process to occur,  and
its emissivity dominates the emissivity of any other process.

\subsubsection{\it Two-loop processes}

The nuclear interaction enters the quasiparticle loop expansion at the 
second order. To compute the emissivity we need a model
of nuclear scattering in background medium. The form of
the nuclear interaction depends, of course, on the nuclear matter model 
one works with. Below we will give a specific example of the computation 
of the neutral current bremsstrahlung process 
$n+n\to n+n +\nu+\anu$~\cite{SEDRAKIAN99b}.
The effective particle-hole interaction can be represented by pion-exchange 
at long distances and contact Fermi-liquid interaction at short 
distances~\cite{FRIMAN_MAXWELL}
\be\label{VBB}
V_{[ph]}(k) = \left(\frac{f_{\pi}}{m_{\pi}}\right)^2
\left(\vecsigma_1\cdot \veck \right)D^{c}(\veck)
\left(\vecsigma_2\cdot \veck \right)+ f_0 
+ f_1 (\vecsigma_1\cdot\vecsigma_2),
\ee
where $f_{\pi}$ is the pion decay constant, $m_{\pi}$ is the pion mass,
$D^{c}(\veck)$ is the on-shell causal pion propagator,
$f_0$ and $f_1$ are the coupling parameters of the Fermi-liquid
theory, $\vecsigma$  is the vector of the Pauli matrices. The form of the 
$ph$ interaction is suitable when the scales in the problem can be separated 
with respect to the  Compton length of the pion $\lambda_{\pi} = 
m_{\pi}^{-1} =1.4$ fm. If the system is characterized 
by scales $L\gg \lambda_{\pi}$ (e.g. is sufficiently dilute) the only 
relevant dynamical degree of freedom is pion and the rest of the nuclear
interaction can be approximated by constants. To obtain values of neutrino
emissivities that are consistent with those computed from the nuclear
$T$-matrix the $\rho$ meson exchange needs to be included explicitly 
in Eq.~(\ref{VBB}).

\begin{figure}[tb]
\begin{center}
\epsfig{figure=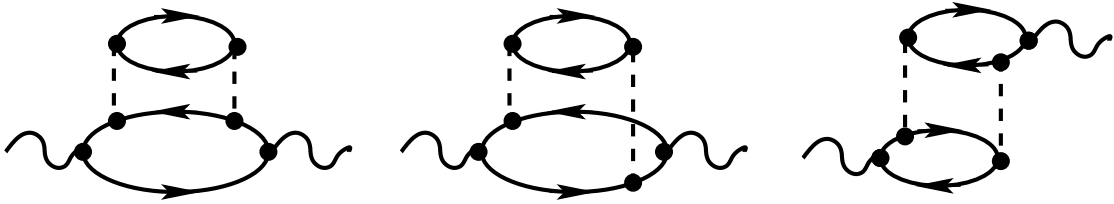,width=14.8cm,angle=0}
\begin{minipage}[t]{16.5 cm}
\caption{ The two-loop polarization tensor. The wavy lines correspond 
to the $W^{+}$  propagators, the solid lines 
to the baryonic propagators, the dashed lines to nuclear interactions.
}\label{fig:two_loop}
\end{minipage}
\end{center}
\end{figure}

The topologically non-equivalent two-loop diagrams of our theory
are shown in Fig.~\ref{fig:two_loop}.
The analytical expression, corresponding to the first 
(from left to right) diagram in Fig.~\ref{fig:two_loop}, is
\bea
i\Pi^{< , \, a}_{\mu \nu}(q) &=&
\int\!\!\prod_{i=1}^4
\left[\frac{d^4p_i}{(2\pi)^4}\right]\frac{dk}{(2\pi)^4} (2\pi)^8
\delta(q+p_4-k-p_3) \delta(k+p_2-p_1)\Tr\left[V(k) G^{<}(p_1)
V(k)G^{>}(p_2)\right]\nonumber\\
&&\Tr\Bigl[\Gamma_{\mu} G^{c}(q+p_4)V(k) D^{c}(k) G^{<}(p_3)
V(k) D^{a}(k)G^{a}(q+p_4)\Gamma_{\nu} G^{>}(p_4)\Bigr],
\label{diag_a}
\eea
where $V(k)$ is the strong interaction vertex, which can be read-off
from Eq. (\ref{VBB}). The contribution of this diagram is readily 
recognized as a {\it propagator dressing} in the $ph$ channel by 
a self-energy which involves an excitation of particle-hole 
collective mode. The second diagram in Fig.~\ref{fig:two_loop}
corresponds to a {\it vertex correction} in the $ph$ channel by an
effective  interaction, which incorporates an intermediate the
particle-hole collective mode excitation.
The third diagram in Fig.~\ref{fig:two_loop}  may be 
interpreted as a particle-hole fluctuation. These diagrams 
for model interaction (\ref{VBB}) show the following features: 
(i) the  vector current contributions from the first 
and second diagrams  mutually
cancel; (ii) the third diagram does not contribute
because the axial-vector contribution involves traces over odd number
of $\sigma$-matrices and the vector-current contribution
is canceled by an equal and of opposite sign contribution from
the diagram which is generated  from the third diagram 
by flipping one of the loops
upside-down; (iii) all contributions due to the Fermi-liquid
interaction cancel after summing the first two diagrams.
Note that the {\it exchange} diagrams are generated from the 
direct ones by means of interchanging the outgoing propagators
in a strong vertex. 

The causal propagator in Eq. (\ref{diag_a}) 
have the following general equilibrium form  
\bea\label{GC}
G^{c}(p)
= \frac{\omega-(\varepsilon_p-\mu)}
{\left[\omega-(\varepsilon_p-\mu)\right]^2+\gamma^2(p)/4}
-\frac{i\gamma(p)/2}
{\left[\omega-(\varepsilon_p-\mu)\right]^2
+ \gamma^2(p)/4}
{\rm tanh}\left(\frac{\beta\omega}{2}\right),
\eea
where ${\rm tanh}\left({\omega}/{2}\right)\equiv [1-2f_F(\omega)]$
and $\epsilon_p = p^2/2m+\Re \Sigma(p)$. 
The arguments of the causal propagators
in the diagrams in Fig.~\ref{fig:two_loop} contain the external four momentum 
$q$ and the propagation between the strong and weak vertex it describes
is off the mass-shell. The off-shell dependence of the propagator 
can be simplified by expanding with respect to small 
parameter $ v\ll 1$ which is the characteristic velocity of 
a baryon (the velocity of light $c= 1$  and we use non-relativistic 
kinematics) 
\be\label{DENOM_ExP}
(\omega +\varepsilon_p)-\varepsilon_{\vec p+\vec q}
\simeq
\omega-\vecp\cdot\vecq/m
-q\frac{\partial}{\partial p}~ \Re\Sigma(p)
-\epsilon_{q} \simeq\omega,
\ee
to the leading order. For off-shell energies not far from the Fermi
energy the quasiparticle damping is an even function of the 
frequency $\gamma(-\omega)=\gamma(\omega)$ 
(this observation is exact in the phenomenological Fermi-liquid theory and
is confirmed by microscopic calculations). Then
\bea\label{GC2}
G^{c}(\pm\omega,\vecp) &=& \pm\frac{\omega}{\omega^2+\gamma(\omega,\vecp)^2/4}
\mp i\frac{\gamma(\omega,\vecp)/2}{\omega^2+\gamma(\omega,\vecp)^2/4}
~{\rm tanh}\left(\frac{\beta\omega}{2}\right).
\eea
Note that the acausal propagator is obtained from the above through
complex conjugation $[G^{c}(p)]^*=-G^{a}(p)$. We conclude that 
the  propagator (\ref{GC2}) is odd under the exchange of the sign
of $\omega$, a property which is important for the vector current 
conservation. We now write down the neutrino emissivity for 
the $n+n\to n+n + \nu + \anu$ process which is the sum of diagrams 
in Fig.~{\ref{fig:two_loop}}
\bea\label{RESULT2}
\epsilon_{\nu\anu} &=&  \frac{32}{5(2\pi)^9} G_F^2 g_A^2
\left( \frac{f_{\pi}}{m_{\pi}} \right)^4\,
\left(\frac{m^*}{m}\right)^4\,  p_F\, I\, T^8
= 5.5 \times 10^{19}\, I_3 \,  T_9^8 ~({\rm erg~cm^{-3}~s^{-1}})
\eea
where $T_9$ is the temperature in units of $10^9$ K, $I_3$ is the
integral $I$ in units $10^3$ defined as
\bea\label{INT1}
I &=&
\int_0^{\infty}d y\, y^6\vert G^{c}(y)\vert^2  {\cal Q}(y)
\int^{\infty}_0 dx  x^4\vert  D^{c}(x)\vert^2
\int_{-\infty}^{\infty} dz ~g(z)~g(y-z)
{L}(z, x) {L}(y-z, x).
\eea
The temperature dependence in Eq. (\ref{RESULT2}) can be understood
by comparing the modified processes to their one-body counterparts 
discussed above. The additional fermion appearing in the initial 
and final state introduces an additional factor $(T/E_F)^2$. 
Since this argument applies equally well to the modified 
Urca process, we conclude that its emissivity  scales with 
temperature as $T^8$.
The scaling of emissivity with the effective mass arises due  to 
its dependence on the density of states of the initial
and final baryons. Each strong interaction vertex introduces a 
power of the pion-nucleon coupling, therefore $\epsilon_{\nu\anu}
\propto f_{\pi}^4$~(see Fig.~{\ref{fig:two_loop}}).
The simplest approximation to the  pion propagator in Eq.~(\ref{INT1})
is to neglect the pion self-energy and approximate it as 
$
D^{c}(k) = [\veck^2+m_{\pi}^2]^{-1}.
$
The free-space approximation should be valid in the vicinity
of the nuclear saturation density and below. The softening
of the one-pion exchange (a precursor of the pion-condensation)
increases the neutrino emissivity by large factors, see for details 
ref.~\cite{VOSKRESENSKY86}. The  Pauli blocking of the final state 
neutrino  at finite temperatures is taken into account by the function
${\cal Q}(y)$; 
In the dilute (anti-)neutrino limit $\beta\mu_{\nu_f}\ll 1$
(where $\mu_{\nu_f}$ is the  chemical potential of neutrinos of
flavor $f$) ${\cal Q}(y) =1$. This is the case below the
temperatures where the neutrinos are trapped. In the low-temperature  
limit  ${\cal L}(z) = z$ and the $z$-integration
decouples from the $x$-integration. Letting 
$\gamma(\omega)\to 0$ (quasiparticle limit)
one finds that  $G^{c}(\omega)=\omega^{-2}$.
Then, the $z$ integration can be carried out analytically upon
dropping the wave-function renormalization contribution:
\be
\int_{-\infty}^{\infty}\!\! dz ~g(z)\, g(y-z)~ z ~(y-z)
 = \frac{y~(y^2+4\pi^2)}{6~(e^y-1)} .
\ee
After these manipulations Eq. (\ref{RESULT2})  reduces to 
the quasiparticle result  of ref.~\cite{FRIMAN_MAXWELL}.
It should be noted here that the OPE approximation to the 
nucleon-nucleon amplitude is not justified in dense matter
from the numerical point of view, and one should include 
other mesons to take into account for the intermediate 
range attraction and short-range repulsion. In particular 
the inclusion of the $\rho$ meson repulsion modifies the
meson propagator to~\cite{ERICSON_MATHIOT_89}
$
D^c(k) =  [\veck^2+m_{\pi}^2]^{-1} - C_{\rho}
[\veck^2+m_{\rho}^2]^{-1},
$
where $C_{\rho} = 1.67$ and $m_{\rho} \simeq 600$ MeV. 
Such a correction substantially improves upon the OPE result
and a quantitative  agreement is achieved between the 
$\pi\rho$-exchange model and full $T$-matrix 
calculations~\cite{VANDALEN03} or  low-momentum 
interactions~\cite{SCHWENK04}.

\subsubsection{\it Landau-Pomeranchuck-Migdal  effect}

In this subsection we explore the effect of the finite width
in the causal propagator (\ref{GC2}) which describes the off-shell
propagation between the weak and strong vertices. 
The off-shell propagations is characterized 
by a length (or time) scale known as the formation 
length (time) first introduced in the context of bremsstrahlung of 
charge particles passing through matter by 
Ter-Mikaelian~\cite{TERMIKA,KLEIN}.  
In our context the gauge boson energy is soft $\omega \ll E_F$ 
and can be associated with the formation length
\be
l_f = \frac{\hbar}{\omega} v_F = \tau_fv_F,
\ee
where $v_F$ is the baryon Fermi-velocity. The formation length is the 
distance that a particle covers during the emission of the gauge boson; 
if $v_F$ is large and  $\omega$ small, $l_f$ can be very long. This 
observation was the basis for the suppression calculation 
by  Landau and Pomeranchuk~\cite{LP} and Migdal~\cite{MIGDAL56}
in the context of high-energy electrons radiating photons. 
\begin{figure}[tb] 
\begin{center}
\epsfig{figure=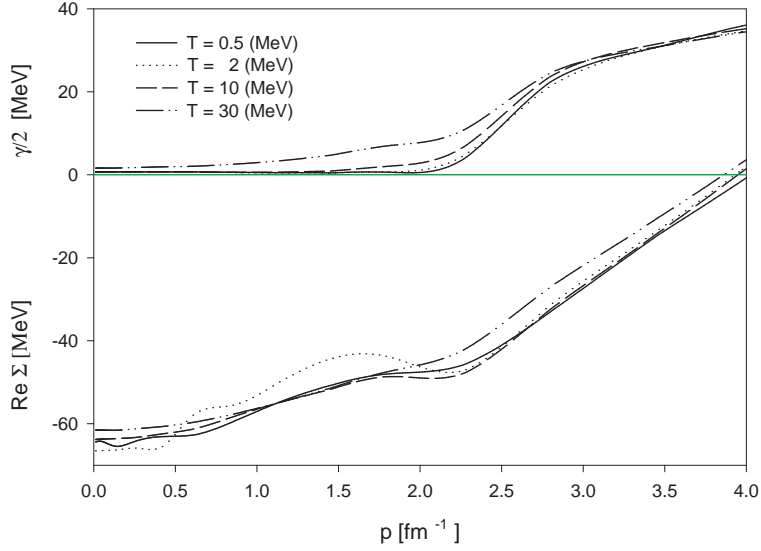,width=12.cm,angle=0}
\begin{minipage}[t]{16.5 cm}
\caption{The on mass-shell damping $\gamma$ ({\it upper panel}) and the real 
part of the self-energy ({\it lower panel}) as a function of the particle 
momentum for neutron matter at density 
$\rho_0 = 0.16$ fm$^{-3}$.
}
\label{fig:SELF_ENERGIES}
\end{minipage}
\end{center}
\end{figure}
There is another scale in the problem - the mean-free-path of a
quasiparticle, 
\be
l_{mfp} = \tau_{mfp}v_F = (n\sigma)^{-1}
\left(\frac{E_F}{T}\right)^2 v_F,
\ee
where $\sigma \sim 40$ mb is the baryon-baryon cross-section 
(mb $= 10^{-26}$ cm$^2$), $n$ is the baryon number density.
In the limit $l_{mfp} \gg l_{f}$ the radiation is from a well defined 
source - the environment has no influence on the radiation. In the
opposite limit $l_{mfp} \ll l_{f}$, the baryon-baryon interaction 
need to be included in the renormalized propagators, which are now
the fundamental degrees of freedom to treat the radiation process.
In the case where $l_{mfp} \sim l_{f}$ the radiation is suppressed 
since the baryon-baryon collisions interfere (destructively) with 
the emission process. 

In our context the Landau-Pomeranchuk-Migdal (LPM) effect is described 
by the finite width of the quasiparticles in Eq.~(\ref{GC2}). The
finite width  not only suppresses the radiation, but also regularizes
the infrared divergence of the radiation cross-section; in the 
case of the neutrino-bremsstrahlung processes this divergences
is absent even in the zero width limit because of high powers 
of $\omega$ appearing in the numerator of the emissivity.

The  width of the quasiparticle propagators can be parameterized
in terms of the reciprocal of the quasiparticle life time in the
Fermi-liquid theory:
\be\label{ZERO_SOUND}
\gamma = a T^2\left[1+\left(\frac{\omega}{2\pi T} \right)^2 \right],
\ee
where $a$ is a density dependent phenomenological parameter. 
Fig.~\ref{fig:SELF_ENERGIES} illustrates the real part of the 
self-energy and the damping in neutron matter computed within 
the finite temperature version of the BBG theory.
In this version of the theory the $K$ matrix is 
complex valued~\cite{SEDRAKIAN99b}.

\begin{figure}[tb]
\begin{center}
\epsfig{figure=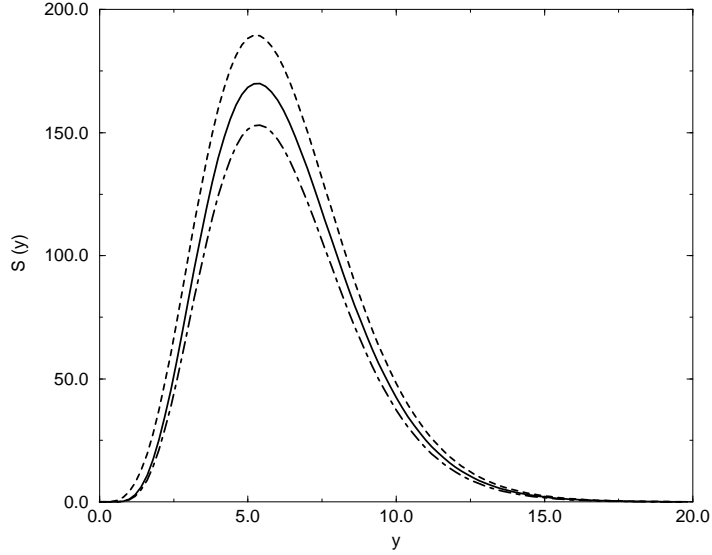,width=8.8cm,angle=-90}
\begin{minipage}[tb]{16.5 cm}
\caption{ The neutrino spectral function 
(\ref{SPECNU}) at the temperature $T=20$ MeV and
density $\rho_0 = 0.16$ fm$^{-3}$.
The dashed curve is the zero width limit, the 
dashed-dotted curve includes only the leading 
order in $\gamma$ contribution from the causal 
propagator, the solid curve is the full non-perturbative result.
}\label{fig:nu_spectrum}
\end{minipage}
\end{center}
\end{figure}

The emergent neutrino spectrum can be characterized by the 
spectral function  
\bea\label{SPECNU}
S(y) &=& \vert G^{c}(y)\vert^2  {\cal Q}(y)
\int^{1}_0 dx  x^4 \vert D^{c}(x)\vert^2
\int_{-\infty}^{\infty} dz ~g(z)~g(y-z){ L}(z, x) { L}(y-z, x) .
\eea
The dependence of the integral on the (dimensionless)
neutrino frequency $y = \beta\omega$ at $T = 20$ MeV
and the saturation density $\rho_0$ is shown in
Fig.~\ref{fig:nu_spectrum} for three cases: the limit  
of vanishing width ({\it dashed line}),
including the leading order in $\gamma$ 
contribution in the width [i.~e. the first 
term in Eq. (\ref{GC2})]  ({\it dashed-dotted line}) and 
the full non-perturbative result ({\it solid line}). 
The energy carried by neutrinos is of order of $\omega \sim 6T$ in 
all three cases since  the average energy carried 
by each neutrino (non-interacting relativistic massless particle) is 
$3T$.  The finite width of propagators  leads to a suppression of 
the bremsstrahlung rate. Keeping the full non-perturbative
expression for the causal propagators enhances the value of the
integral with respect to the leading order perturbative 
result. The LPM effect sets in  when
$\omega \sim \gamma$. As neutrinos are produced thermally,
the onset temperature of the LPM effect is
of the order of $\gamma$. Equation (\ref{ZERO_SOUND})
shows that the value of the parameter $a$ controls  the onset temperature
which turns out to be of the order of 5 MeV.

An alternative to the two-loop calculation outlined above is the computation
based on a one-loop polarization tensor with fully off-shell propagators, 
as suggested in the case of photoemission in ref.~\cite{KNOLL} and 
neutrino emission in ref.~\cite{SD99}. However, dressed propagators 
need to be supplemented with dress vertex functions that satisfy 
the Ward identities in self-consistent manner; the problem of dressed
vertices is discussed in ref.~\cite{MARGUERON}. Multi-pair
excitations processes, that are relevant to the description of the LPM 
effect within the Landau-Fermi liquid theory are discussed in 
Refs.~\cite{PETHICK1,PETHICK2}.

\subsubsection{\it Soft-neutrino approximation}

\begin{figure}[tb]
\begin{center}
\epsfig{figure=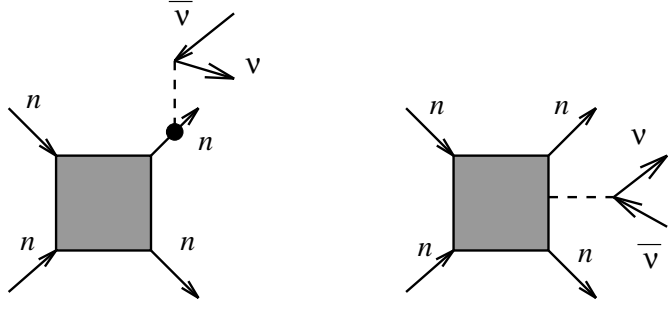,width=8.8cm,angle=0}
\begin{minipage}[t]{16.5 cm}
\caption{ Illustration of the external ({\it left graph}) and internal
({\it right graph}) contributions to the neutrino bremsstrahlung.  
The shaded block corresponds to the $T$-matrix, the dashed line 
to the weak interaction, the arrows to the participating particles. 
}\label{fig:TM}
\end{minipage}
\end{center}
\end{figure}

The {\it soft neutrino approximation} arises in the models
that are based on either the free-space or medium modified 
$T$-matrix theories~\cite{TIM02,HANHART1,HANHART2}. 
We have seen that the typical energy 
carried by neutrinos is of the order of several MeV, 
which is small on the nuclear energy scales of tens of MeV.
The neutrino bremsstrahlung process is therefore called ``soft". 
Note that this is not the case for the Urca processes, where 
the electron energy is of the same order of magnitude as the 
neutron energy. In the case of the bremsstrahlung one can 
apply the ideas underlying the Low theorem for the photon bremsstrahlung
which states that the bremsstrahlung amplitude to the leading $O(\chi^{-1})$ 
and next-to-leading $O(1)$ order in the expansion with respect to 
$\chi = \omega/E_f$, where $\omega$ and $E_F$ are the characteristic energies
of neutrinos and baryons, is determined by the non-radiative
cross-section. The weak matrix element is written as
\be \label{WEAK_M}
{\cal M}_{\mu}^a = T(p_1\,p_2';p_1-q,p_2) 
G(p_1-q) \Gamma_{\mu} +  \gamma_{\mu}G(p_1'+q)T'(p_1'+q,p_2';p_1,p_2)
+ (1\leftrightarrow 2),
\ee
where $T$ is the scattering $T$-matrix, $G(p) = m\Lambda^{+}/(p\cdot q)$ 
is the free-space Green's function with the positive energy 
projector defined as $\Lambda^+(p) = (\sla p+m)/(2m)$, 
$ \Gamma_{\mu} = (G_F/\sqrt{2})
\gamma_{\nu}(c_V-c_A)(\tau_a/2)$ is the weak interaction vertex and 
$\tau^{a}$ is the isospin operator. Next, expand the 
$T$-matrices in (\ref{WEAK_M}) around their on-shell matrix $T_0$
\be 
T_1 = T_0 -q\cdot \partial_{p_1} T_0+\dots\quad 
T_1' = T_0 -q\cdot \partial_{p_1'} T_0+\dots
\ee
The expansion of Eq.~(\ref{WEAK_M}) can be divided into the external 
and internal contributions, depending whether the neutrino 
is emitted from the external 
leg of the $T$-matrix or from the internal interaction line 
(see Fig.~\ref{fig:TM}). For the vector 
current the external term can be computed explicitly while the form of the
internal term is fixed from the requirement of the vector current conservation
$q^{\mu}{\cal M}_{\mu}^{V,a} = 0$. For the axial vector case one needs to
take into account the concept of partially conserved axial currents (PCAC)
which leads to the condition~\cite{TIM02}
\be 
q^{\mu}{\cal M}_{\mu}^{A,a} = \frac{f_{\pi}}{2}
\frac{m_{\pi}^2}{q^2+m_{\pi}^2} {\cal M}_{\pi}^a,
\ee
where ${\cal M}_{\pi}^a$ is the pion emission matrix  element.
To the leading order in the expansion with respect to the 
small parameter $\chi$, i. e. $O(\chi^{-1})$ the vector and axial 
vector matrix elements read 
\bea\label{VECTOR}
{\cal M}_{\mu}^{V,a}  &=& \frac{G_Fc_V}{2\sqrt{2}} 
\left(-T_0 \frac{p_{1\mu}}{p_1\cdot q}\tau^a  
+ \tau^a \frac{p'_{1\mu}}{p_1'\cdot q}T_0\right) + \{ 1\leftrightarrow 2\},
\\\label{AxIAL_VECTOR}
{\cal M}_{\mu}^{A,a}  &=& \frac{mG_Fc_V}{\sqrt{2}} 
\left(-T_0 \frac{\Lambda^+(p_1)}{p_1\cdot q}\gamma_{\mu}\gamma_5\tau^a  
+ \gamma_{\mu}\gamma_5\tau^a \frac{\Lambda^+(p'_1)}{p_1'\cdot q}T_0\right) 
+ \{ 1\leftrightarrow 2\}.
\eea
It can be verified that the vector current is conserved. For matrix
elements with the nucleons on the mass shell, the most general
nonrelativistic charge-independent nucleon-nucleon amplitude contains
central, spin-spin, tensor, spin-orbit, and quadratic spin-orbit terms.
For fixed isospin it can be written
\be\label{TEXP}
T = T_C + T_S \vecsigma_1\cdot \vecsigma_2 + T_TS_{12} + T_{SO}
{\vecL}\cdot {\bm S} + T_Q Q_{12}
\ee
where the tensor and quadratic spin-orbit operators are defined
as $S_{12} = 3 {\vecsigma_1\cdot \vecr}{\vecsigma_2\cdot \vecr}-
\vecsigma_1\cdot\vecsigma_2 $ and $Q_{12} = ({\vecsigma_1\cdot \vecL}
{\vecsigma_2\cdot \vecL}+ {\vecsigma_2\cdot \vecL}
{\vecsigma_1\cdot \vecL})/2$.

Consider the process $n+n\to n+n+\nu+\anu$. In the non-relativistic 
limit $p\cdot q = \vecp\cdot \vecq -m\omega \simeq m\omega$ to leading
order in $v_F/c$ and the vector current contribution (\ref{VECTOR})
vanishes. The contribution from the axial-vector current is given 
by the commutator 
\be \label{AXIAL_COMMUTATOR}
{\cal M}_{0}^{A} = 0,\quad {\bm M}^{A} = \frac{G_F}{2\sqrt{2}}\frac{g_A}{\omega}
[T^{nn},{\bm S}],
\ee
where ${\bm S} = (\vecsigma_1+\vecsigma_2)/2$, ${\bm M}^{A}$ is the 
spatial component of the vector ${\cal M}^{A}_{\mu}$, $T^{nn}$ is the
neutron-neutron scattering $T$-matrix and the axial coupling 
constant $g_A  = 1.26$. The commutator
(\ref{AXIAL_COMMUTATOR}) is non-zero only for the tensor, 
spin-orbit and quadratic spin-orbit terms in the expansion 
(\ref{TEXP}), therefore only spin-triplet nucleon-nucleon partial
waves contribute. The dominant tensor force is much stronger 
in the $np$ system and despite the fact that the proton fraction 
is small the process $n+p\to n+p+\nu+\anu$ can gain significance.
The vector and axial-vector matrix elements of this process 
in the non-relativistic and soft-neutrino approximations are
\bea
{\cal M}^V_0 = -\frac{\vecq}{\omega}\cdot {\bm M}^V, \quad 
{\bm M}^V = -\frac{G_F}{2\sqrt{2}} \frac{c_V^n-c_V^p}{\omega}
\frac{\veck}{m} T^{np},\\
{\cal M}^A_0 = 0, \quad
{\bm M}^A= \frac{G_F}{\sqrt{2}} \frac{g_A}{\omega} 
[T^{np},{\bm S}^{-}],
\eea
where $\veck = \vecp-\vecp'$, 
${\bm S}^{-} = (\vecsigma_1-\vecsigma_2)/2$, $c_V^n = -1$, 
$c_V^p = 1-4{\rm Sin}^2\theta_W$. The soft neutrino approximation 
is not valid for the modified Urca process (\ref{MOD_URCA}) as
the energy transfer in the reaction could be large, of the order
of the neutron Fermi-energy. The neutrino emissivities evaluated with
the matrix elements quoted above (which thus include the 
full $T$-matrix) are reduced by factor 4-5 compared to the
results obtained with the one-pion-exchange 
amplitudes~\cite{TIM02,HANHART1,HANHART2,VANDALEN03}, but are 
of the same magnitude if the OPE is supplemented by repulsive 
$\rho$-meson exchange.

\subsection{\it Graviton emission in Kaluza-Klein theories}

We have seen that the charged current two-loop process - the modified
Urca reaction - is dominant among these type of reactions.  There
is a distinct physical situation where only  the charge-neutral current
processes contribute - the case of 
gravitational bremsstrahlung~\cite{HANHART2,CULLEN}. This 
process arises within the models where the standard model fields
live on a four-dimensional manifold, while gravity is allowed to 
propagate in ($n+4$) dimensions, where $n$ is the number of extra 
dimensions. Such schemes provide a natural explanation of the hierarchy 
problem in particle physics; 
the gravity appears to be much weaker that the remaining three
fundamental forces because it is diluted by its extension in extra
dimensions. This picture fits into the traditional Kaluza-Klein (KK) 
theories which contain the usual four dimensional space-time manifold
plus additional compact dimensions which form an unobservable, small
manifold (which until recently
was believed to be Planck size $\sim M_P^{-1}$, where
$M_P = 1.2 \times 10^{19}$ GeV). Recent models of extra dimensional gravity 
propose that there are $n$ extra compact dimensions, all of which are 
about the same size $R$ and $R$ is much larger than the Planck scale,
possibly as large as a millimeter~\cite{HAMED1,HAN}. 
The size $R$ of the extra dimensions
is given in terms of the Planck mass $M_P$ and an effective mass $M$, 
which is taken of the order of 1 TeV, as 
\be
R^n \sim \frac{M_P^2}{M^{n+2}}.
\ee
At scales of the order of $R$ the Newtonian gravity is expected to fail.
If $n =1 $ then for $M\sim 1$ TeV one finds $R\sim 10^{10}$ km, which 
implies that there must be deviations from Newtonian gravity over 
solar system distances. However, if $n=2$ then $R \le 1$ mm. Since gravity 
has not been tested at distances smaller than millimeter,
large extra dimensions are consistent with present experimental
knowledge. An interesting consequence of this theory is that 
a $4+n$ dimensional graviton can propagate in extra dimensional space, 
while the standard model particles are confined to the four 
dimensional space. 

The size of the extra dimensions can be constrained if there is an
evidence of missing
energy in astrophysical processes such as  supernova explosions. Since 
the models of supernovae based on the standard physics explain 
the duration and energy of the neutrino pulse observed from SN1987A,
any mechanism that drains sufficient energy from the core of the supernova
will destroy the agreement. The bremsstrahlung of gravitons in the
nucleon-nucleon collision was suggested as such a mechanism~\cite{HAMED1}.
The interaction between gravitons and dilatons and the standard model 
particles is described by the  Lagrangian density 
\be
{\cal L} = -\frac{\kappa}{2}\sum_{\vec j}\left[
h^{\mu\nu,\vecj}T_{\mu\nu} +\left(\frac{2}{3(2+n)}\right)^{1/2}\phi^{\vecj}
T_{\mu}^{\nu}\right],
\ee
where $h^{\mu\nu,\vecj}$ and $\phi^{\vecj}$ are the graviton and dilaton 
fields, $\vecj$ is an $n$-dimensional vector representing the momentum of the 
mode in the extra dimensions. The differential rate at which the KK particles 
escape into extra dimensions can be related to the on-shell $T$-matrix
in neutron matter and is given by the expression~\cite{HANHART2}
\be \label{KK_RATES}
\frac{d\epsilon_{KK}}{d\omega} = \frac{8G_N}{5\pi}\frac{k}{\omega}
\left(\frac{4\overline{p^2}}{m_N}\right)^2 \Sin^2\theta_{cm} \vert T\vert^2
\left\{\begin{array}{cc}
{2}\left(1-\gamma_{\vecj}\right)^2/[9(2+n)]\\
{19}/{18}+{11}\gamma_{\vecj}^2/9 +2\gamma_{\vecj}^4/9
\end{array}\right.   ,
\ee
where the upper line corresponds to extra-dimensional gravitons, 
the lower line to dilatons, 
$\gamma_{\vecj} \equiv m_{\vecj}^2/{\omega^2}$ and $m_{\vecj} = \vecj^2/R^2$
is the effective mass of KK particles on the four dimensional brane; here
$G_N$ is Newton's constant,  $\overline{p^2} = p^{'2}+p^2$ and 
$\Cos \theta_{cm} = \hat p\cdot \hat p'$, where $\vecp$ and $\vecp'$
are the momenta of colliding neutrons, $k$ and $\omega$ are the momentum
and energy of radiated particles. The gravitational emissivity is 
obtained upon taking the phase-space integrals over the rates
(\ref{KK_RATES}) and summation over the momenta $\vecj$. 
The bounds obtained with the 
one-pion-exchange approximation and the full $T$-matrix (and an 
analog of the soft neutrino approximation) are 
\be
R < \left\{\begin{array}{ll}
7 \times 10^{-4} {\rm mm} \quad (n = 2), & 9 \times 10^{-7 }{\rm mm}
 \quad (n =3), \\
3 \times 10^{-4} {\rm mm} \quad (n = 2), & 4 \times 10^{-7}{\rm mm}
\quad (n =3),
\end{array}\right.
\ee
where the upper and lower lines differ mainly in the treatment of the nuclear
interaction: the upper line corresponds to the free space interaction between
nucleon in terms of a $T$-matrix~\cite{HANHART2}, 
the lower line to the one-pion-exchange~\cite{CULLEN} 
interaction. These bounds provide one of the most important constraints on 
the size of extra dimensions, assuming that our current understanding 
of the supernova mechanism and energetics is correct.

\subsection{\it The role of pairing correlations in neutrino 
                radiation rates}
 \label{sec:CPBF}

\begin{figure}[tb]
\begin{center}
\epsfig{figure=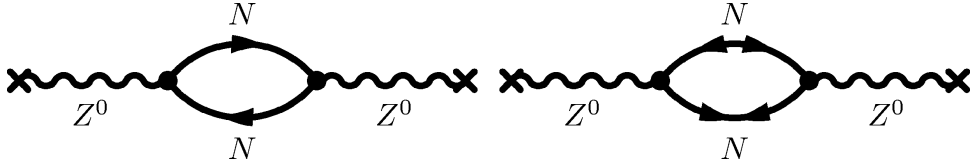,width=12.8cm,angle=0}
\begin{minipage}[t]{16.5 cm}
\caption{ The one-loop contribution to the polarization tensor
in the superfluid matter; solid lines refer to the baryon propagators, 
wavy lines to the (amputated) $Z^0$ propagator.
}\label{fig:CPBF}
\end{minipage}
\end{center}
\end{figure}

 Pairing correlations play a twofold role in the thermal evolution 
 of neutron stars. At asymptotically low temperatures they suppress
 the neutrino emission processes exponentially, because the number 
 of excitations vanishes as exp$(-\Delta/T)$, where $\Delta$ is the 
 gap in the quasiparticle spectrum. At moderate temperatures $T\le T_c$
 the pairing field lifts the constraints on the one-body (quasiparticle) 
 bremsstrahlung (\ref{BREMS}) and opens a 
new channel of neutrino 
radiation~\cite{RUDERMAN78,VOSK87}. The corresponding 
 charge-neutral current diagrams are shown in Fig.~\ref{fig:CPBF}. 
 These diagrams are associated with the following polarization tensor
 \be\label{POL_TENS_NN}
 i\Pi_{V/A}^{<} (q ) = -2g_B(q_0)\Im \Pi_{V/A}(q ) = \sum_{\sigma}\int\frac{d^4p}{(2\pi)^4}
\left[G^<(p+q)G^>(p)\mp F^<(p+q)F^{>\dagger}(p) \right],
 \ee
where $G(p)$ and $F(p)$ are the normal and anomalous propagators,
defined in Subsec.~\ref{sec:SUPT}, the $\sigma$-summation is over the spins.  
The superfluid in a neutron star can be considered as a two-component
system, which, for a fixed density and temperature, consists of paired
quasiparticles in the condensate and elementary excitations above the
condensate. Their quasi-equilibrium densities are
controlled by Cooper pair formation and pair breaking processes.
The rate of these reactions are non-exponential in the vicinity 
of $T_c$, however, they are suppressed at asymptotically low
temperatures exponentially, because of an exponential 
decrease of excitations above the condensate. These
processes can proceed with the emission of neutrino pairs via the
reactions $\{NN\}\rightarrow N + N + \nu +\bar{\nu}$ and $N +
N\rightarrow \{NN\} + \nu +\bar{\nu}$, where $\{NN\}$ denotes the
Cooper pair, $N$ an excitation. Neutrinos of all three flavors 
can be emitted in such a process. The corresponding emissivities are 
~\cite{VOSKRESENSKY_LNP}
\bea\label{NPFB}
  \epsilon_{\nu\anu} = \frac{12}{15\pi^5}G_F^2
                   C_N p_{FN}m_N^*T^7
		  \left(\frac{\Delta_N}{T}\right)^7  
		   I\left(\frac{\Delta_N}{T}\right)
		   \theta(T_{cN}-T),\quad N\in (n,p)
		    \eea
where 
$C_n = \xi_1\gamma_n^2 + \xi_2 g_A^2\gamma^2_{n,\sigma}$
and $C_p = \xi_0\gamma_p^2 + \xi_3$. 
Here $\xi_1 = 1$, $\xi_2 = 0$ if neutrons 
are paired in the $^1S_0$ state and $\xi_1 = 2/3$
$\xi_2 = 4/3$ if they are paired in the $^3P_2$ state. 
Because of their relative low density protons  pair in
the $^1S_0$ channel and there is only
a vector current contribution from protons; the factor
$\xi_3\sim 1$ takes into account the electromagnetic
correlations in the weak vertex. The functions
$\gamma_{n/p}$ and $\gamma_{n/p,\sigma}$ take into
account the correlations due to the strong force in
the vector and axial vector vertices respectively.
The temperature dependence of the emissivity
$\epsilon_{\nu\anu}\propto T^7$ is  familiar from
the analysis of the reaction $n\to n + \nu+\anu $  in magnetic fields 
[Eq. (\ref{BBREMS})] and is characteristic for 
the one-body bremsstrahlung process.  
The remaining parameters in Eq. (\ref{NPFB}) enter
through the density of state of a single fermion. The integral 
in Eq. (\ref{NPFB}) is defined as 
\bea\label{INT2}
  I(x) = \int^{\infty}_{0} \frac{(\mbox{cosh}~y)^5
  dy}{[\mbox{exp}(x~\mbox{cosh}~y)+1]^2} 
  \simeq\sqrt{\frac{\pi}{4x}}  e^{-2x}.
\eea
It can be seen that in the limit $\Delta /T \gg 1$, the rates of the CPBF
process are exponentially suppressed, as it is the case for the competing
two-nucleon processes. However, because of mild phase
space restrictions (phase space volume of a single nucleon)
these processes considerably contribute to the total 
neutrino emissivity at moderate temperatures $T\le T_c$. The
magnitude of the CPBF processes is  $\epsilon_{\nu\anu}\sim 
 10^{21} \times T_9^7$.

We now turn to the charged current weak decay 
Urca process $n\to p + e+ \anu$ in the superfluid matter and 
concentrate on the one-loop approximation~\cite{SEDRAKIAN05}. 
This process is described by the  first diagram in 
Fig.~(\ref{fig:CPBF}) with the $Z^0$ replaced by $W^+$ and 
$NN = n,p$. The second diagram does not 
contribute at one-loop.~\footnote{Note that 
ref.~\cite{SEDRAKIAN05} treats also the second diagram. While
this diagram contributes at the second order 
in the strong interaction, and in the general case where 
the loops are summed up to all orders, it is strictly 
zero at one-loop order.}
The vector and axial-vector one-loop polarization tensors 
have the form
\be\label{POL_TENS_URCA}
 i\Pi_{V/A}^{<} (q )  = \sum_{\sigma}\int\frac{d^4p}{(2\pi)^4} G^<(p+q)G^>(p),
 \ee
i.e. are identical for the vector and axial vector vertices; 
explicit evaluation of this expression leads to
 \bea \label{POL_TENS_URCA2}
 i\Pi_{V/A}^{<} (q )  &=& \sum_{\sigma}\int\frac{d^3p}{(2\pi)^3}
 \Biggl\{
\left(\frac{u_p^2u_k^2}{\omega+\ep_p-\ep_k+i\delta}
-\frac{v_p^2v_k^2}{\omega-\ep_p+\ep_k+i\delta}
\right)\left[f(\ep_p)-f(\ep_k)\right]\nonumber\\
&&\hspace{-1.5cm}+
u_p^2v_k^2\left(\frac{1}{\omega-\ep_p-\ep_k+i\delta}
-\frac{1}{\omega+\ep_p+\ep_k+i\delta}
\right)\left[f(-\ep_p)-f(\ep_k)\right]\Biggr\},
 \eea
where $u_p^2 = (1/2)\left(1+\xi_p/\ep_p\right)$, $u_p^2+v_p^2 = 1$ with 
$\ep_p = \sqrt{\xi_p^2+\Delta^2_p}$ and $\xi_p$ being
the proton single particle spectra in the superfluid and unpaired
states; the quantities with the 
index $k$ refer to the same functions for neutrons.
Inspection of the denominators of four terms contributing
in Eq.~(\ref{POL_TENS_URCA2}) shows that the first two terms
correspond to  excitations of a particle-hole pair while the
last two to excitation of particle-particle and hole-hole
pairs. The last term does not contribute to the neutrino
radiation rate $(\omega > 0)$. We identify the first two
terms as scattering ($SC$) terms, while the third
term as pair-braking ($PB$) term.
For unpaired neutrons and protons $u_{p,k} = 1$ and $v_{p,k} = 0$,
only the first term survives and one recovers 
the polarization tensor of non-superconducting matter.
Upon evaluating the phase space integrals,
the neutrino emissivity is written as
\bea\label{SUP_EMISSIVITY4}
\epsilon_{\anu}= \epsilon_0 J,\quad
J =
-\frac{1}{6}\int_{-\infty}^{\infty}\!dy~g_B(y)
\left[I^{SC}+ I^{PB}\right]\int_{0}^{\infty} dz z^3
f_e(z-y).\nonumber  ,
\eea
where $\epsilon_0$ is defined in Eq.  (\ref{URCA_EMISSIVITY}). This
result differs from its normal state counterpart by the
sum of the integrals $I^{SC}+ I^{PB}$ which reduces to the logarithm
in Eq.  (\ref{URCA_EMISSIVITY}) for $T\ge T_c$.
\begin{figure}[tb]
\begin{center}
\epsfig{figure=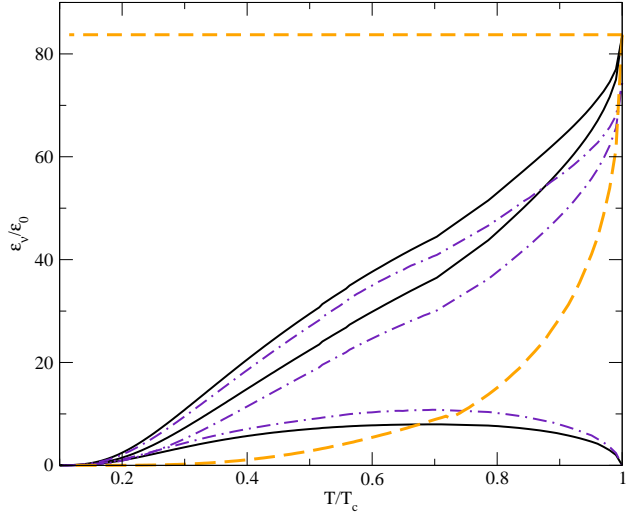,width=8.8cm,angle=-90}
\begin{minipage}[t]{16.5 cm}
\caption{ The neutrino emissivity in units
of $\epsilon_0$ versus temperature (solid lines $\Delta_n(0)=\Delta_p(0) =
0.5$ MeV, dashed-dotted lines $\Delta_n(0)=0.5,$ $\Delta_p(0) = 2$ MeV).
The scattering, pair-breaking
contributions and their sum are shown by  dashed and
dashed-dotted and solid lines. The upper short-dashed line is the
extrapolation of the rate for unpaired matter to low temperatures, the lower one corresponds to the exponential suppression
as discussed in the text.
}\label{fig:SUPURCA}
\end{minipage}
\end{center}
\end{figure}
These integrals are defined as \be
\left( \begin{array}{c}
I^{SC}\\
I^{PB}
\end{array}\right) =
\int_{-\infty}^{\infty}dx~ \left(\begin{array}{c}
u_p^2u_k^2\\
u_k^2v_p^2\\
\end{array}\right) \Xi^{SC/PB} \left[f(\pm\sqrt{x^2+w_p^2})-f(\sqrt{x^2+w_p^2}+y)\right]
\theta (1-\vert x_0^{\pm}\vert) ,
\ee
where $x = \beta \xi_p$, $w_i = \beta\Delta_i$ ($i\in n,p$),
$y = \beta\omega$
and $\Xi^{SC/PB} = 
(\omega \pm \ep_p) /\sqrt{\vert(\omega\pm\ep_p)^2-\Delta_n^2\vert}
$; the explicit form of 
the functions $x^{\pm}_0$ is given in ref.~\cite{SEDRAKIAN05}.
Fig.~\ref{fig:SUPURCA} shows the temperature dependence of the direct 
Urca neutrino emissivity in the range  $0.1\le T/T_c\le 1$.
An important feature seen in this figure is the nearly 
linear dependence of the emissivity on the temperature in the range
$0.1\le T/T_c\le 1$; the commonly assumed exponential decay
- a factor exp($-\Delta/T$) with $\Delta = {\rm max}(\Delta_n,
\Delta_p)$  -  underestimates the emissivity. 
(Similar conclusion concerning the suppression of the 
direct Urca process by pair correlations were reached in
ref.~\cite{YAKOVLEV} which treated the scattering contribution to 
the emissivity). The contribution of the pair-breaking 
processes  becomes substantial
in the low-temperature range $0.1\le T/T_c\le 0.4$ where it 
is about  half of the scattering contribution at  
$T/T_c\sim 0.1$. For unequal values of pairing gaps (e.g. $\Delta_p = 2$,
 $\Delta_n = 0.5$ MeV) the emissivity of the scattering
process is suppressed, since at a given temperature the 
phase space accessible to the excited states is reduced. 
The pair-breaking processes are almost unaffected since 
they are related to the scattering of particles in and out-of 
the condensate. 

So far our discussion was confined to the case where the nucleons
are paired in the  $^1S_0$ channel. At high densities, as is well know,
neutrons are paired in the $^3P_2$ channel. In this case
the gap function  is not isotropic in 
general and may have nodes on the Fermi-surface.
The suppression of the phase space of the Urca process was studied in 
terms of suppression factors  ${\cal R} = \epsilon_{\nu}/\epsilon_0$.
In the case where the gap function has nodes at the Fermi-surface the 
suppression is a power law~\cite{PAGE_REVIEW,YAKOVLEV}. 
Another possible implication 
of the $P$-wave superfluidity is the existence of Goldstone modes -
collective excitations - which are due to breaking of the rotational 
invariance by the anisotropic $P$-wave condensate. These Goldstone boson 
or ``angulons''($\alpha$) may couple to 
the weak neutral current~\cite{BEDAQUE}.
The process $\alpha+\alpha \to \nu+\anu$ leads to an emissivity which
scales as $\epsilon_{ang} \propto 10^{17} T_9^9 (0.15/v)^3 $ 
erg cm$^{-3}$ s$^{-1}$
where $v$ is the angulon speed. The high power of temperature in the 
process is due to the thermal nature of angulons. At moderate temperature 
$T\le T_c$ the CPBF processes are more efficient than the angulon
annihilation; at extreme low temperatures and in the cases where CPBF 
processes are suppressed exponentially, the angulon annihilation processes
may become important.

%\subsection{\it Other aspects of neutrino physics}
%label{other_CHAP3}

\section{Cooling of neutron stars}

The equation of state of dense hadronic matter and the
neutrino emissivities discussed in the previous chapters
are the key ingredients of cooling simulations of neutron stars.
The results of such simulations are combined with the experimental 
measurements of surface temperatures of neutron stars to gain 
information on properties of compact stars. In this chapter we 
review  these simulations and their comparison with 
observational surface temperatures.

Neutrino emissivities control the cooling rate of a neutron star 
during the first $10^4-10^5$ yr of their evolution. For later times 
the photon emission from the surface dominates and the heating in 
the interior can be a significant factor in maintaining the surface 
temperature above the observational limit. Depending on the dominant 
neutrino emission process in the neutrino emission era $t\le 10^5$ 
yr the cooling proceeds according to the slow (standard) or the fast 
(nonstandard) scenario~\cite{PAGE_REVIEW,VOSKRESENSKY_LNP}. 

The slow cooling scenario is based on neutrino cooling via the
modified Urca and bremsstrahlung processes, modified appropriately to
take into account the superfluidity of the star's interior. 
The fast cooling scenarios invoke ``exotic''
cooling mechanism(s), such as the pion/kaon decay processes, 
the direct Urca process on nucleons or hyperons, as well as their 
counterparts in deconfined quark phase(s). We have seen 
that the phase-space arguments are the key to understanding the 
relative importance of different processes; those leading to 
fast cooling are characterized by a one body phase space and hence 
temperature dependences are $\sim T^6$ while those responsible for 
slow cooling originate from two-body processes and their emissivities
scale with the temperature as $T^8$.

An inspection of the observational data on neutron star surface temperatures, 
which is commonly presented on a plot of photon-luminosity (or surface
temperature) vs age (see Fig.~\ref{fig:SCHAAB1} and \ref{fig:SCHAAB2}
below)
shows that the data cannot be described by a single
cooling track; there must be an effective mechanism that regulates 
the cooling rate in a manner that some of the stars cool faster than 
others. In particular the Vela pulsar 0833-45 and the radio-silent compact star
Geminga have temperatures that are far too low to be accommodated within 
a model based on slow cooling agents. It is reasonable to 
assume that the heavier
stars cool via some fast mechanism, while the lighter stars cool slowly
via the modified processes. The  fast cooling agents are effective 
above a certain density threshold, which could be either the density 
of phase transitions 
to a novel state of matter or the density at which kinematical 
constraints are lifted, as is the case for the Urca process.
Such an approach allows one to spread the cooling tracks within 
a range that can accommodate the currently available data.
While such a strategy is a plausible ansatz one should keep in 
mind that neutron star cooling is a complex,  multi-parameter problem 
and the processes which are actually responsible for fast 
(and to some extent slow) cooling of neutron stars are not known with 
certainty.

The thermal evolution is governed by the coupled system of
equations for energy 
balance~\cite{Thorne77,Schaab95,Schaab96,Schaab98,PAGE_MINIMAL,GRIG},
\bea\label{COOLING1}
  \frac{\partial(L e^{\nu})}{\partial r} &=& 
    4\pi r^2 e^\Lambda\left(-\epsilon_\nu e^{\nu}
      +h e^{\nu} -c_v\frac{\partial(Te^\nu/2)}{\partial t}
    \right)\, , \label{eq:ebal}
\eea
and thermal energy transport,
\bea\label{COOLING2}
  \frac{\partial(T e^{\nu/2})}{\partial r} &= &
    - \frac{(L e^{\nu}) e^{(\lambda-\nu)/2}}{4\pi r^2\kappa} \, ,
\label{eq:etran}
\eea
where $L$ is the luminosity, $T$ the temperature.
This system requires as a microphysical input the neutrino emissivity
$\epsilon_\nu(\rho,T)$, the heating rate $h(\rho,T,\Omega,\dot\Omega)$,
with $\Omega$ and $\dot\Omega$ being the spin frequency and its derivative,
the heat capacity $c_\mathrm{v}(\rho,T)$, and the thermal conductivity,
$\kappa(\rho,T)$, where $\rho$ is the local density. The boundary 
conditions for Eqs. (\ref{COOLING1}) and (\ref{COOLING2}) are
$L(r = 0) = 0$ and $T(r=r_e) = T_{e}$, where $r_e$ and $T_e$ are
radius and temperature of the envelope. The last condition matches
the temperature of the star to that of the envelope, since  to a good
approximation the temperature gradients and heat transport within the 
envelope are independent of the cooling of the ``thermal core'' of the 
star which extends from densities $10^{10}$ g cm$^{-3}$ to its center.
After the first $10^2-10^3$ years, the thermal core of the star is
isothermal and the cooling of the star is described by Eq. 
(\ref{COOLING1}) which can be written in terms of quantities 
which are integrals of the microscopic parameters over the volume 
of the isothermal core
\be \label{COOLING3}
C_v \frac{dT}{dt} = -L_{\nu} - L_{\gamma} + H,
\ee
where $L_{\nu}$ and $L_{\gamma}$ are the neutrino and photon luminosities, 
$H$ is the heating rate, and $C_v$ is the integrated specific heat of the 
core. The photon luminosity is given by the black-body radiation formula
$L_{\gamma} = 4\pi R^2 \sigma T_e^4$, where $\sigma$ is the Stephan-Boltzmann
constant, $R$ is the radius of the star, $T_e$ is the surface temperature.
The neutrino cooling era is characterized by the condition 
$\vert L_{\nu}\vert \gg  \vert L_{\gamma}\vert,  \vert H\vert$.
The specific heat of the star
is mainly due to degenerate fermions and 
scales with the temperature as $C_v\sim T$. If there 
is a single dominant neutrino emission process with known 
temperature dependence Eq. (\ref{COOLING3}) can be integrated; for
typical slow processes (modified Urca, etc) $L_{\nu}\propto T^8$ and 
one obtains $t\propto T^{-6}$; for rapid cooling processes (e.g. direct Urca)
$L_{\nu}\propto T^6$ and $t\propto T^{-4}$. The photon
cooling era is characterized by the condition 
$\vert L_{\nu}\vert \ll  \vert L_{\gamma}\vert,  \vert H\vert$. 
If the heating processes are ignored, the photon luminosity balances 
the thermal inertia term on the r. h. side of Eq. (\ref{COOLING3}). 
In more realistic cases the thermal inertia terms can be ignored and 
the photon luminosity is fully determined by the heating rate, 
which is a function of time.

\subsection{\it Observational data}

Only a small fraction of radio pulsars that are visible through their radio
emission have measurable photon fluxes from their surfaces or magnetospheres.
A representative sample of 27 pulsars that will be used below 
to illustrate the constraints on the cooling theories by the 
observational data is given in Table~\ref{tab:observation}. 
The effective surface temperatures are specified together 
with their $2\sigma$ error range.

\begin{table*} \centering
\small
\renewcommand{\baselinestretch}{1.0}
\begin{tabular}{lccccccl}
\hline
Pulsar            & $P$     & $\dot P$ & $\log(\tau)$ & $\log(K)$
       & $\log(T_{\rm eff}^\infty)$    & Category &  \\
                  & [ms]    & [$10^{-15}{\rm\,ss}^{-1}$] & [yr] & [s] 
       & [K]                         &       & \\
\hline
0531+21           &   33.40 &   420.96 & 2.97\dag & -13.9
       & $6.18^{+0.19}_{-0.06}$ & B    &  \\
(Crab) &&&&&&& \\
1509-58           &  150.23 &  1540.19 & 3.19 & -12.6
       & $6.11\pm 0.10$         & B    &  \\
0540-69           &   50.37 &   479.06 & 3.22 & -13.6
       & $6.77^{+0.03}_{-0.04}$ & B    &  \\
0002+62           &  241.81 &          & $\sim 4$\dag & $\sim -13$
       & $6.20^{+0.07}_{-0.27}$ & A,bb &  \\
0833-45           &   89.29 &   124.68 & $4.3\pm 0.3$\dag & -14.0
       & $6.24\pm0.03$                 & A,bb &  \\
(Vela) &&&&&    $5.88\pm 0.09$         & A,mH &  \\
1706-44           &  102.45 &    93.04 & 4.24 & -14.0
       & $6.03^{+0.06}_{-0.08}$ & B    &  \\
1823-13           &  101.45 &    74.95 & 4.33 & -14.1
       & $6.01\pm 0.02$         & C    &  \\
2334+61           &  495.24 &   191.91 & 4.61 & -13.0
       & $5.92^{+0.15}_{-0.09}$ & C    &  \\
1916+14           & 1181    &   211.8  & 4.95 & -12.6
       & $5.93$                 & D    &  \\
1951+32           &   39.53 &     5.85 & 5.03 & -15.7
       & $6.14^{+0.03}_{-0.05}$ & B    &  \\
0656+14           &  384.87 &    55.03 & 5.05 & -13.7
       & $5.98\pm0.05$ 		& A,bb &  \\
&&&&&    $5.72^{+0.06}_{-0.05}$ & A,mH &  \\
0740-28           &  167    &    16.8  & 5.20 & -14.6
       & $5.93$                 & D    &  \\
1822-09           &  769    &    52.39 & 5.37 & -13.4
       & $5.78$                 & D    &  \\
0114+58           &  101    &     5.84 & 5.44 & -15.2
       & $5.98\pm 0.03$         & C    &  \\
1259-63           &   47.76 &     2.27 & 5.52 & -16.0
       & $5.88$                 & C    &  \\
0630+18           &  237.09 &    10.97 & 5.53 & -14.6
       & $5.76^{+0.04}_{-0.08}$ & A,bb &  \\
(Geminga) &&&&&    $5.42^{+0.12}_{-0.04}$ & A,mH &  \\
1055-52           &  197.10 &     5.83 & 5.73 & -14.9
       & $5.90^{+0.06}_{-0.12}$ & A,bb &  \\
0355+54           &  156.38 &     4.39 & 5.75 & -15.2
       & $5.98 \pm 0.04$        & C    &  \\
0538+28           &  143.15 &     3.66 & 5.79 & -15.3
       & $5.83$                 & C    &  \\
1929+10           &  226.51 &     1.16 & 6.49 & -15.6
       & $5.52$                 & B    &  \\
1642-03           &  388    &     1.77 & 6.54 & -15.2
       & $6.01\pm 0.03$         & C    &  \\
0950+08           &  253.06 &     0.23 & 7.24 & -16.3
       & $4.93^{+0.07}_{-0.05}$ & B    &  \\
0031-07           &  943    &     0.40 & 7.56 & -15.4
       & $5.57$                 & D    &  \\
0751+18           &    3.47 & $7.9\times 10^{-4}$ & 7.83 & -20.6
       & $5.66$                 & C    &  \\
0218+42           &    2.32 & $8.0\times 10^{-5}$ & 8.66 & -21.7
       & $5.78$                 & C    &  \\
1957+20           &    1.60 & $1.7\times 10^{-5}$ & 9.18 & -22.6
       & $5.53$                 & C    & \\
0437-47           &    5.75 & $3.8\times 10^{-5}$ & 9.20 & -21.7
       & $5.94\pm 0.03$         & B    &  \\
\hline
\end{tabular}
\caption{Sample of observed data.
The entries are: rotation period $P$ and period derivative
$\dot P$, spin-down age $\tau = P/2\dot P$, $K = P\dot P$ and effective
(redshifted) surface temperature $T_{\rm eff}^\infty$; bb and 
mH refer to blackbody and magnetized hydrogen atmosphere fits; $\dagger$ 
refers to known true age rather than the spin-down age.}
\end{table*}
The ages of the pulsars listed in
Table~\ref{tab:observation}  are their spin-down 
ages, $\tau=P/2\dot P$, except the PSRs 0531+21 (Crab), 0833-45 (Vela) and
0002+62, for which the age is know either through historical records
(Crab pulsar) or the age of the supernova remnant they are embedded 
in. The spin-down age  assumes that
the star decelerates under the action of magnetic dipole radiation
with a constant, time independent rate.  If we write the star's spin-down 
rate as
\be
\dot\Omega(t) =  K(t) \Omega^n(t),
\ee 
where  $\Omega$ is the spin frequency and
$n$ is the breaking index, the assumption above translates to 
$K = $ const and $n = 3$. The spin-down age approximates
the true age of a pulsar within an accuracy of a factor of 3 or so. 
The sample of pulsars in Table~\ref{tab:observation} 
is divided in four  categories depending on a
number of observational and fitting features.
Four pulsars that belong to {\it category D}
have not been detected in the soft $X$-ray range.
The sensitivity of instruments sets an upper limit for
the surface temperature.  
The data from ten pulsars of {\it category C} contain too 
few photons for spectral fits. The surface temperatures for these
objects were  obtained by using the totally detected
photon flux.  %These pulsars are marked with filled triangles.
The spectra of eight pulsars of {\it category B}, which include
the Crab pulsar 0531+21, can be fitted either (i) by a power-law
spectrum or (ii) by a blackbody spectrum with high 
temperature and small effective area,  much smaller than a 
neutron star surface. Presumably, their $X$-ray
emission is dominated by magnetospheric 
emission. Therefore, 
the temperatures, determined from the spectral fits, are 
probably higher than the actual surface temperatures. 
%Pulsars of this type are marked with arrows.
Finally, there are five pulsars of {\it category A}, 
0833-45 (Vela), 0656+14, 0002+62,
0630+18 (Geminga), and 1055-52, 
whose spectrum can be fitted by a two-component fit. The 
soft blackbody component is attributed to the 
surface  emission, while the hard blackbody (or power-law) 
component is attributed to the magnetospheric emission.  
These pulsars are marked with error bars in Figs.~\ref{fig:SCHAAB1}
and \ref{fig:SCHAAB2}.

\subsection{\it Cooling simulations }
\label{Sec:cooling}

\begin{figure}[tb]
\begin{center}
\epsfig{figure=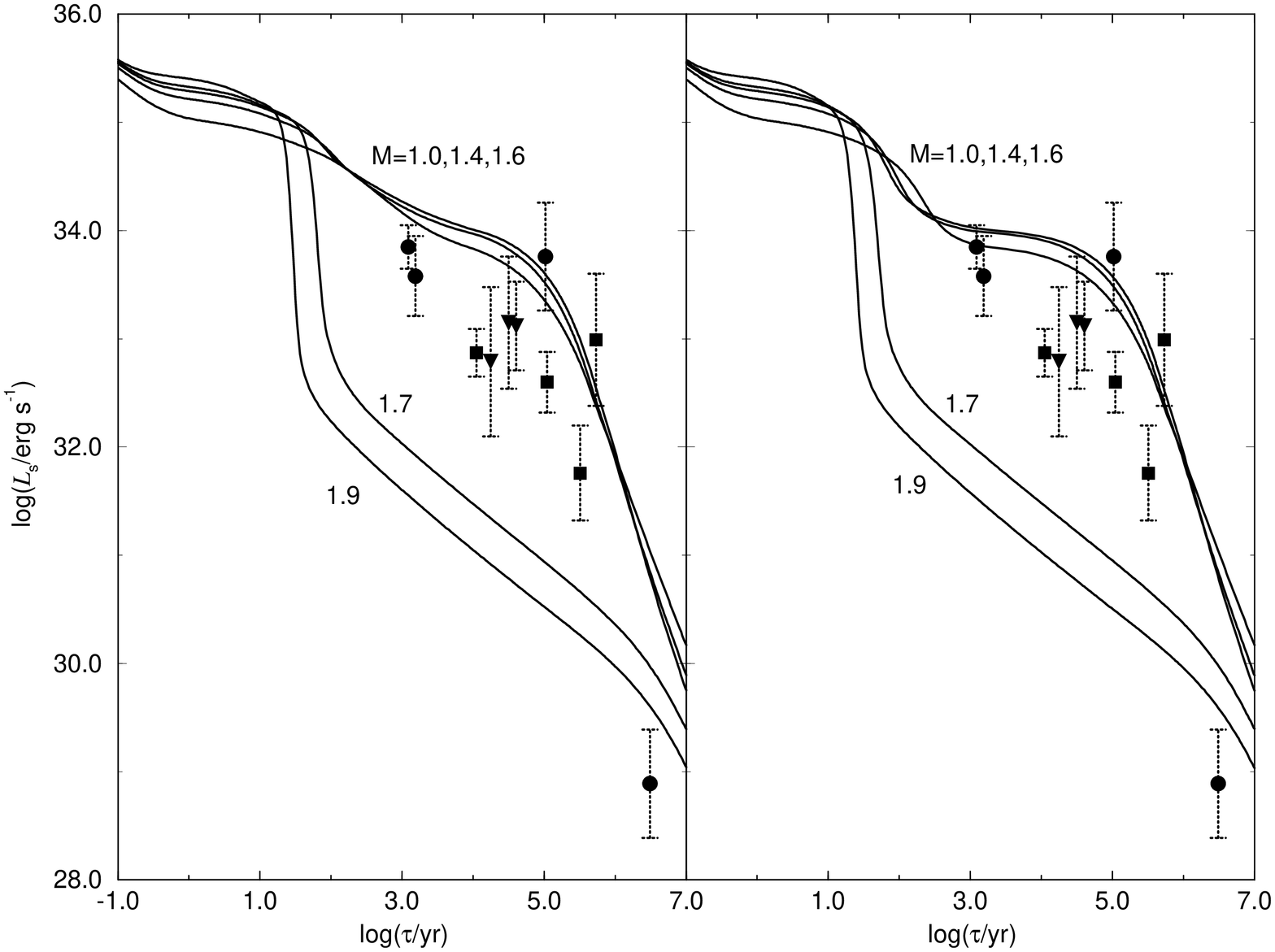,width=8.cm,height=10.0cm, angle=-90}
\begin{minipage}[t]{16.5 cm}
\caption{Dependence of surface photon luminosity on the age~\cite{Schaab96}. 
The  model includes pion condensation at high densities; 
the CPBF processes are switched off. The right panel includes  
pion mode softening effects in the modified Urca process~\cite{VOSKRESENSKY86} 
while the left panel is based on OPE result of ref.~\cite{FRIMAN_MAXWELL}.
}\label{fig:SCHAAB1}
\end{minipage}
\end{center}
\end{figure}

Figure~\ref{fig:SCHAAB1} shows the cooling of a family of neutron stars
with masses in the range 1.0 to 1.9 $M_{\odot}$ featuring 
the same microphysical ingredients~\cite{Schaab96}.
Baryonic matter is paired below the critical temperatures of the superfluid
phase transition which are of the order of $10^9$ K; a phase transition
to the pion condensed phase at the density $n_c = 3 n_0$ is assumed.
The Cooper pair-breaking/formation processes (CPBF)
are not included in the simulation. The initial phase of the cooling for
$ 0\le t\le 10^2$ yr is independent of the assumed value of the
temperature at $t=0$. At this stage, the star supports temperature gradients
throughout the thermal core. The first kink in the cooling 
curves at $t\sim 10^{2}$ marks the point where the cooling 
wave, which propagates outwards from the center of the star, 
reaches the surface. During the entire subsequent evolution
the thermal core is isothermal and the slope of the cooling curves within
the era $10^2\le t\le 10^5$ yr is determined by the dominant neutrino 
emission mechanism(s).
The second kink at $t\simeq 10^5$ yr marks the point where the transition
from the neutrino dominated cooling to photon dominated cooling occurs. 
The change in the slope is due to the difference in the temperature
dependence of the photon and neutrino cooling rates. The neutrino 
dominated cooling era is independent of the initial temperature assumed 
at the instance when the star become isothermal (unless this temperature 
is unrealistically low); in contrast, the initial conditions for the 
photon cooling dominated era strongly depends on the cooling pre-history 
of the star.
\begin{figure}[tb]
\begin{center}
\psfig{figure=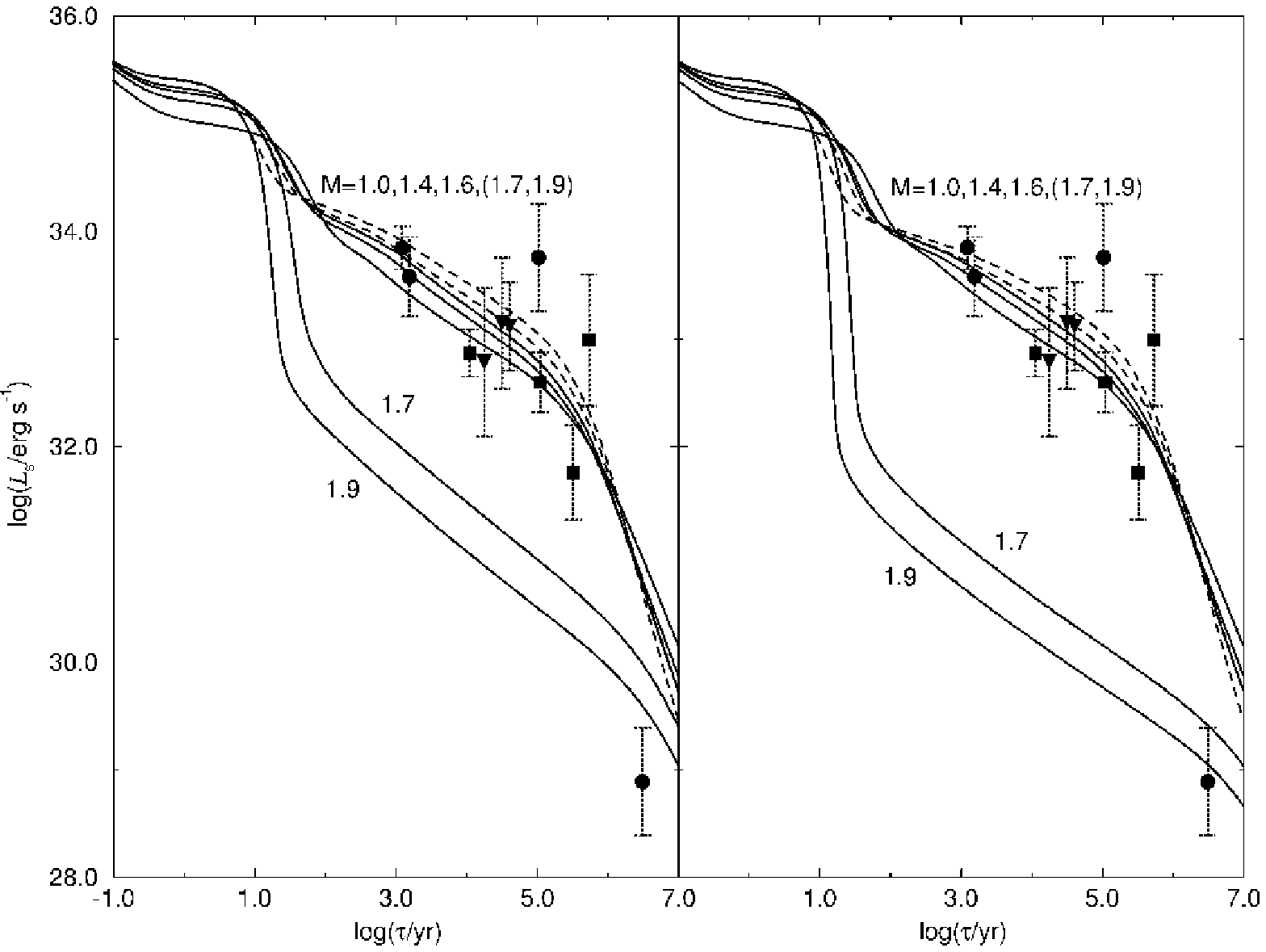,width=10.cm,angle=0}
\psfig{figure=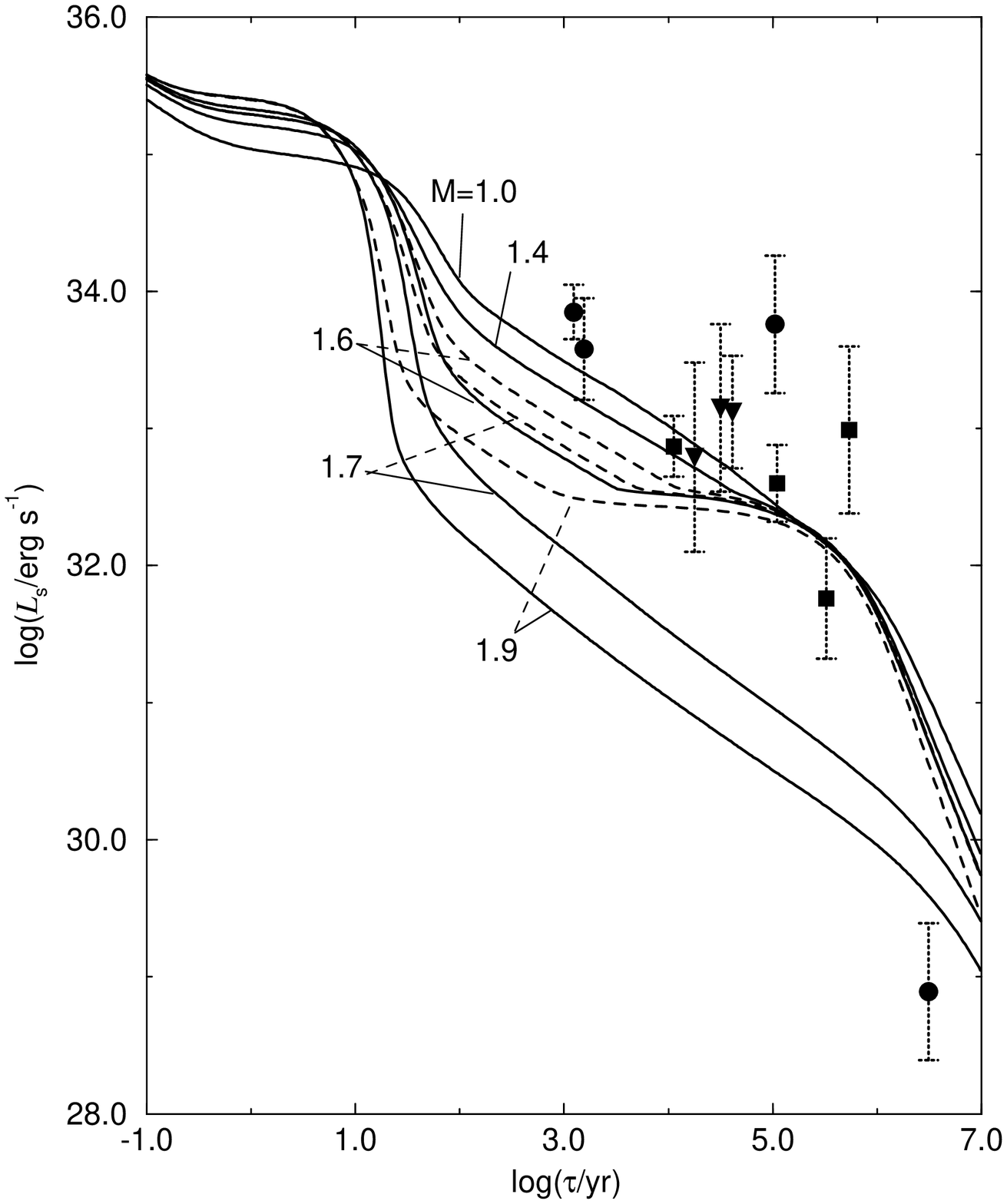,width=6.cm,angle=0}
\begin{minipage}[t]{16.5 cm}
\caption{Dependence of surface photon luminosity on the age. The CPBF 
processes are included in all graphs. From left to right: the 
model includes pion condensation, the model includes in addition 
the Urca process;  the model features neutron gap reduced 
by factor 6.~\cite{Schaab96,Schaab98}.
}\label{fig:SCHAAB2}
\end{minipage}
\end{center}
\end{figure}
The late time asymptotics of cooling in the photon dominated 
era is mainly determined by the models of the non-isothermal 
envelope which relate the temperature of the
isothermal core to the surface temperature; the 
core temperature drops across the envelope by 
roughly two orders of magnitude. The late time asymptotics of cooling
is likely to be dominated by the heating in the 
interior of the star. 

Let us now concentrate on the neutrino 
dominated cooling era. The qualitatively different
behavior of stars with masses $M > 1.6 M_{\odot} $ 
and $M < 1.6 M_{\odot} $ arises
from the fact that the former feature a pion condensate, since their central
densities are larger than the threshold density for pion condensation. This
segregation of the cooling tracks into high-temperature and low-temperature
ones is common to the cooling theories with 
slow and fast cooling agents.
An inspection of the observational data, shows that the neutrino dominated era
cannot be fitted by a single track and a mechanism is needed to provide
a smooth crossover from slow to rapid cooling.  One possibility 
is the  ``threshold mechanism"  in which some of the (low mass) 
stars have central densities below a threshold for a fast processes, 
and thus cool slow, while others, more massive, above 
this threshold and thus cool fast. Such mechanism is unsatisfactory, 
since one will need an extreme fine tuning of masses of 
neutron stars to accommodate the data (see e.~g. Fig.~\ref{fig:SCHAAB1}).
It should be noted that any phase transition 
(as, e.g., pion or kaon condensation)
or kinematically constrained process (such as the direct Urca process) will
have precursors, where the fluctuations below 
the transition density will transform
the sharp transition into a smooth crossover. The effects of such
a precursor for the case of pion condensation, the so called
pion mode softening, and its effect on cooling is studied in
refs.~\cite{Schaab96,GRIG}.

A more interesting (and realistic) mechanism of the crossover from 
slow to fast cooling
is offered by the CPBF processes. The variations
in the cooling rates introduced by these processes arise due to the fact
that the density profiles of pairing gaps for baryons map differently on the
density profiles of light and massive stars. This effect is demonstrated in Fig.~\ref{fig:SCHAAB2} where the models presented in Fig.~\ref{fig:SCHAAB1} are supplemented by the CPBF processes. The leftmost panel shows the effect of
including the CPBF process in the presence of pion condensate, while the next
plot to the right shows the effect of adding the direct Urca process. While
the segregation between the high and low mass objects remain, the cooling
tracks of low mass objects are now spread in a certain range.

A wider range of surface temperatures can be covered upon suppressing
(artificially) the neutron $^3P_2$ gaps by a factor of a few. In this case,
the matter is unpaired at high-densities and the star cools  
as partially superfluid object. One may conclude that not only 
the magnitudes of the gaps are important, but also the density profiles 
over which they are spread~\cite{PAGE_REVIEW,PETHICK,PAGE_MINIMAL,GRIG}.

Cooling simulations that do not include the heating processes
in the interiors of compact stars fail to account for the highest temperatures
in the sample of $X$-ray emitting pulsars. It is very likely that the late
time evolution of compact stars including the entire photon cooling era
and partially the neutrino cooling era are significantly affected by
the heating processes in the star's interior. The heating processes can
be roughly divided into three categories: (i) the heating due to the
frictional motion of neutron and proton vortices in the superfluids; (ii) the
heating that arises due to the local deviations of matter composition
from $\beta$-equilibrium; (iii) the heating due to mechanical processes in
the solid regions of the star (crust cracking, etc.). Describing these
processes will require a detailed account of the physics of neutron
star interiors on the mesoscopic scales and is beyond the 
scope of this article (see ref.~\cite{Schaab98} for details).

\section{Concluding remarks}

This review covered a number of topics that are relevant to the theory
of compact stars, in particular the physics of hadronic matter with
baryonic degrees of freedom and weak interaction processes involving
hadrons. It should be clear from the presentation that despite the enormous
progress achieved during the past decade, the theory is
far from being completed and a large number of exciting topics remain
to be studied in the future.

Compact stars continue to pose an enormous intellectual challenge to the
physics and astrophysics communities. Rapid progress
during the last four decades since the discovery of the first
pulsar is an evidence of the vitality of this field. Current and planned
observational programs continue the exploration
of compact objects in the electromagnetic spectrum; studies of compact
stars through gravity waves may become possible in the near future.
All this provides an excellent basis for future theoretical studies of
compact stars.

\section*{Acknowledgments}

I am grateful to my colleagues and collaborators for their 
important contributions and insight to the material included 
in this review. This research was supported through a Grant 
from the SFB 382 of the Deutsche Forschungsgemeinschaft.

\end{document}